\documentclass[
 aip,
 amsmath,amssymb,
 reprint,
]{revtex4-1}

\usepackage{graphicx}
\usepackage{dcolumn}
\usepackage{bm}
\usepackage{xcolor}
\usepackage{url}

\usepackage[utf8]{inputenc}
\usepackage[T1]{fontenc}
\usepackage{mathptmx}
\usepackage{etoolbox}
\usepackage[most]{tcolorbox}

\newcommand{\degrees}{$^\circ$}
\newcommand{\apjs}{Astrophys. J. Supp.}
\newcommand{\aap}{Astron. \& Astrophys.}

\makeatletter
\def\@email#1#2{
 \endgroup
 \patchcmd{\titleblock@produce}
  {\frontmatter@RRAPformat}
  {\frontmatter@RRAPformat{\produce@RRAP{*#1\href{mailto:#2}{#2}}}\frontmatter@RRAPformat}
  {}{}
}
\makeatother
\begin{document}

\preprint{AIP/123-QED}

\title{A Millimeter-Wave Photometric Camera for Long-Range Imaging Through Optical Obscurants Using Kinetic Inductance Detectors}

\author{J. Sayers}
 \email{Jack@caltech.edu}
 \affiliation{Department of Physics, Mathematics, and Astronomy, California Institute of Technology, Pasadena, California 91125, USA}
 
\author{D. Cunnane}
\affiliation{NASA Jet Propulsion Laboratory, California Institute of Technology, Pasadena, California 91011, USA}

\author{S. Crystian}
\affiliation{Department of Physics, Harvard University, Cambridge, MA 02138, USA}

\author{P. K. Day}
\affiliation{NASA Jet Propulsion Laboratory, California Institute of Technology, Pasadena, California 91011, USA}

\author{F. Defrance}
\affiliation{NASA Jet Propulsion Laboratory, California Institute of Technology, Pasadena, California 91011, USA}

\author{B. H. Eom}
\affiliation{NASA Jet Propulsion Laboratory, California Institute of Technology, Pasadena, California 91011, USA}

\author{J. Greenfield}
\affiliation{School of Earth and Space Exploration, Arizona State University, Tempe, Arizona 85287, USA}

\author{M. Hollister}
\affiliation{Cryogenic Design Solutions LLC, Mahopac, New York 10541, USA}

\author{B. R. Johnson}
\affiliation{Department of Astronomy, University of Virginia, Charlottesville, VA 22904, USA}

\author{H. G. LeDuc}
\affiliation{NASA Jet Propulsion Laboratory, California Institute of Technology, Pasadena, California 91011, USA}

\author{P. Mauskopf}
\affiliation{School of Earth and Space Exploration, Arizona State University, Tempe, Arizona 85287, USA}
\affiliation{Department of Physics, Arizona State University, Tempe, Arizona 85287, USA}

\author{N. McNichols}
\affiliation{Department of Physics, Mathematics, and Astronomy, California Institute of Technology, Pasadena, California 91125, USA}

\author{C. Roberson}
\affiliation{School of Earth and Space Exploration, Arizona State University, Tempe, Arizona 85287, USA}

\author{M. C. Runyan}
\affiliation{NASA Jet Propulsion Laboratory, California Institute of Technology, Pasadena, California 91011, USA}

\author{A. B. Sriram}
\affiliation{School of Electrical, Computer and Energy Engineering, Arizona State University, Tempe, Arizona 85287, USA}

\author{S. Stanton}
\affiliation{Department of Physics, Mathematics, and Astronomy, California Institute of Technology, Pasadena, California 91125, USA}
 
\author{R. C. Stephenson}
\affiliation{Department of Physics, Mathematics, and Astronomy, California Institute of Technology, Pasadena, California 91125, USA}

\author{L. C. Walters}
\affiliation{Department of Astronomy, University of Virginia, Charlottesville, VA 22904, USA}

\author{E. Weeks}
\affiliation{School of Earth and Space Exploration, Arizona State University, Tempe, Arizona 85287, USA}

\date{\today}

\begin{abstract}
Passive imaging through optical obscurants is a promising application for mm-wave sensing. We have thus developed the Superconducting Kinetic Inductance Passive Radiometer (SKIPR), a 150 GHz polarization-sensitive photometric camera optimized for terrestrial imaging using a focal plane array with 3,840 kinetic inductance detectors (KIDs). We present a full description of the instrument design, with a particular emphasis on the cryogenic system based on a Gifford-McMahon cryocooler with a two-stage Adiabatic Demagnetization Refrigerator and a dedicated 1.59 m crossed Dragone telescope with an altitude/azimuth mount. We include a detailed lab-based characterization of the KIDs, which results in a determination of their superconducting resonator parameters and optical properties. We also present \textit{in situ} measurements from the telescope, including point-spread functions and noise characterization. In sum, we find that SKIPR performs as expected, providing diffraction-limited imaging with detector noise performance set by the random arrivals of photons from the ambient background. There is minimal variation in detector characteristics over the full SKIPR focal plane array, and the overall detector yield is 92 per cent.
\end{abstract}

\maketitle

\section{\label{sec:level1}Introduction\protect}

Traditional imaging at optical wavelengths can be degraded or even prevented due to the presence of obscurants in the atmosphere \cite{Chen1975,Kokhanovsky2004,Warren2019}. Within the atmosphere, optical visibilities as low as 10~m are relatively common, severely limiting the utility of such imaging in those conditions. At radio wavelengths, transmission through these obscurants is significantly better, with little to no attenuation near the low-frequency end of the radio window at tens of MHz \cite{Liebe1985,Wentz2016}. However, diffraction effects significantly degrade angular resolution at those wavelengths, even for imaging systems with very large physical apertures. As a result, there is strong motivation to pursue imaging at the high-frequency end of the radio window, in the mm-wave band at hundreds of GHz. In this regime, meter-scale apertures can provide arcminute-scale angular resolution while still allowing for long-range imaging through obscurants that prevent the use of optical techniques \cite{Liebe1985,Pardo2001,Paine2019}. Examples of this general approach have been described elsewhere \cite{Samluk2012}.

Historically, it has been difficult to obtain mm-wave detectors with sufficient sensitivity for operation in the atmosphere where the fundamental noise limit is set by the random arrival of photons from the ambient temperature background near 300~K. As recently as the turn of the century, state of the art receivers based on either coherent detection technologies or bolometers with semi-conductors contained of order ten total detectors \cite{Jarosik2003,Crill2003}. Over the next decade, these technologies were scaled to focal planes with hundreds of detectors \cite{Bischoff2013,Yoon2006}. Since then, the introduction of superconducting detectors, both bolometric and those based on pair breaking, now allow for single focal planes with several thousand (or more) mm-wave detectors \cite{Everett2018,Westbrook2018,Gordon2020}.

\begin{figure}[t]
  \centering
  \includegraphics[width=\columnwidth]{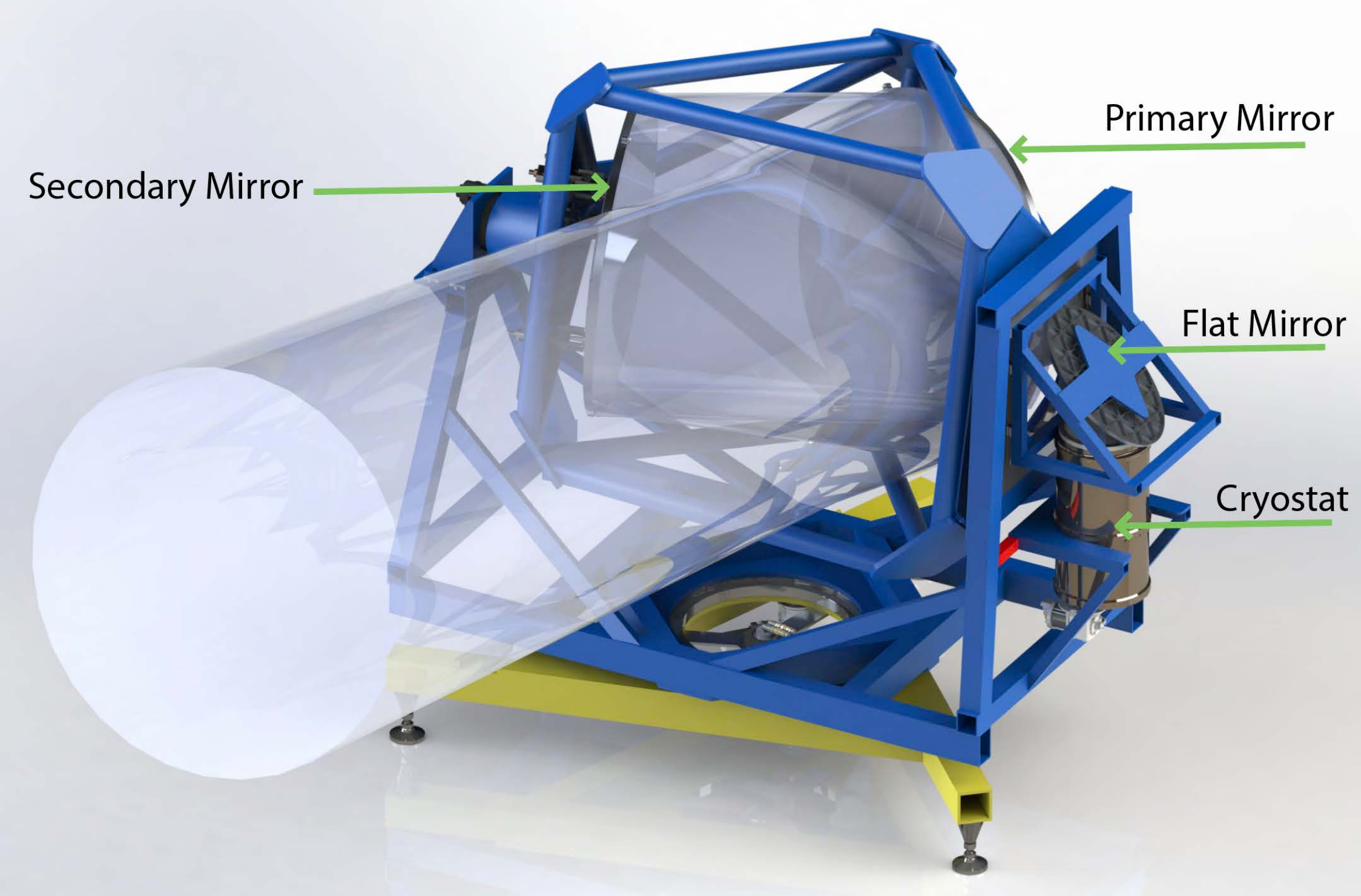}
  \caption{Isometric view of the solid model of the as-built SKIPR system. The three mirrors are labeled, along with the cryostat that houses the KID FPA at the prime focus of the telescope. A see-through cylindrical solid indicates the optical coupling path to the primary mirror, which is 1.59~m in diameter.}
  \label{fig:schematic}
\end{figure}

With the advent of these multi-kilopixel focal plane arrays (FPAs), it is now practical to obtain mm-wave data that provides sufficient spatial fidelity for imaging. Within this context, we have developed the Superconducting Kinetic Inductance Passive Radiometer (SKIPR) \cite{Sayers2020}, a 150~GHz polarization sensitive imager with an FPA populated with 3840 pair-breaking kinetic inductance detectors (KIDs). In total, the FPA contains four separate detector tiles, with each tile comprised of 960 KIDs fabricated on a single Si wafer. The FPA is paired with a custom-built 1.59~m diameter crossed Dragone telescope with an altitude/azimuth (alt/az) mount to provide 5$'$ angular resolution over a 3.5\degrees\ field of view (FOV). In total, the telescope includes two powered mirrors along with a flat tertiary mirror to fold the optical path. The KID FPA, which is housed within a cylindrical cryostat, is then placed at the prime focus of the telescope. See Figure~\ref{fig:schematic}. Currently, SKIPR is fully operational from an outdoor location within NASA's Jet Propulsion Laboratory (JPL).

In this paper, we provide a detailed description of the design of SKIPR, our lab-based characterization of the detectors, and \emph{in situ} measurements from the telescope. In Section~\ref{sec:cryogenic}, we describe the cryogenic system used to cool the detectors below the superconducting transition temperature. While not the main focus of this paper, we then detail the readout electronics used to record the detector signals in Section~\ref{sec:readout}. The telescope design and operation is then discussed in Section~\ref{sec:telescope}, and the performance of the detectors in both the lab and on the telescope is provided in Section~\ref{sec:detectors}. A summary, along with considerations for future development, is then given in Section~\ref{sec:summary}.

\begin{figure*}[t]
  \includegraphics[width=\textwidth]{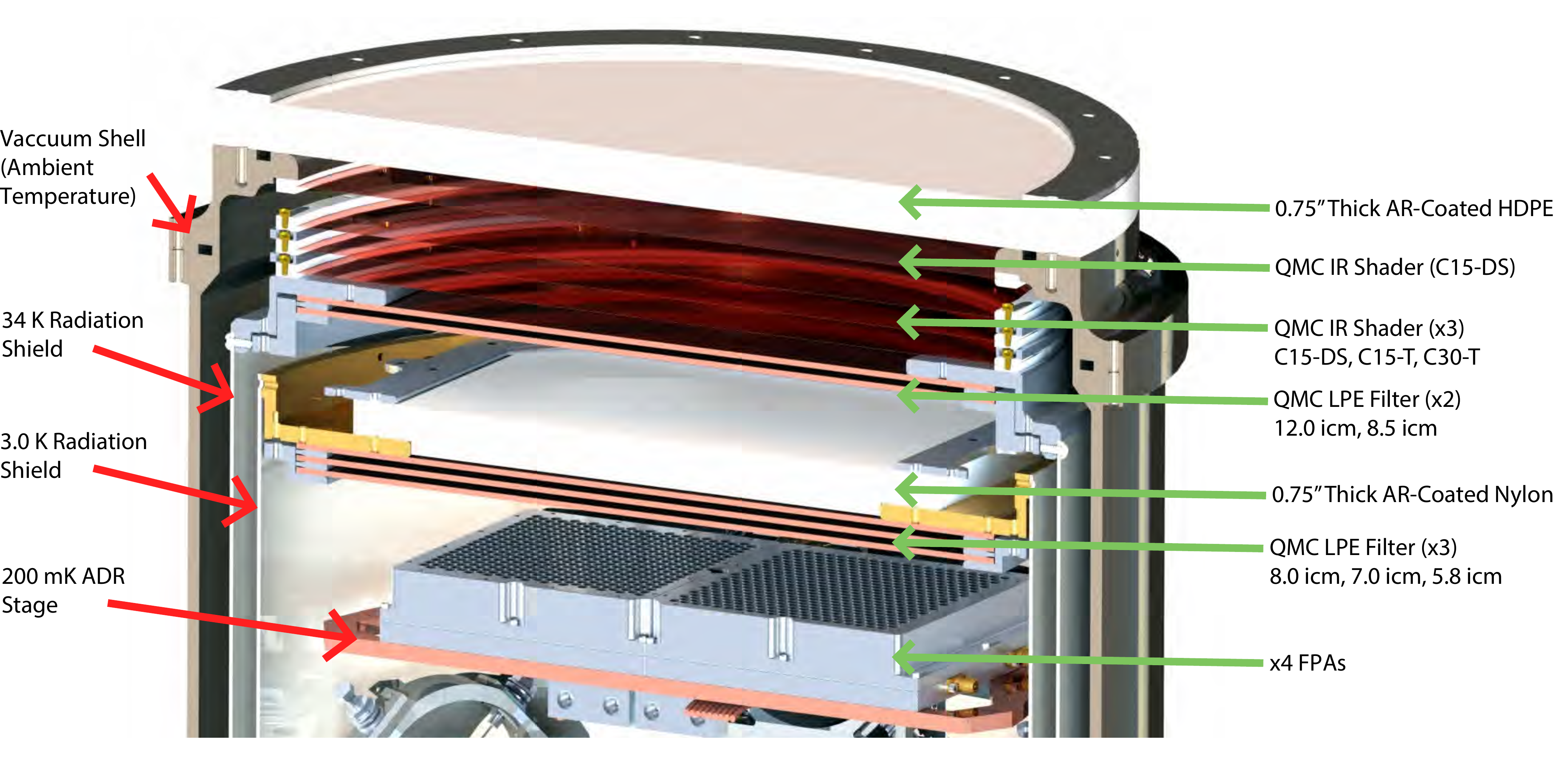}
  \caption{Cutout view of the optical shaders in the imager cryostat (see~Table~\ref{tab:filters} for more details).}
  \label{fig:filters}
\end{figure*}

\section{\label{sec:cryogenic}Cryogenic System\protect}

To cool the KIDs to the desired operating temperature of 200~mK, we use a DRC-200 multi-stage closed-cycle cooling system supplied by Star Cryoelectronics\footnote{https://starcryo.com}. The first two cooling stages, which continuously operate at 34~K and 3.0~K when our system is in a 300~K ambient environment, are provided by a Sumitomo RDK-415D2 Gifford-McMahon (GM) cryocooler paired with an FA-70 indoor/outdoor compressor unit from Shi Cryogenics Group\footnote{https://www.shicryogenics.com}. The subsequent two cooling stages are provided by the two-stage Adiabatic Demagnetization Refrigerator (ADR) supplied with the DRC-200 unit. The first ADR stage is a gadolinium gallium garnett (GGG) paramagnet that operates at approximately 700~mK, while the second ADR stage a ferric ammonium allum (FAA) paramagnet that is actively controlled at 200~mK during normal operation. At this temperature, the typical hold time is six hours. Cycling the ADR requires a little more than one hour, and the FAA stage of the ADR is able to reach an ultimate base temperature of 105~mK when not actively controlled.

\subsection{GM Cryocooler}

The operating temperature of the first stage of the GM cryocooler implies a total load of approximately 15~W based on the manufacturer's specifications. We estimate the conductive load due to $\times$4 tubular G10-CR supports connected to the cryostat shell, each with a cross sectional area of 62~mm$^2$ and a length of 610~mm, to be 0.6~W \cite{Fickett1979}. There is an additional conductive load from $\times$8 760~mm long stainless steel coaxial cables, Micro-Coax part number UT-085B-SS\footnote{https://www.carlisleit.com/brands/micro-coax/}, which are clamped to the stage approximately 400~mm from the cryostat feed-thru, equal to 0.7~W \cite{Marquardt2002}. There are also $\times$2 size 18 American Wire Gauge (18-AWG) copper wires, $\times$32 36-AWG phospher-bronze wires, and $\times$28 36-AWG manganin wires, all approximately 350~mm in length, contributing a total load of 0.3~W. The dominant load, however, is radiative. If fully transmissive, our 300~mm diameter optical window would present almost 20~W of loading on the first stage of the GM cryocooler. We are able to significantly reduce this load by installing a double-sided IR shader from QMC Instruments\footnote{http://www.qmcinstruments.co.uk/} on the cold side of the cryostat window (see Table~\ref{tab:filters} and Figure~\ref{fig:filters} for the full details of our optical filtering). The difference in stage temperature with and without the shader suggests that it reduces the load by approximately 75 per cent, to 5~W, comparable with other results in the literature \cite{Gudmundsson2015}. The sides of the 34~K radiation shield, with a total area of 0.52~m$^2$, are covered in multi-layer insulation (MLI), which we assume to have an effective emissivity of 0.02 \cite{Miyakita2014}, resulting in a radiative load of 4.8~W from the cryostat shell. The remaining load, approximately 4~W, is attributed to radiation on the bottom plate of the 34~K radiation shield, which has a total area of 0.10~m$^2$, implying it has an average emissivity of 0.08. Given that the plate lacks MLI, and it also contains a large number of irregularly shaped attachments such as heat straps, this emissivity value is reasonable.

\begin{table}[t]
    \begin{tabular}{c|c|c|c|c}
        Filter & Notes & Thickness & Stage & Transmission \\ \hline
        Cryostat Window & HDPE & 0.75 inches & 300~K & 97 per cent \\
        QMC IR Shader & C15-DS & N/A & 300~K & 97 per cent \\
        QMC IR Shader & C15-DS & N/A & 34~K & 97 per cent \\
        QMC IR Shader & C15-T & N/A & 34~K & 99 per cent \\
        QMC IR Shader & C30-T & N/A & 34~K & 99 per cent \\
        QMC LPE Filter & 12.0 icm & N/A & 34~K & 95 per cent \\
        QMC LPE Filter & 8.5 icm & N/A & 34~K & 97 per cent \\
        Dielectric & Nylon & 0.75 inches & 3.0~K & 91 per cent \\
        QMC LPE Filter & 8.0 icm & N/A & 3.0~K & 97 per cent \\
        QMC LPE Filter & 7.0 icm & N/A & 3.0~K & 99 per cent \\
        QMC LPE Filter & 5.8 icm & N/A & 3.0~K & 96 per cent \\
    \end{tabular}
    \caption{Optical filters to minimize radiative load and to define the high frequency edge of the detector passband. The filters are listed in the order they are encountered by photons incident from outside the cryostat. The cryostat window and QMC IR Shaders have clear circular apertures with a diameter of 300~mm, all other filters have square clear apertures with 206~mm sides. The overall transmission of the full stack is approximately 68 per cent.}
    \label{tab:filters}
\end{table}

The operating temperature of the second stage of the GM cryocooler implies a total load of approximately 200~mW based on the manufacturer's specifications. This stage is supported from the first stage of the GM cryocooler with a single tubular G10-CR support with a cross sectional area of 590~mm$^2$ and a length of 170~mm, resulting in a conductive load of 19~mW. The $\times$8 stainless steel coaxial cables, with lengths of approximately 400~mm from the thermal clamp connected to the first stage of the GM cryocooler, produce an additional conductive load of 32~mW \cite{Marquardt2002}. There are also $\times$26 36-AWG phospher-bronze wires and $\times$22 36-AWG manganin wires, each approximately 250~mm in length, resulting in a total load of less than 1~mW. A hollow stainless steel shaft, coupled to the heat switch on the sub-K stage, is estimated to contribute a total load of 5~mW. In addition, we have $\times$4 SiGe low noise amplifiers, Cosmic Microwave model CITLF2\footnote{https://www.cosmicmicrowavetechnology.com/}, each of which dissipates approximately 22~mW of power when operated at its optimal bias conditions, for a total of 88~mW. The remaining load, approximately 55~mW, is thought to be primarily radiative via the optical window, which is further shielded by three QMC IR shaders and two QMC LPE filters installed at the 34~K stage. Based on measured stage temperatures with and without the window open, we estimate this load to be approximately 100~mW, consistent with results in the literature for similar instruments \cite{Gudmundsson2015}. Our total calculated load is thus 245~mW, which is slightly larger than the measured total load, and indicates there are no other significant sources of loading on the second stage of the GM cryocooler.

\subsection{ADR}

\begin{figure}[t]
  \includegraphics[width=\columnwidth]{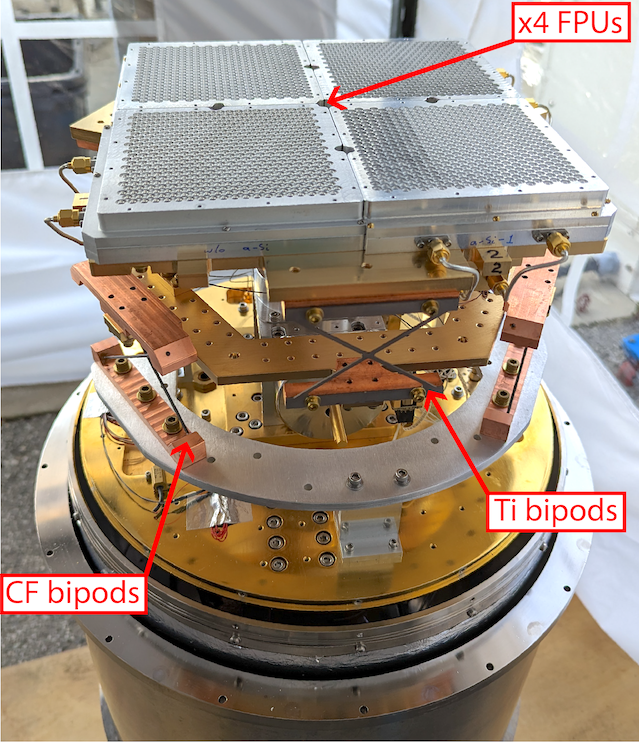}
  \caption{View of the imager with the cryogenic radiation shields removed. The four focal plane units, each with 960~KIDs, are shown near the top of the image and are connected to the ultracold FAA stage of the ADR. This stage is supported by three sets of Ti bipods connected to the intermediate GGG stage of the ADR. The intermediate stage is in turn supported by three sets of carbon fiber bipods connected to the second stage of the GM cryocooler.}
  \label{fig:cold_stage}
\end{figure}

In order to achieve the longest possible hold time with the ADR, we have configured the cryogenic system to minimize the heat load on it. Originally, the cryostat was outfitted with tensioned kevlar strings to support the sub-K stages. These supports were later redesigned by Cryogenic Design Solutions based on bipods to increase rigidity, as detailed below. The load on the GGG stage is dominated by thermal conduction from the 3~K GM stage via $\times$3 bipod supports made from Graphlite carbon fiber rods \cite{Kelleris2014}, shown in Figure~\ref{fig:cold_stage} along with $\times$8 0.047-inch NbTi coaxial cables from Keycom\footnote{https://keycom.co.jp/}. Each of the $\times$6 carbon fiber legs have a cross sectional area of 2.5~mm$^2$ and a length of 104~mm, resulting in a load of 14.5~$\mu$W on the ADR GGG stage \cite{Kelleris2014}. The coaxial cables have a typical length of 200~mm between the second stage of the GM cryocooler and the thermal clamp on the ADR GGG stage, resulting in an additional load of 12.3~$\mu$W \cite{Daal2019, Drobizhev2017}. Finally, there is an additional conductive load of approximately 0.4~$\mu$W from a total of $\times$16 36-AWG manganin wires, each approximately 200~mm in length \cite{Peroni1999}. 

\begin{table}[t]
    \begin{tabular}{lr}
        Component & Load \\ \hline
        \multicolumn{2}{c}{{\bf 34 K Stage}} \\ 
        $\times$4 Tubular G10-CR Supports & 0.6~W \\
        $\times$8 SS Coaxial Cables & 0.7~W \\
        Other Wiring & 0.3~W \\
        Radiative (Window) & 5$\phantom{.0}$~W \\
        Radiative (Cylindrical Shield) & 4.8~W \\
        Radiative (Shield Bottom) & 4$\phantom{.0}$~W \\
        {\bf Stage Total} & {\bf 15$\phantom{.}$~W} \\ \hline 
        \multicolumn{2}{c}{{\bf 3.0 K Stage}} \\ 
        G10-CR Support Tube & $\phantom{1}$19~mW \\
        $\times$8 SS Coaxial Cables & $\phantom{1}$32~mW \\
        Other Wiring & $\phantom{11}$1~mW \\
        Heat Switch Shaft & $\phantom{11}$5~mW \\
        $\times$4 SiGe Amplifiers & $\phantom{1}$88~mW \\
        Radiative & 100~mW \\
        {\bf Stage Total} & {\bf 245~mW} \\ \hline 
        \multicolumn{2}{c}{{\bf 700 mK ADR GGG Stage}} \\ 
        $\times$3 Graphlite Carbon Fiber Bipod Supports & 14.5~$\mu$W \\
        $\times$8 NbTi Coaxial Cables & 12.3~$\mu$W \\
        Other Wiring & $\phantom{1}$0.4~$\mu$W \\
        {\bf Stage Total} & {\bf 27.2~$\mu$W} \\ \hline 
        \multicolumn{2}{c}{{\bf 200 mK ADR FAA Stage}} \\
        $\times$3 Timet Titanium Bipod Supports & 0.16~$\mu$W \\
        $\times$8 NbTi Coaxial Cables & 0.44~$\mu$W \\
        Other Wiring & 0.02~$\mu$W \\
        Intrinsic Conduction Within ADR & 0.22~$\mu$W \\
        Radiative & 3.3$\phantom{1}$~$\mu$W \\
        {\bf Stage Total} & {\bf 4.1$\phantom{1}$~$\mu$W} \\ \hline 
    \end{tabular}
    \caption{Summary of the Load on each cryogenic temperature stage due to relevant sources.}
    \label{tab:loading}
\end{table}

Thus, the total conductive load on the ADR GGG stage is estimated to be 27.2~$\mu$W. The specification for the stage is 1.2~J of cooling capacity at 1~K. Using the formalism from \citet{Bartlett2015}, we estimate that the cooling capacity is approximately 0.8~J in our system where the ADR GGG stage is at a temperature of 700~mK. Therefore, the total hold time of the ADR GGG stage is calculated to be approximately eight hours. This is slightly longer than our achieved hold time of six hours, which is likely limited by the ADR FAA stage (see below).

The ADR FAA stage is mechanically supported from the ADR GGG stage with $\times$3 bipods made from Timet\footnote{https://www.timet.com/} Ti 15-3-3-3 (see Figure~\ref{fig:cold_stage}). Each bipod leg has a cross sectional area of 1.9~mm$^2$ and a length of 79~mm, resulting in a total thermal load of 0.16~$\mu$W from all $\times$6 legs \cite{Daal2019}. There is an additional conductive load from $\times$8 NbTi coaxial cables, which we calculate to be 0.44~$\mu$W assuming a typical cable length of 100~mm between the thermal clamp on the ADR GGG stage and the ADR FAA stage \cite{Daal2019,Drobizhev2017}. Finally, there are a total of $\times$8 36-AWG manganin wires approximately 100~mm in length connected to the ADR FAA stage resulting in a load of 0.02~$\mu$W \cite{Peroni1999}. The specification sheet for the ADR implies an additional intrinsic conductive load of 0.22~$\mu$W. 

Thus, the total conductive load on the ADR FAA stage is estimated to be 0.84~$\mu$W. For the achieved temperatures in our system, the total cooling capacity of the ADR FAA stage is approximately 90~mJ. Given our measured hold time of six hours, this implies a total load of 4.1~$\mu$W, suggesting there is a radiative load of 3.3~$\mu$W on the ADR FAA stage. This is comparable to the total incident optical power transmitted through the optical filter stack, which is calculated to be 10.5~$\mu$W, indicating that approximately 30 per cent of the total incident power is absorbed by the stage. 

To better assess the accuracy of our estimated conductive load on the ADR FAA stage, we have also performed hold time tests in a modified configuration. The optical filters were replaced with aluminum covers at each temperature stage to largely eliminate the radiative load, the NbTi coaxial cables were removed, and the carbon fiber bipods supporting the ADR GGG stage were replaced with less conductive (but less rigid) kevlar supports, which are estimated to contribute a load of 0.11~$\mu$W. In this configuration, our achieved hold time was 30 hours, suggesting a total load of 0.83~$\mu$W. If we assume that the radiative load is negligible, then the remaining load from our nominal model including the kevlar supports is 0.35~$\mu$W from the combination of the manganin wires and the intrinsic conductive load of the ADR. This test thus sets an upper limit of 0.5~$\mu$W for any excess conductive load on the ADR FAA stage that is not included in our model. Alternatively, this can be converted to a lower limit of 2.8~$\mu$W for the total radiative load on that stage when the system is operated in its default state without aluminum covers over the optical openings.

A full summary of the loading contributed by sources relevant to each stage in the cryogenic system is provided in Table~\ref{tab:loading}.

\section{\label{sec:readout}Readout\protect}

\begin{figure}[t]
  \centering
  \includegraphics[width=\columnwidth]{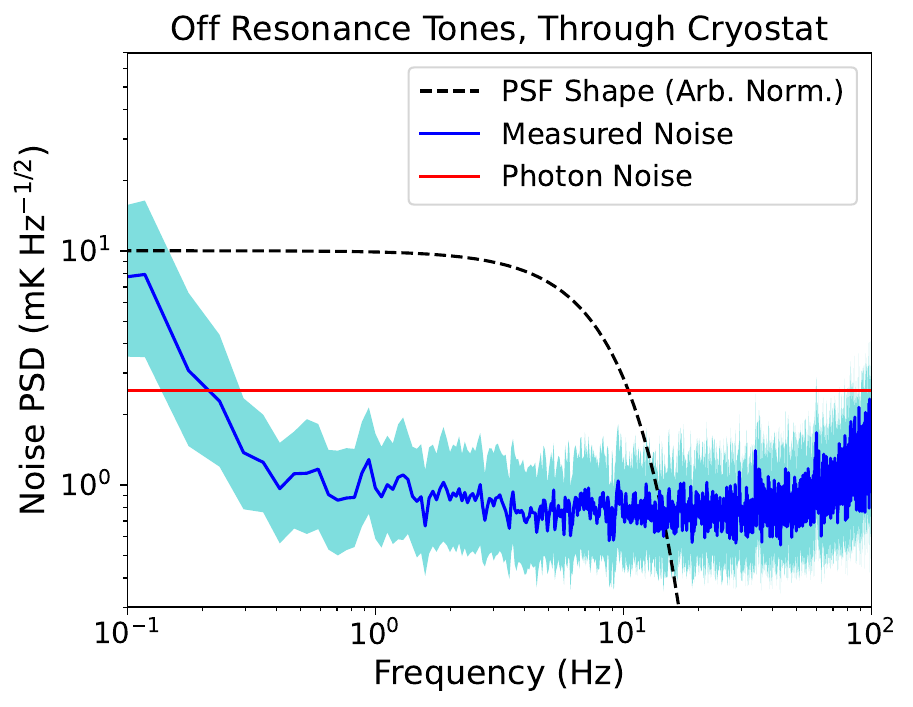}
  \caption{Measured noise power spectrum from off-resonance tones, illustrating the noise due solely to the readout electronics when the system is operated in its nominal configuration. The power level of the off resonance tones has been matched to the power level of the on resonance tones at the location of the cryogenic SiGe amplifier, and has been referenced to the on-sky calibration of the KIDs in units of mK~Hz$^{-1/2}$. The approximate signal band of interest is indicated by the dashed line, which corresponds to the instrument response to a point-like object at our nominal dither speed of 1.5\degrees~s$^{-1}$ (see Section~\ref{sec:optical_performance} for more details).}
  \label{fig:readout_noise}
\end{figure}

The original readout system for SKIPR was based on the room-temperature Reconfigurable Open Architecture Computing Hardware 2 (ROACH2) and is described in detail elsewhere \cite{Gordon2016}. In brief, the field-programmable gate array (FPGA) on the ROACH2 is programmed to playback a set of up to 1024 individual tones, each of which is nominally centered on a KID resonance, with specified frequencies and complex-valued amplitudes within two 256~MHz sidebands. The DAC board designed for the MUSIC instrument converts the digital playback to an analogue signal \cite{Duan2010}, which is then mixed up to the desired KID resonance band with a controllable local oscillator (LO). Next, the signal is sent through the feedline on the detector tile, before being mixed back down to the baseband using the same LO. After digitization with the MUSIC ADC board, the FPGA uses a two-stage channelization algorithm to digitally down convert and record the complex-valued data streams associated with each tone at a cadence of 488.28~Hz. Combinations of attenuators and low noise amplifiers are inserted between the various components of the readout system to achieve the desired power levels for optimal performance, corresponding to approximately $-$25~dBm/tone at the DACs/ADCs and $-$75~dBm/tone at the detector tile.

Recently, the readout has been upgraded to a new system, which replaces the ROACH2 and MUSIC ADC/DAC with the Xilinx ZCU111 Radio Frequency System on a Chip (RFSoC) module\footnote{https://www.xilinx.com/products/boards-and-kits/zcu111.html}. General details of this type of readout implementation can be found elsewhere \cite{Sinclair2022,Sinclair2024}. Functionally, this new readout operates almost identically to the original ROACH2-based system, with the primary benefits related to a significant reduction in the overall size, weight, and power consumption. Because the implementation, including the number of tones per channel and noise performance, of the two readout systems is approximately identical, we do not distinguish between them in this manuscript when describing the performance of the imager.

The fundamental noise floor set by the random arrival of photons from the ambient environment on the detector is approximately 2.5 mK~Hz$^{-1/2}$ (corresponding to approximately $-$90~dBc~Hz$^{-1}$ given the desired readout power, see Section~\ref{sec:detectors} for more details), and thus sets the performance requirement for the system. By centering tones away from detector resonant frequencies, we find that the noise for data streams from the entire readout chain typically varies between $-$98~dBc~Hz$^{-1}$ and $-$105~dBc~Hz$^{-1}$ with an approximately flat spectral shape above 0.5~Hz, thus satisfying our requirement. See Figure~\ref{fig:readout_noise}. Given the typical tone power, along with the typical resonance depth of 5--10~dB, this corresponds to between $-$178~dBm~Hz$^{-1}$ to $-$190~dBm~Hz$^{-1}$. We note that this is higher than the expected noise from the cryogenic SiGe amplifier, which has a noise temperature of 2~K in our detector band, corresponding to $-$195~dBm~Hz$^{-1}$, suggesting it is not the limiting noise source in our system. This is further confirmed by noise measurements that bypass the cryogenic system, which yield similar noise levels, and we also note our measured noise is consistent with the room-temperature noise performance of $-$98~dBc~Hz$^{-1}$ obtained using an almost identical system for the BLAST-TNG instrument \cite{Gordon2016}.

In practice, we utilize the readout system as follows to collect imaging data. First, the LO is swept with a step size of 1~kHz over a range spanning 50~kHz above and below its nominal value to measure the complex-valued transmission of each detector resonance as a function of frequency. These data are then used to re-center the playback tones, since the resonators are found to shift by as much as a couple 10s of kHz based on changes in ambient optical loading and rotations within the Earth's magnetic field. In practice, re-centering is generally not required for relatively small moves, i.e., $\lesssim 20$\degrees\ in azimuth (AZ) or $\lesssim 5$\degrees\ in zenith angle (ZA), which shift the tones by less than a couple of kHz. After re-centering, a second LO sweep is performed over a shorter range, and the resulting data are used to calibrate the conversion factor for each tone from its complex-valued data streams to a fractional frequency shift for the KID resonator. Observational data are then streamed for the desired duration, after which all of the recorded detector data streams are converted to fractional frequency shift, which is subsequently converted to brightness temperature via the optical calibration measurements described in Section~\ref{sec:detectors}.

\section{\label{sec:telescope}Telescope\protect}

We have designed and fabricated a custom telescope for SKIPR based on a crossed Dragone geometry \cite{Dragone1978} and an alt/az mount. Below, we separately provide the details of both the optical and mechanical design.

\begin{figure}[t]
  \centering
  \includegraphics[width=\columnwidth]{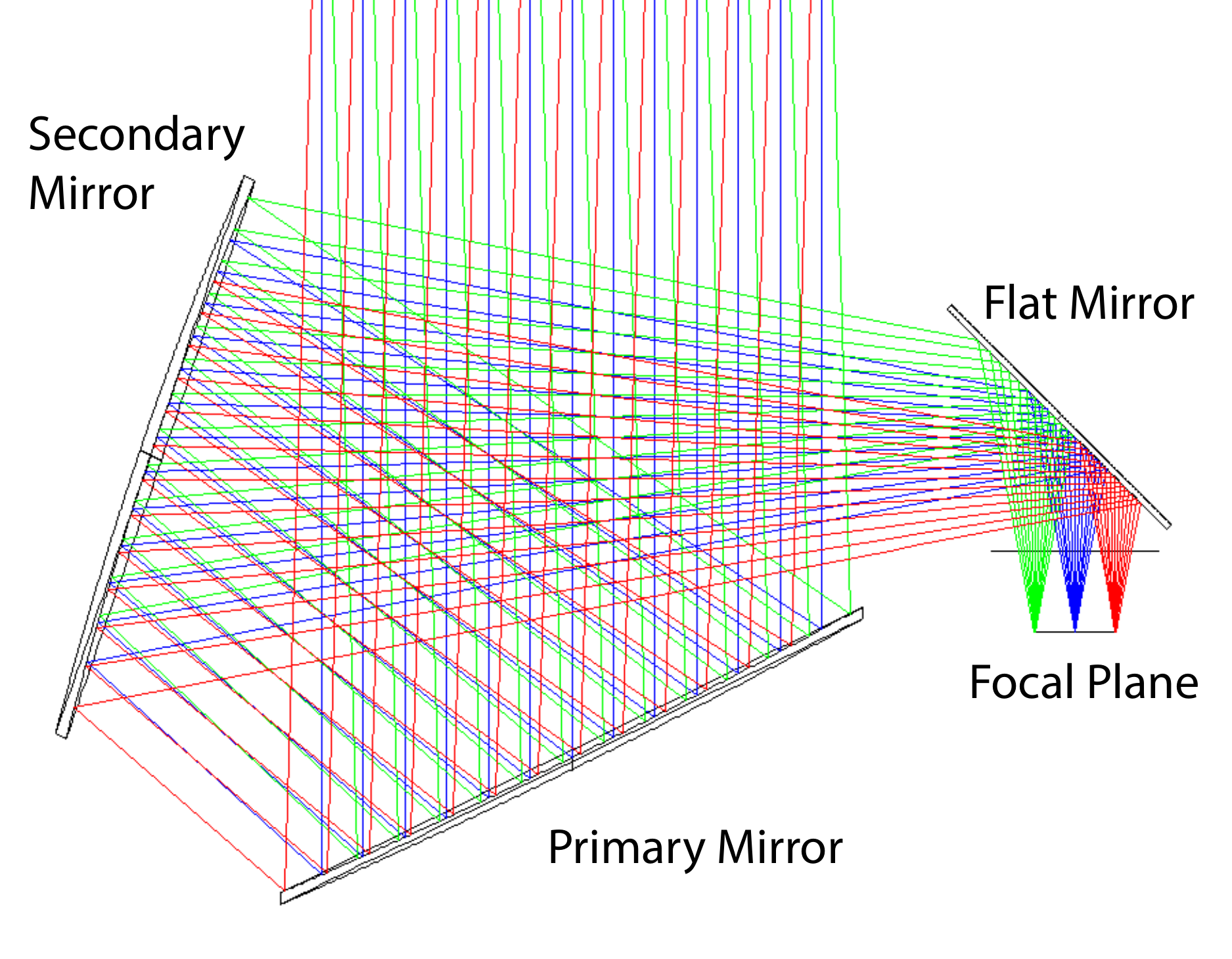}
  \caption{Ray-trace of the SKIPR telescope optics.}
  \label{fig:telescope}
\end{figure}

\begin{figure}[t]
  \centering
  \includegraphics[width=\columnwidth]{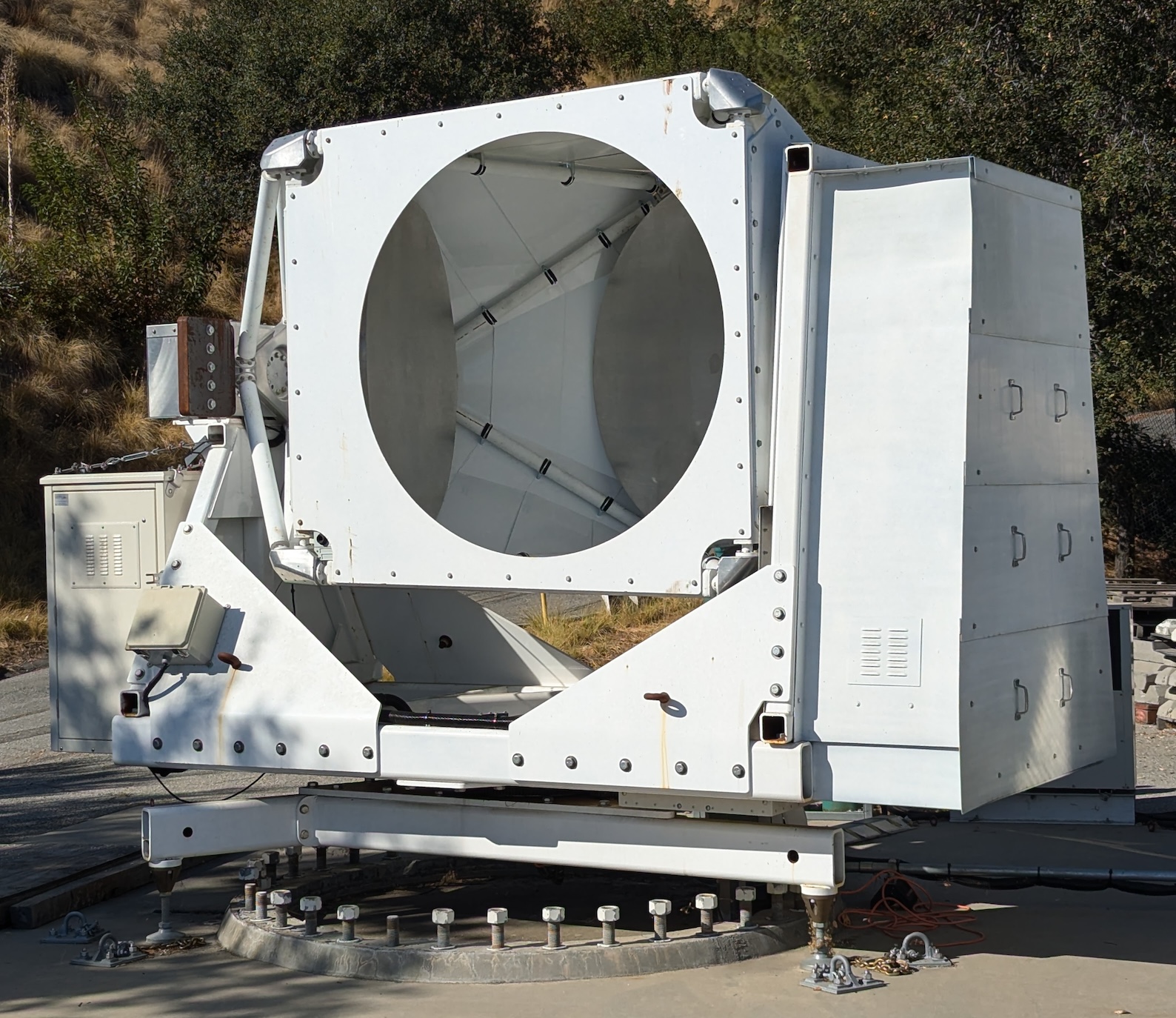}
  \caption{Photograph of the SKIPR telescope at JPL. The edges of both the primary and secondary mirrors are visible within the circular opening in the baffle structure. The white enclosure to the right of the mirrors is the receiver cabin, which houses the imager cryostat. The total height of the telescope system is approximately 3~m and the total width is approximately 4~m.}
  \label{fig:telescope_picture}
\end{figure}

\begin{figure}
    \centering \includegraphics[width=\columnwidth]{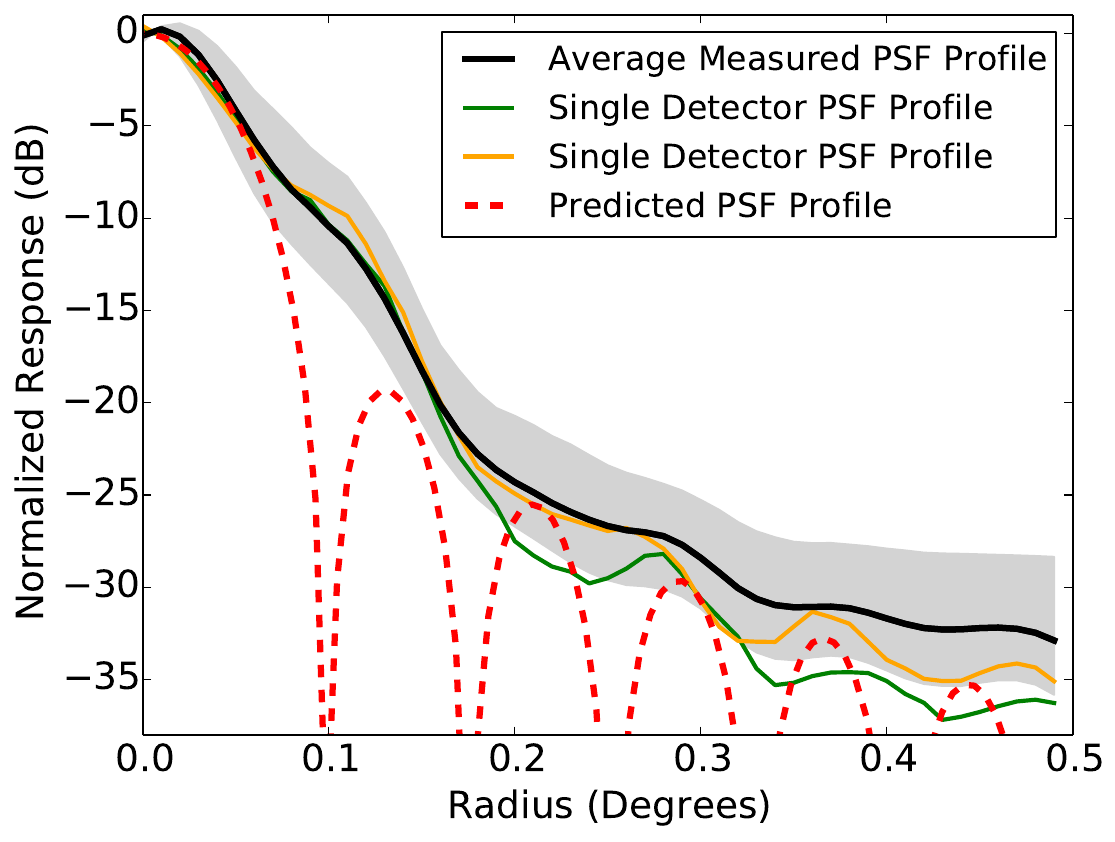}
    \caption{Radial profile of the average measured PSF (solid black line), along with the region spanning 68 per cent of the individual PSFs. Prior to averaging, all of the individual profiles were normalized to a value of one at the center. Overlaid as a dashed red line is the predicted PSF profile. The agreement between the measured and predicted PSF shape is generally good, with the exception of a slight shoulder in the measured PSF with an amplitude near $-$10~dB at a radius of approximately 0.1\degrees. By averaging over all detectors, the diffraction peaks at larger radii are smoothed over due to slight offsets in radius between detectors, as illustrated by the two individual detector PSF profiles shown in thin green and yellow lines. The diffraction peaks are more easily seen in the two-dimensional PSFs presented in Figure~\ref{fig:beam_shapes}.}
    \label{fig:beam_profile}
\end{figure}

\subsection{Optical Design and Performance}
\label{sec:optical_performance}

 \begin{figure*}
    \centering 
    \includegraphics[width=0.32\textwidth]{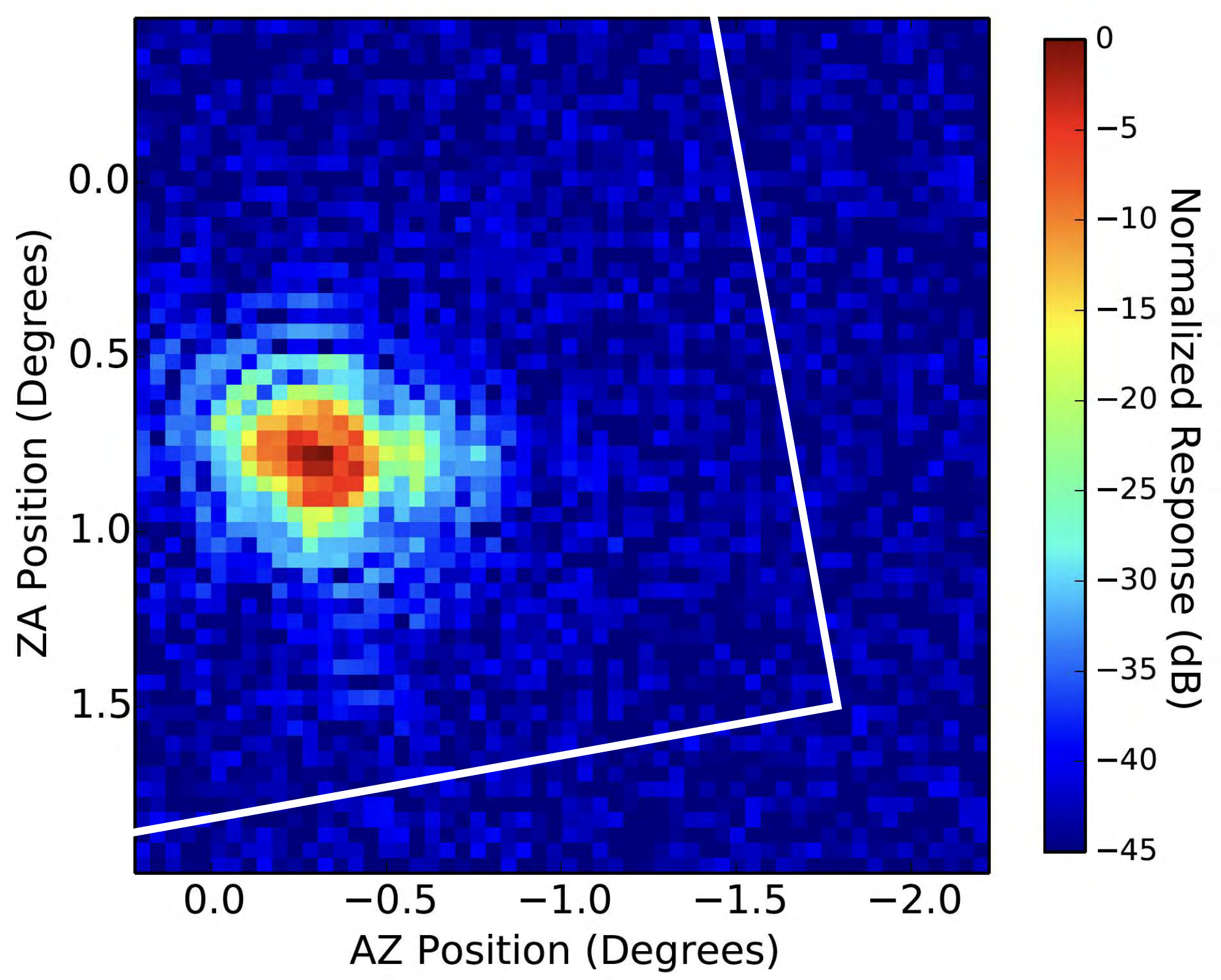}
    \hspace{0.005\textwidth}
    \includegraphics[width=0.32\textwidth]{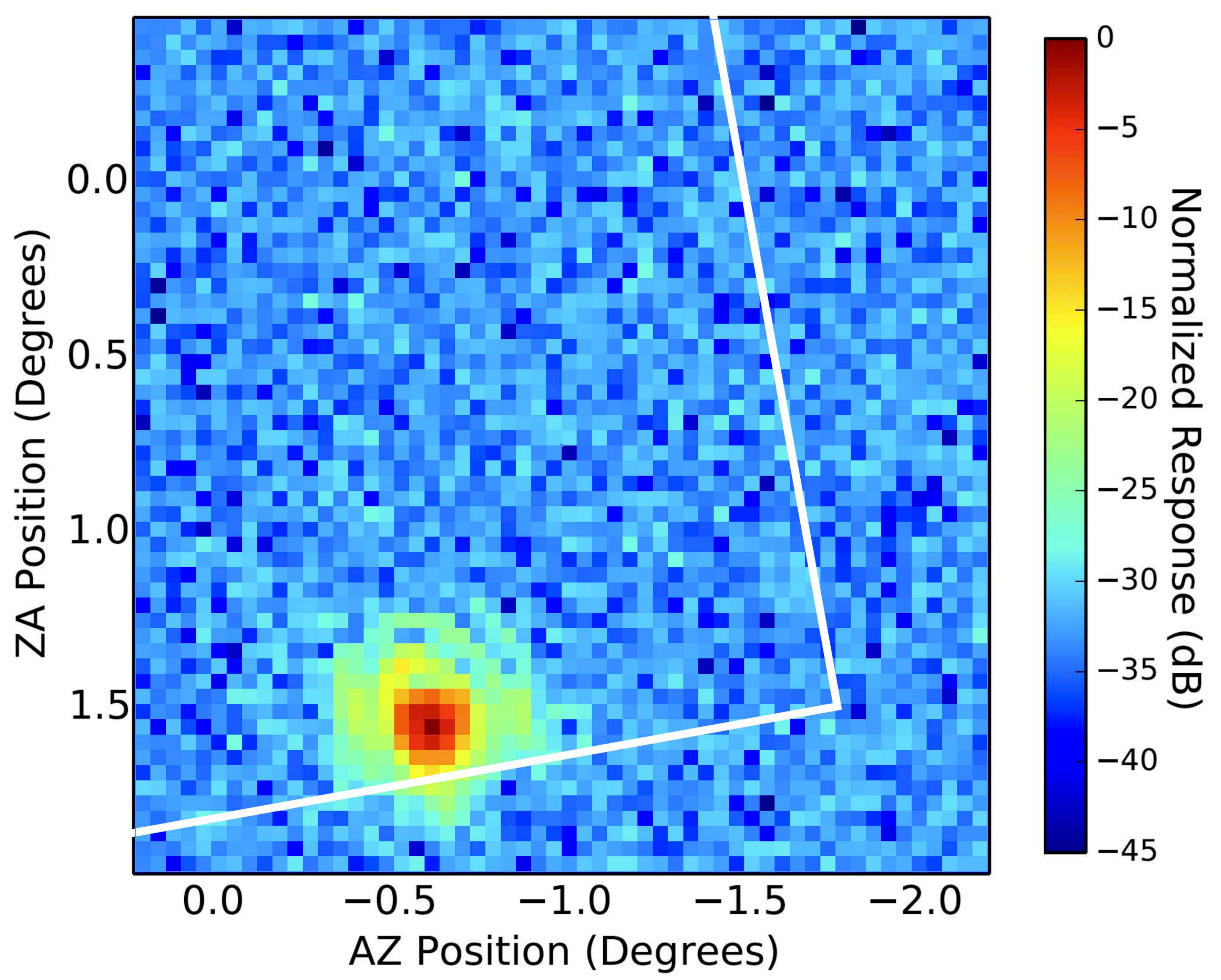}
    \hspace{0.005\textwidth}
    \includegraphics[width=0.32\textwidth]{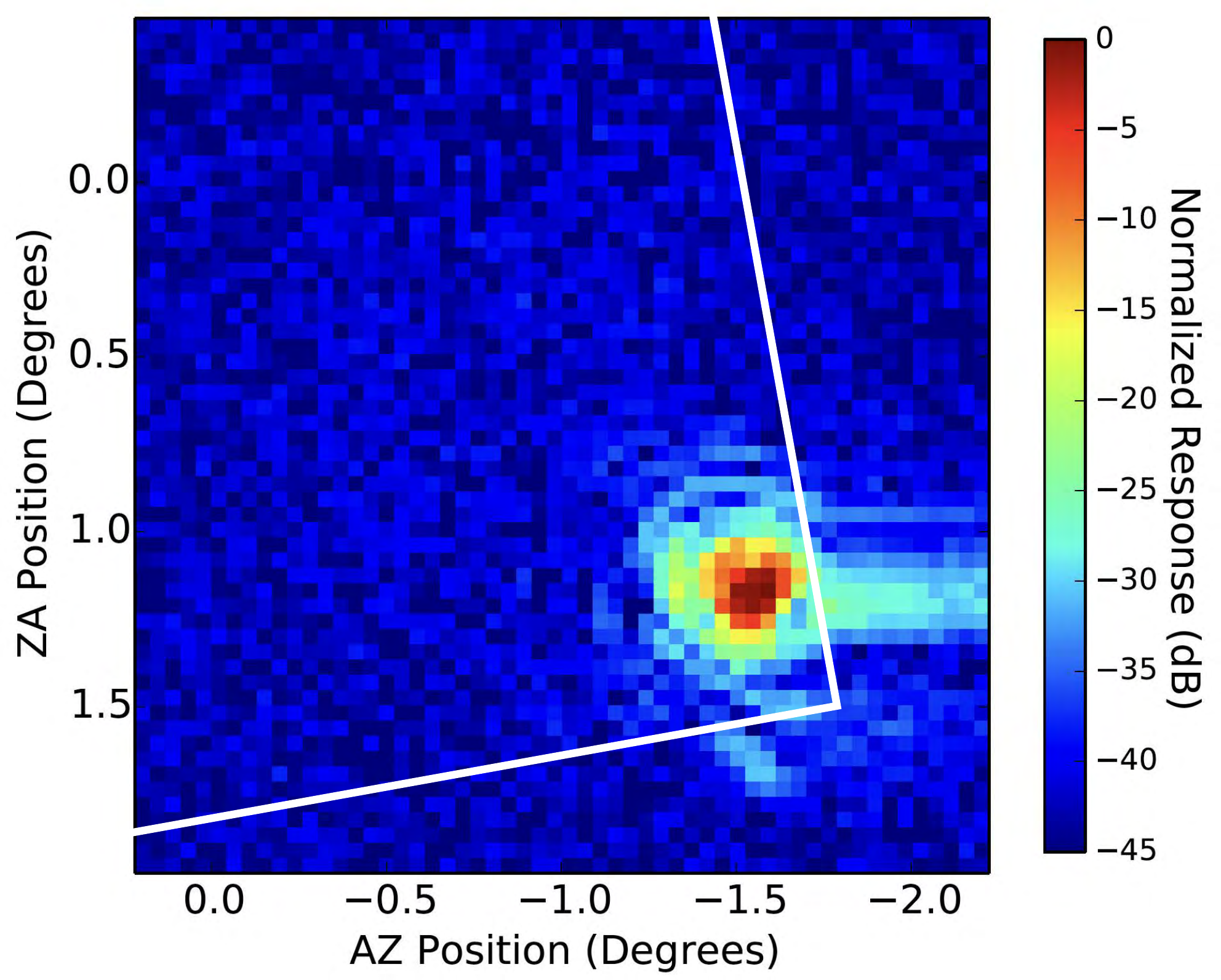}

    \centering 
    \includegraphics[width=0.32\textwidth]{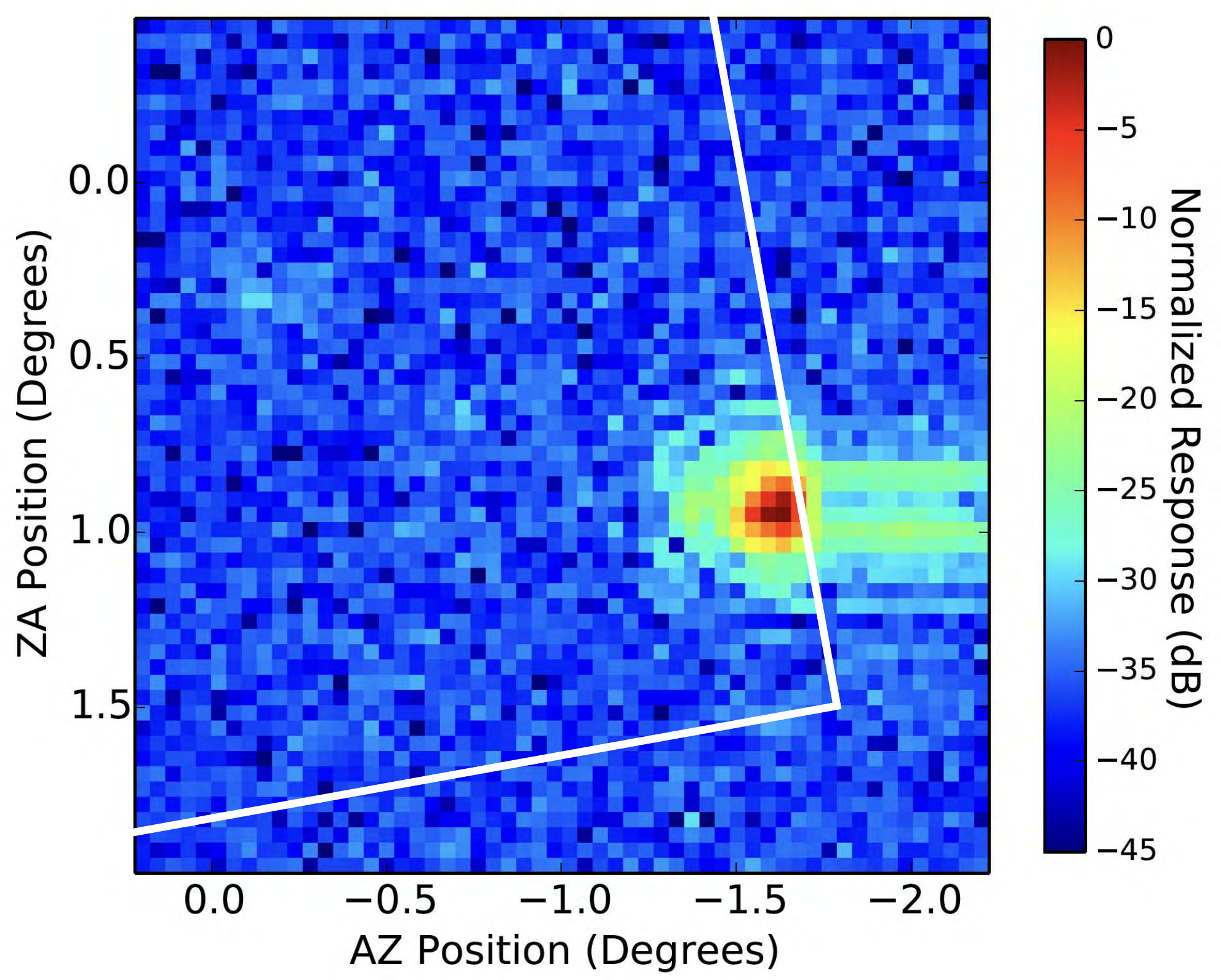}
    \hspace{0.005\textwidth}
    \includegraphics[width=0.32\textwidth]{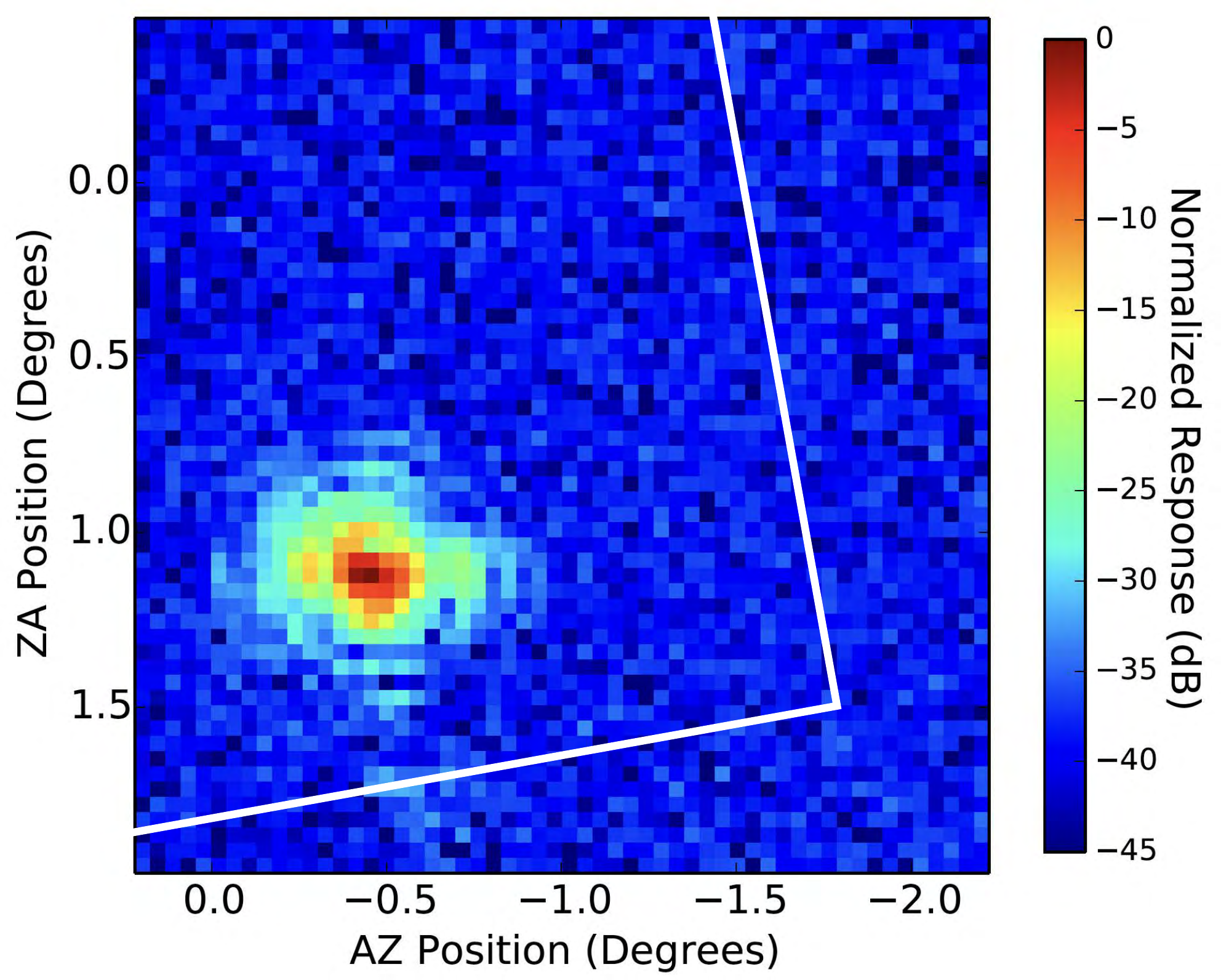}
    \hspace{0.005\textwidth}
    \includegraphics[width=0.32\textwidth]{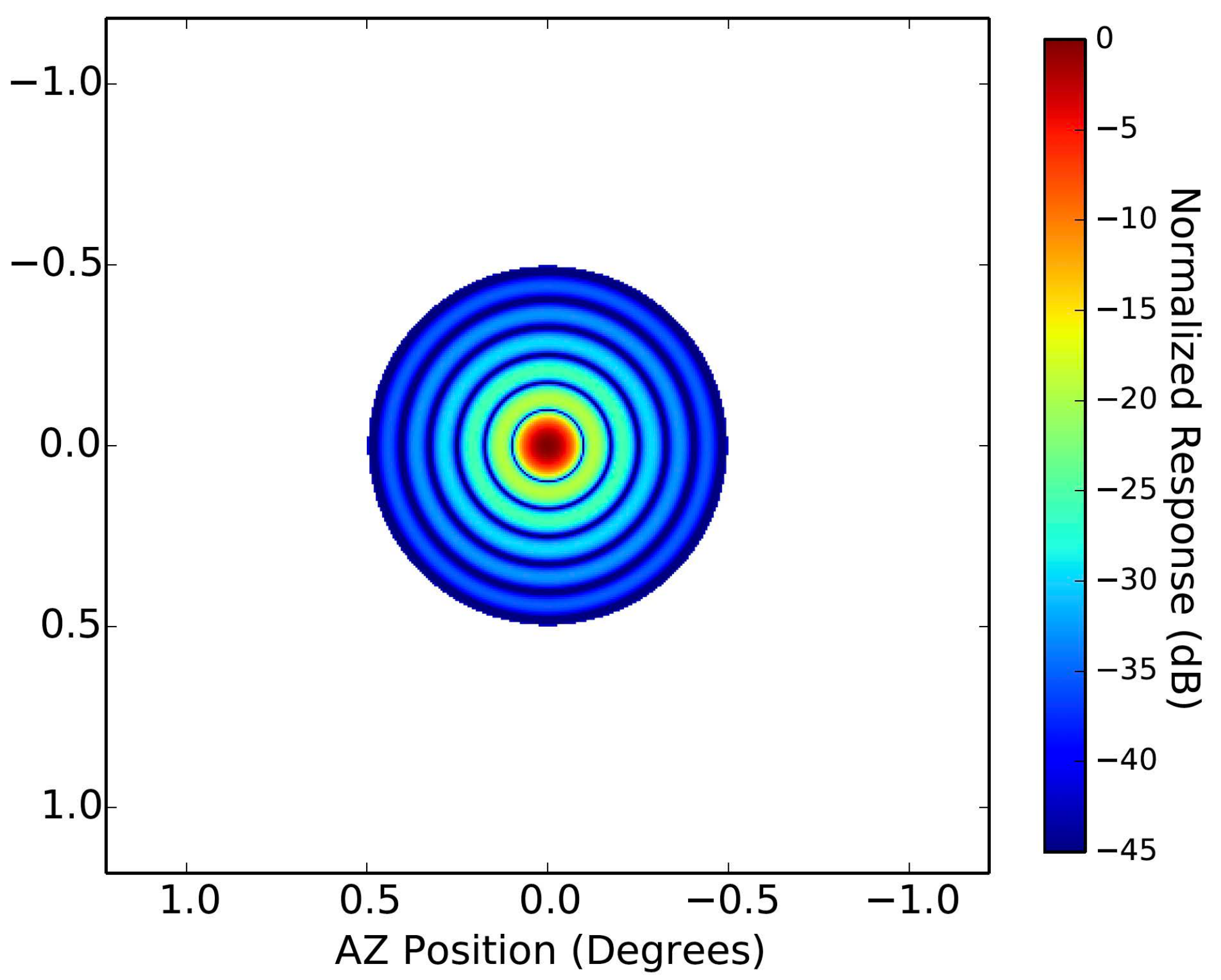}
    \caption{Top row: Measured PSFs for three detectors sensitive to the horizontal polarization direction. Bottom Row: Measured PSFs for two detectors sensitive to the vertical polarization direction, along with the predicted on-axis PSF shape. In all cases the azimuth (AZ) and zenith angle (ZA) coordinates are referenced to the optical axis of the telescope, and the approximate edge of the FPA is shown as a white line. Most of the PSF shapes are similar to the one shown in the upper left, and it is only the detectors near the edge of the FPA which have additional artifacts that we attribute to vignetting and/or scattering in the optical path.}
    \label{fig:beam_shapes}
 \end{figure*}

 The primary mirror has a projected diameter of 1.59~m and a focal length of 26.8~m, and the secondary mirror has a projected diameter of 1.42~m and a focal length of 17.4~m. The ray bundle from the telescope passes through an opening in the elevation bearing of the mount, where it is folded at 90\degrees\ by a flat mirror with a projected diameter of 0.60~m. The prime focus of the telescope is formed 0.50~m below the fold mirror. The imager cryostat is attached to the telescope mount on a translating stage that allows for precise alignment of the FPA at the prime focus. See Figure~\ref{fig:telescope}. The final focal ratio is 2.30, and the corresponding plate scale at the prime focus is 60$''$ mm$^{-1}$, resulting in the four FPA units subtending an approximately square footprint of 3.43\degrees~$\times$~3.43\degrees\ in size. The center-to-center spacing of the feedhorns used to couple radiation to the KIDs, equal to 5.0~mm, corresponds to approximately 1.1(f/\#)$\lambda$. Image quality over the entire 3.43\degrees\ field of view is excellent, with a Strehl ratio above 0.99 even in the extreme corners. The system is also highly telecentric to efficiently couple with the feedhorns, with an exit pupil 11.9~m beyond prime focus. This is much larger than the projected size of the FPA, which is approximately 206~mm per side, indicating that the exit pupil is located close to the value of infinity required for perfect telecentricity.

 Our design does not include an aperture stop, and so the telescope primary mirror serves as the only such aperture in the system. This choice was motivated by several considerations, including the operational optical load of approximately 300~K from the ambient environment, which mitigates the potential performance degradation from optically coupling to ambient temperature and/or reflective surfaces beyond the perimeter of the primary mirror. Another consideration was the available space within the existing cryostat, which is not sufficient to accommodate the optics required to produce a cold (3~K) aperture stop. We have performed idealized simulations of our optical system under the assumption that the primary mirror is the only limiting aperture in the system, which yield an approximately Gaussian far-field point spread function (PSF) with a full-width at half-maximum (FWHM) of 0.082\degrees\ at the effective band center of 148~GHz \cite{Sayers2020}. Outside of the primary PSF, there are a series of diffractive rings with amplitudes that decrease with angular offset and are $\lesssim 20$~dB below the peak of the PSF. See Figures~\ref{fig:beam_profile} and \ref{fig:beam_shapes}.

\begin{figure*}
    \centering 
    \includegraphics[width=0.45\textwidth]{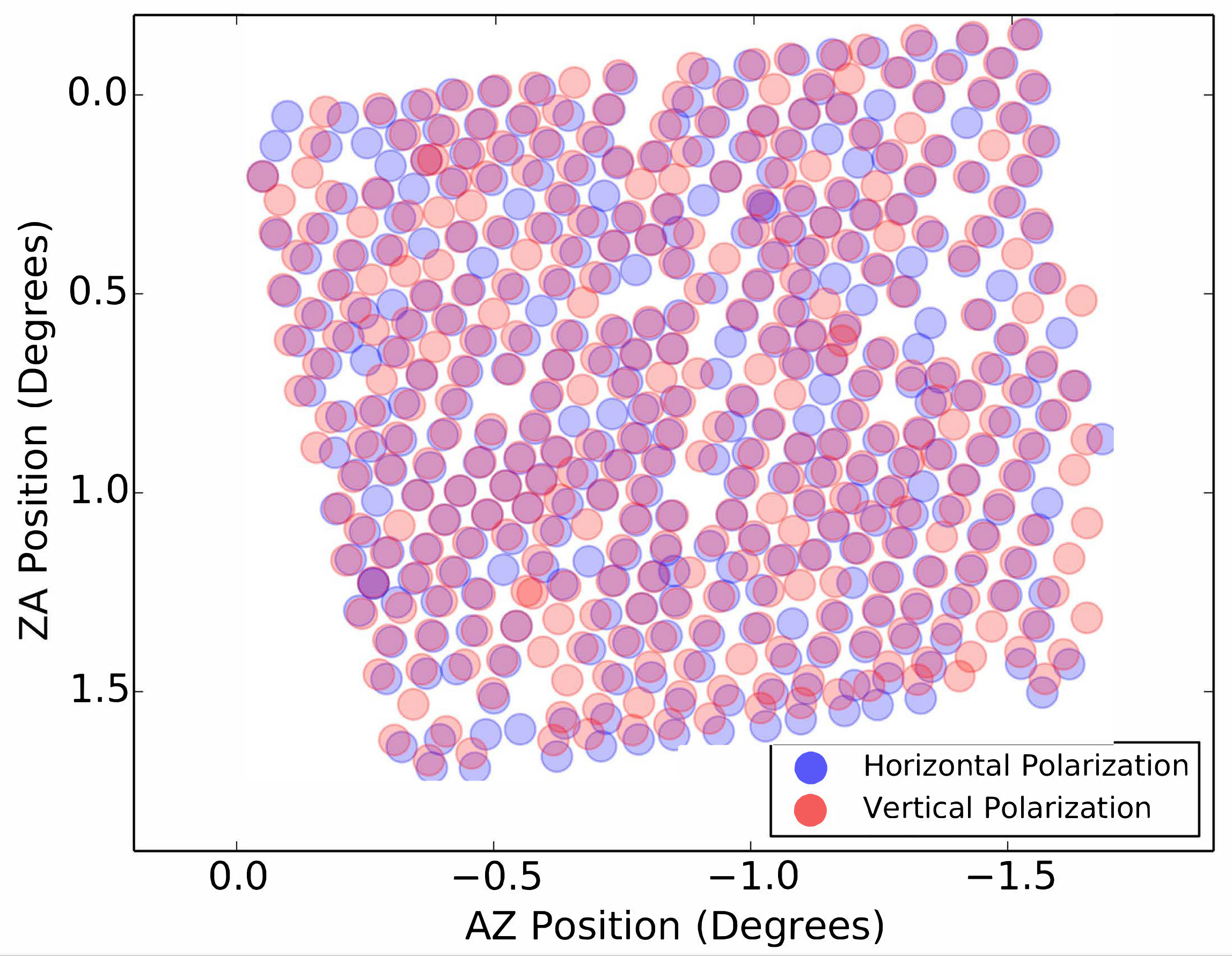}
    \hspace{0.005\textwidth}
    \includegraphics[width=0.51\textwidth]{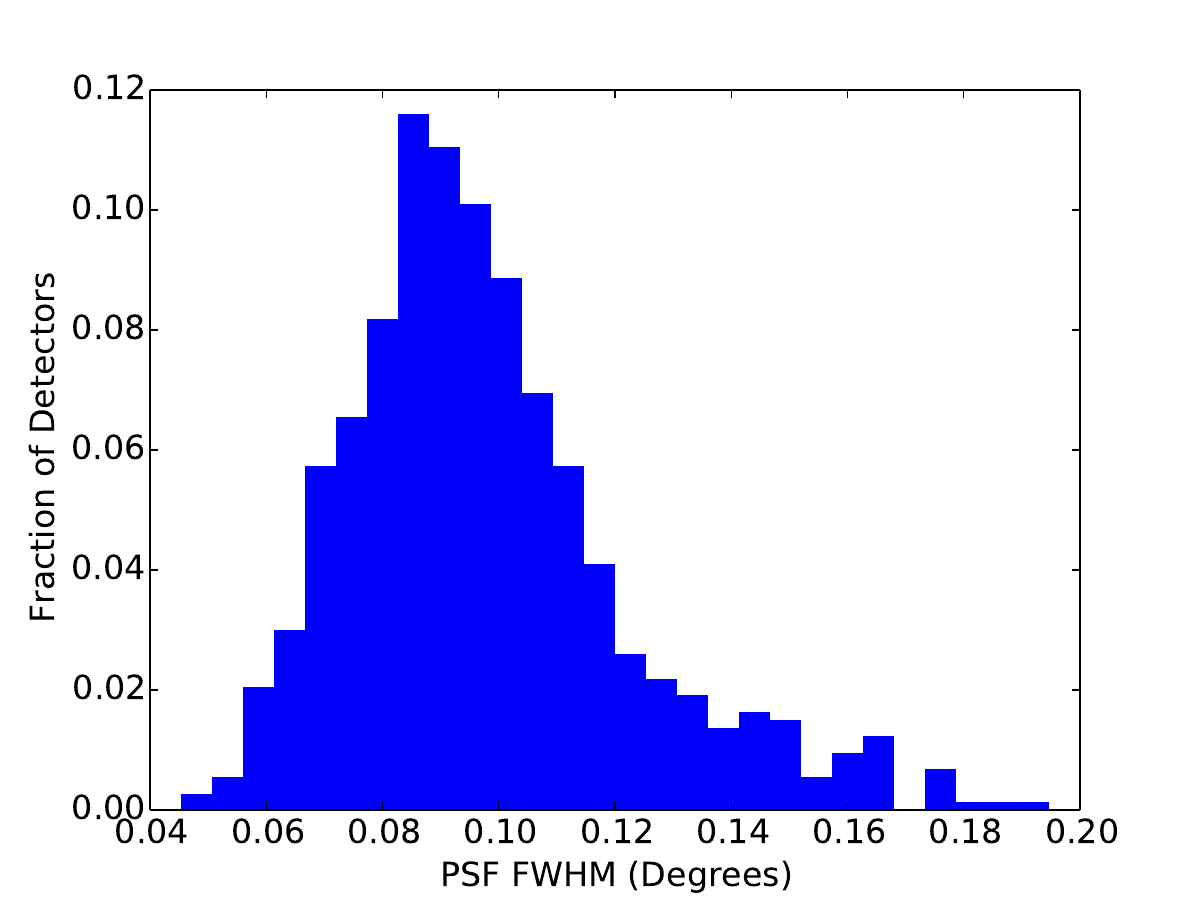}
    \caption{Left: Measured positions and polarizations for detectors in one focal plane tile relative to the optical axis of the telescope. Each circle corresponds to the predicted FWHM of the PSF. There is minimal distortion and a measured plate scale of 58$''$ mm$^{-1}$, in good agreement with the prediction of 60$''$ mm$^{-1}$. Right: The distribution of fitted PSF FWHMs for the detectors in Tile 2, with a median of 0.094\degrees\ and an interquartile range of 0.028\degrees. For the 140~GHz source used to obtain the measurements, the predicted FWHM is 0.087\degrees.}
    \label{fig:beam_hist}
 \end{figure*}

 To measure the PSF in the far field for each detector, we placed a coherent 140~GHz source with a power of approximately 0.1~mW at a distance of 2.42~km from the telescope. A rectangular horn with 24~dB of gain was used, resulting in an effective source temperature of approximately 2000~K. While this provided high signal-to-noise PSF measurements in reasonable integration times, a slight correction is required for the non-linear detector response when it was closely aligned with the source. The setup emits in a single polarization, and two separate measurements were made with the horn roughly aligned with the orientations of the two polarization directions selected by the detectors.  Figures~\ref{fig:beam_profile} and \ref{fig:beam_shapes} show the measured PSF profiles for detectors in a range of positions within the focal plane for both polarizations. The agreement with the predicted shape is generally quite good, including the appearance of diffractive rings at the expected amplitude and angular offset in the two-dimensional plots shown in Figure~\ref{fig:beam_shapes}. In addition, the angular positions of the PSF centers are in good agreement with expectations and indicate a plate scale of 58$''$ mm$^{-1}$ with minimal distortion. See Figure~\ref{fig:beam_hist}. Furthermore, there is a high degree of co-alignment between the two detectors coupled to orthogonal polarizations within each feedhorn. Also shown in Figure~\ref{fig:beam_hist}, the distribution of fitted FWHM values has a median value of 0.094\degrees, in reasonable agreement with the predicted value of 0.087\degrees\ for 140~GHz radiation. Referenced to the effective band center of 148~GHz, this indicates a median FWHM of 0.089\degrees. We note that the interquartile range of FWHM values is 0.028\degrees, with a modest tail in the distribution extending to larger FWHM values. Fitted PSF FWHMs that are smaller than the expected value of 0.087\degrees\ are attributed to noise fluctuations. While the per-detector fitting uncertainty is generally between 0.003\degrees~--~0.008\degrees, some detectors have much larger uncertainties in the range 0.01\degrees~--~0.03\degrees.

 We note from Figure~\ref{fig:beam_shapes} that additional artifacts are present in the PSFs near the edge of the focal plane, particularly in the AZ direction, and generally with an amplitude of $\lesssim 20$~dB below the peak of the PSF. We attribute these artifacts in part to the square aperture presented by the optical filters on the 34~K and 3~K stages, which increasingly vignette the optical path when moving off-axis in the focal plane. However, given the symmetry of these square apertures combined with the higher level of artifacts in detectors near the AZ edge of the focal plane compared with the ZA edge of the focal plane, we suspect that non-idealities in the ambient temperature optics of the telescope also contribute. In particular, the relative separations of the chief rays for different off-axis field positions within the focal plane are highest at the folding flat mirror, making it a leading candidate for such positionally-dependent artifacts.

\subsection{Mechanical Design and Performance}

Our mechanical design is based on the requirement that deformations due to gravity or the maximum planned acceleration of the structure itself while actively collecting imaging data, 1\degrees\ s$^{-2}$, negligibly impact the optical performance. For gravity, we consider the most extreme scenario of complete rotations in ZA, where its impact is typically two orders of magnitude larger than that expected from motion of the telescope during a typical imaging dither. For local deformations of the mirror surfaces, we aimed for a maximum deflection of $\lambda$/125, equal to 16~$\mu$m, which would at most result in a Ruze scattering efficiency loss of 1 per cent \citep{Ruze1966}. Due to its orientation relative to horizontal, this specification is more readily achieved for the secondary mirror, which we find has a maximum deflection of 2.5~$\mu$m. The primary mirror is more prone to gravitational deformations, and while the majority of its surface has maximum deformations of less than 16~$\mu$m, we find that deformations of $\gtrsim20$~$\mu$m can occur at the extreme edges.

In addition to local distortions of the mirror surface, we also consider global translations and rotations of the mirrors due to deformations of the overall support structure. We again find that the primary mirror is more impacted, with its mount points translated by up to 0.3~mm. For overall mirror translations along or orthogonal to the optical axis, we find that this has a negligible impact on image quality, corresponding to a degradation of the Strehl ratio by 0.01 per cent. However, these translations can also cause an overall tilt of the mirror, with a maximum rotational deviation of 0.04\degrees\ when moving the telescope from zenith to the horizon. This is approximately half of the PSF FWHM, and thus indicates that a non-negligible change in telescope pointing can occur. In practice, this pointing offset should not have a significant impact while dithering to collect a single image, which we expect to typically occur over a relatively small range of ZA. Furthermore, a lookup table as a function of ZA can be readily calibrated to allow for minimal pointing variation in any telescope orientation.

The primary and secondary mirrors were fabricated by N.A.B. Precision Tooling in the United Kingdom, with a requested surface accuracy of $\lambda$/40, corresponding to 50~$\mu$m. This specification was chosen as a reasonable compromise between manufacturing costs and optical performance, and ensures Ruze scattering losses of $\le 10$ per cent. Following fabrication, inspections of both mirrors were performed on 50~$\mu$m grids over the entire surface, yielding rms deviations of 41~$\mu$m for the primary mirror and 29~$\mu$m for the secondary mirror. These rms values correspond to Ruze scattering losses of 6 per cent and 3 per cent, respectively.

\begin{figure*}
    \centering 
    \includegraphics[height=0.36\textwidth]{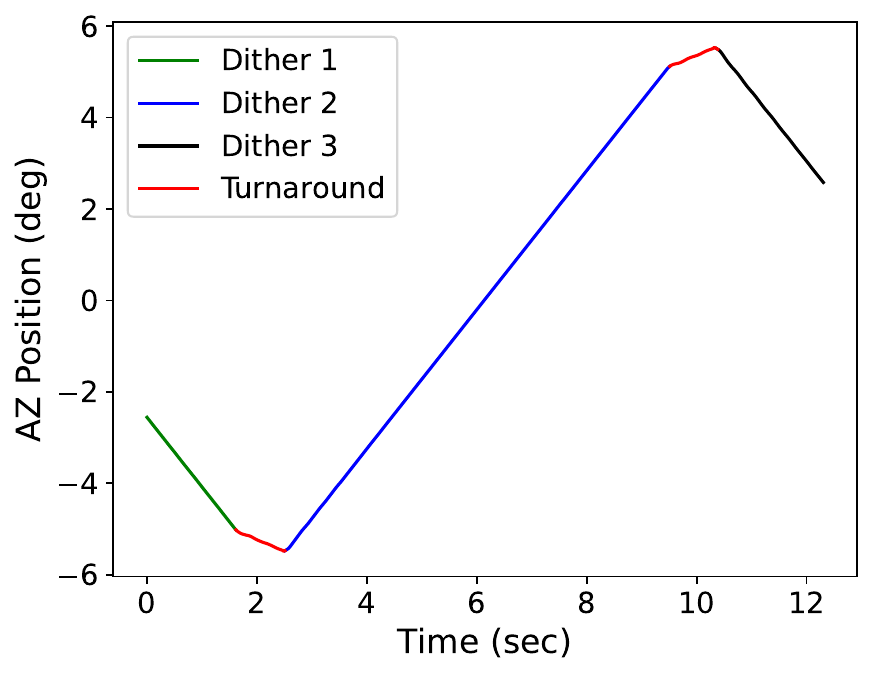}
    \hspace{0.005\textwidth}
    \includegraphics[height=0.36\textwidth]{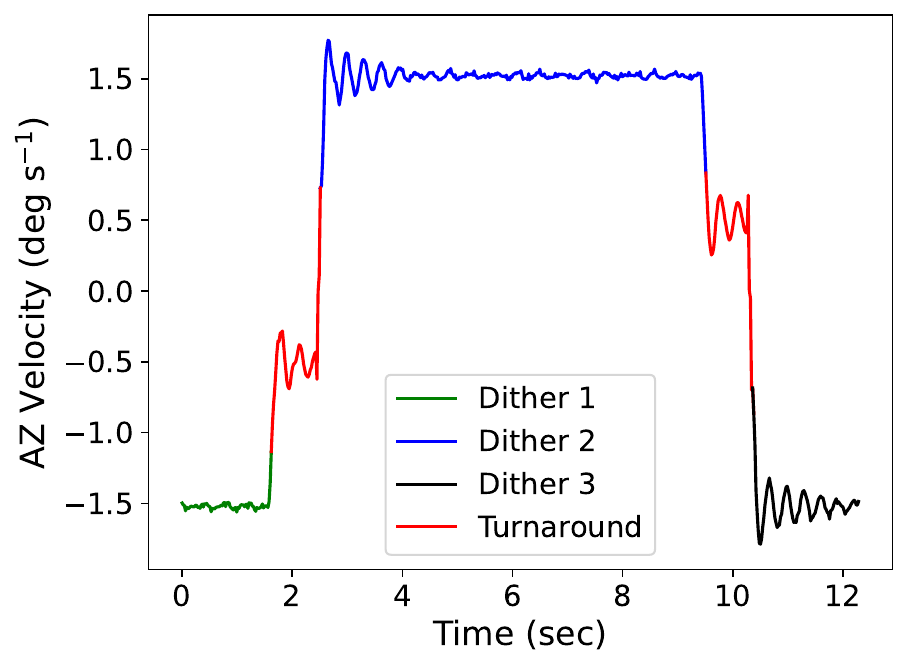}
    \caption{Example telescope motion during a set of dithers. Left indicates the AZ position and right indicates the AZ velocity. At the start of a dither, the velocity deviates from the set value of 1.5\degrees\ s$^{-1}$ by as much as 0.2\degrees\ s$^{-1}$. However, the deviations dampen after approximately 1~s, after which the rms deviations are 0.02\degrees\ s$^{-1}$.}
    \label{fig:dither}
\end{figure*}

The telescope ZA is driven by a Kollmorgen AKM54K motor and AKD motor controller with a gearing ratio of 270 from the motor to the telescope. Our measurement of the ZA position is accurate to approximately 0.00014\degrees, limited by our readout setup and not the intrinsic specifications of the controller. Control of motion is achieved using a programmable analog voltage output and readout of ZA position is obtained via an ethernet communication with the AKD motor controller at a rate of 50~Hz. This setup ensures that the difference between requested and actual ZA is less than 0.0019\degrees, corresponding to approximately 2 per cent of the PSF FWHM.  The telescope AZ is driven by a pair of Kollmorgen AKM33H motors and S706 Motor Controllers. The second AZ motor is maintained at a constant and relatively small torque to minimize backlash while moving the telescope. The AZ gearing ratio is 1169. As with the ZA motors, motion is controlled via an analog voltage output, and AZ position information is delivered through a serial connection with the S706 motor controller at a rate of 50~Hz. We record the AZ position to a precision of 0.00003\degrees, and the maximum difference between requested and actual AZ is 0.0011\degrees, corresponding to approximately 1 per cent of the PSF FWHM. 

Currently, we dither in the AZ direction while collecting imaging data, holding the ZA coordinate fixed. Initially during this dither, the AZ velocity can fluctuate by as much as 0.2\degrees\ s$^{-1}$ from the nominal set velocity of 1.5\degrees\ s$^{-1}$. However, these deviations dampen within approximately 1~s of the start of motion, after which the rms of the velocity is approximately 0.02\degrees\ s$^{-1}$ for the remainder of the dither. See Figure~\ref{fig:dither}. We have not identified any degradation in performance during the initial 1~s settling period, and so we do not currently attempt to correct for this non-ideality (e.g., by discarding the data at the start of the dither).

\begin{figure*}
    \centering 
    \includegraphics[width=0.49\textwidth]{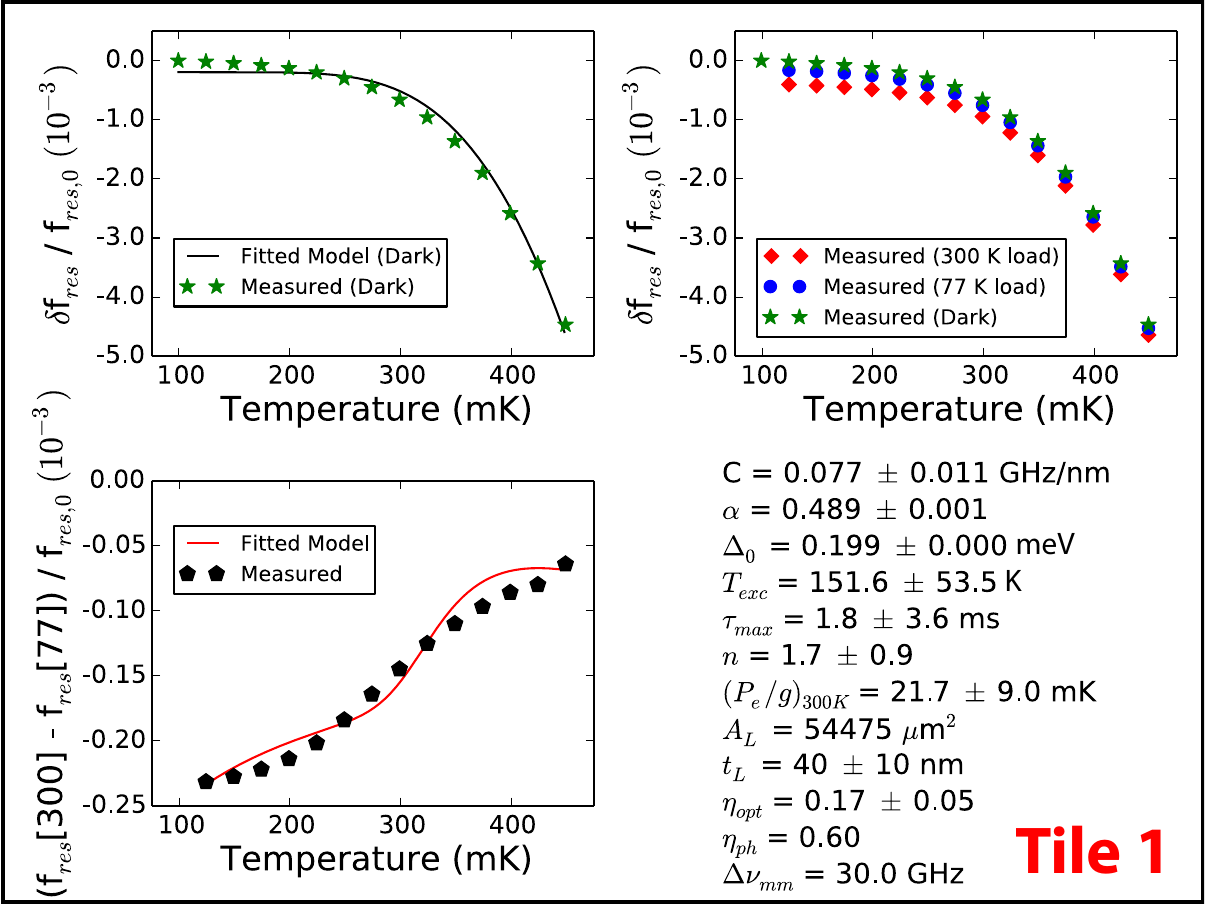}
    \hspace{0.005\textwidth}
    \includegraphics[width=0.49\textwidth]{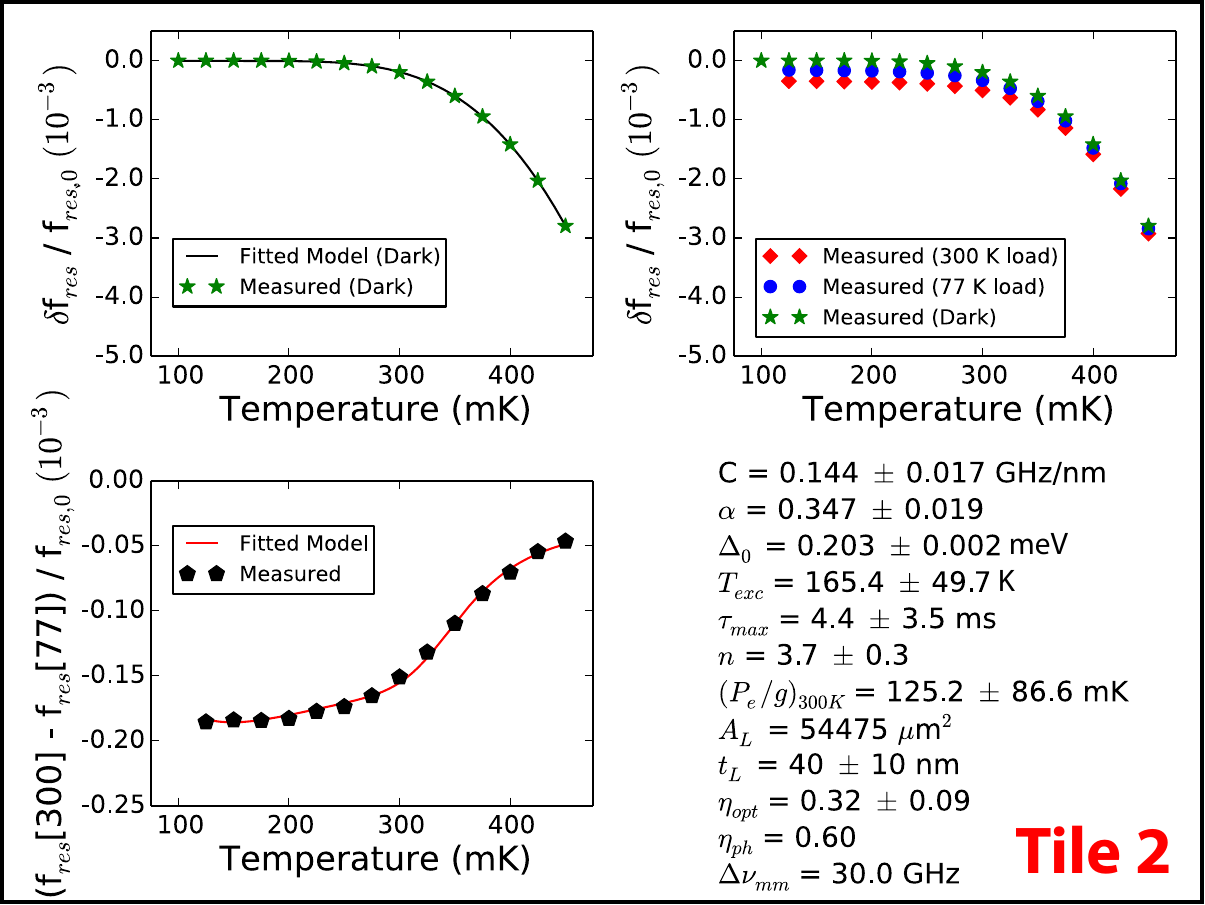}

    \centering 
    \includegraphics[width=0.49\textwidth]{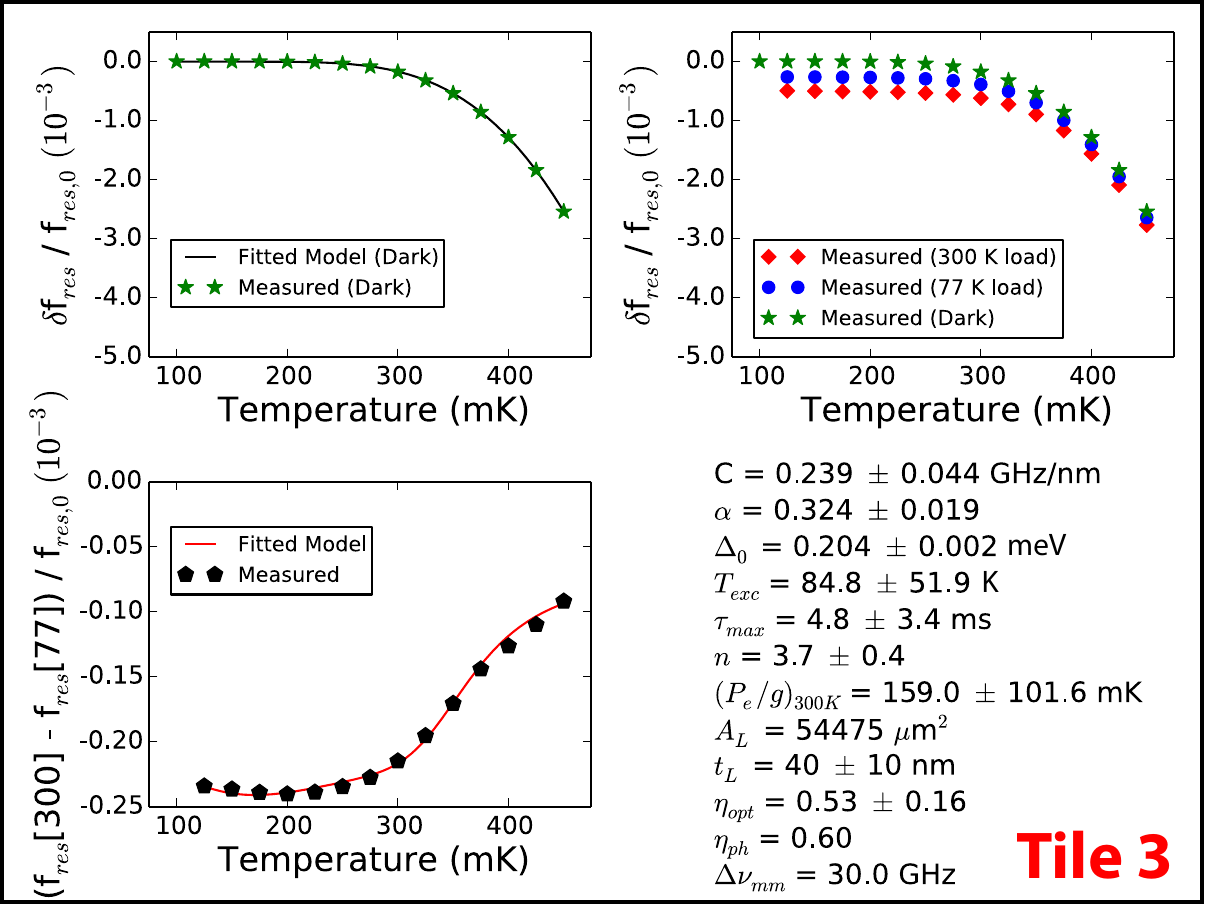}
    \hspace{0.005\textwidth}
    \includegraphics[width=0.49\textwidth]{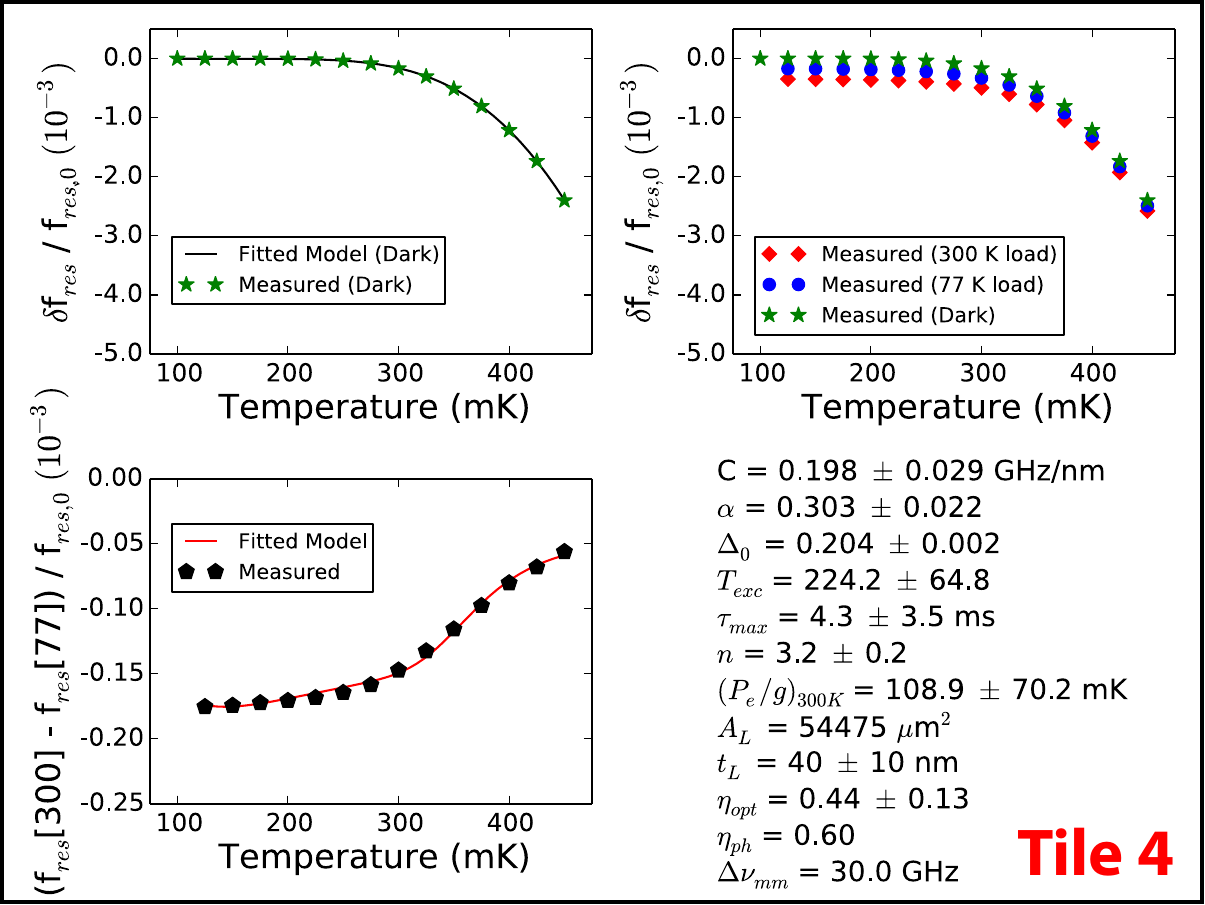}
    \caption{Fits to our superconducting resonator model for example KIDs from each of the four detector tiles. See text for more details. The upper left plot within each panel shows data collected in a dark configuration, with an associated fitted model that does not contain any of the optical parameters. The lower left plot within each panel shows the difference in resonant frequency when the detectors are exposed to two blackbody sources outside of the cryostat window, one at the ambient temperature of approximately 300~K and one immersed in liquid nitrogen at a temperature of approximately 77~K. Overlaid is a fit containing the optical parameters, using priors on the non-optical parameters previously constrained by the dark data. The parameter values from the fit are shown in the bottom right with, [$C$] the conversion from incident optical power to detector response in GHz nm$^{-1}$; [$\alpha$] the kinetic inductance fraction; [$\Delta_0$] the superconducting bandgap parameter in meV; [$T_{exc}$] the excess optical load in K; [$\tau_{max}$] the maximum quasiparticle lifetime in ms; [$n$] the power law index describing a possible elevation in quasiparticle temperature compared to superconductor temperature; [$(P_e/g)_{300K}$] the effective transition when the quasiparticle temperature is elevated from the superconductor temperature in mK; [$A_L$] the area of the aluminum KID inductor; [$t_L$] the thickness of the aluminum KID inductor; [$\eta_{opt}$] the optical efficiency; [$\eta_{ph}$] the efficiency of converting incident power to quasiparticle excitations; [$\Delta \nu_{mm}$] the detector optical bandwidth in GHz. The KIDs on tiles 2--4 are well described by this model, while the KIDs on tile 1 show small but measurable deviations. Tile 1 was fabricated separately from the others, and the aluminum detector layer on that tile may have some contamination.}
    \label{fig:resonator_fits}
\end{figure*}

A novel approach was taken to route the flexible He lines associated with the GM cryocooler from the compressor, which is fixed to the ground nearby the telescope, to the cryostat, which is mounted on the telescope and moves in the AZ direction. Previous instruments have often employed a rotary feedthrough, which then requires connection to a He reservoir to compensate for the continuous loss of He gas at the rotary interface \cite{Crumrine2018,Lee2020}. Instead, our system uses 4~foot lengths of 3/8~inch stainless steel flex line with an external spiral wrap of polyethylene. These lines are routed through the center of the AZ rotation bearing via a polyethylene tube with a length of 1~foot and an inner diameter of 7-7/8~inch. The polyethylene reduces friction and also prevents wear on the outside of the He flex lines. The lines are rigidly clamped to the telescope base on one end, and to the movable AZ yoke on the other end. The smallest bend radius in our installation is a factor of $\simeq 2$ larger than the 5~inch minimum bend radius of the lines. We find that AZ rotations of $\pm 90$\degrees\ relative to the nominal installation position do not cause noticeable strain on the lines, and this range of motion provides access to the full sky given the 360\degrees\ of ZA rotation allowed by our mount.

\section{\label{sec:detectors}Detectors\protect}

\begin{figure*}
    \centering 
    \includegraphics[width=0.49\textwidth]{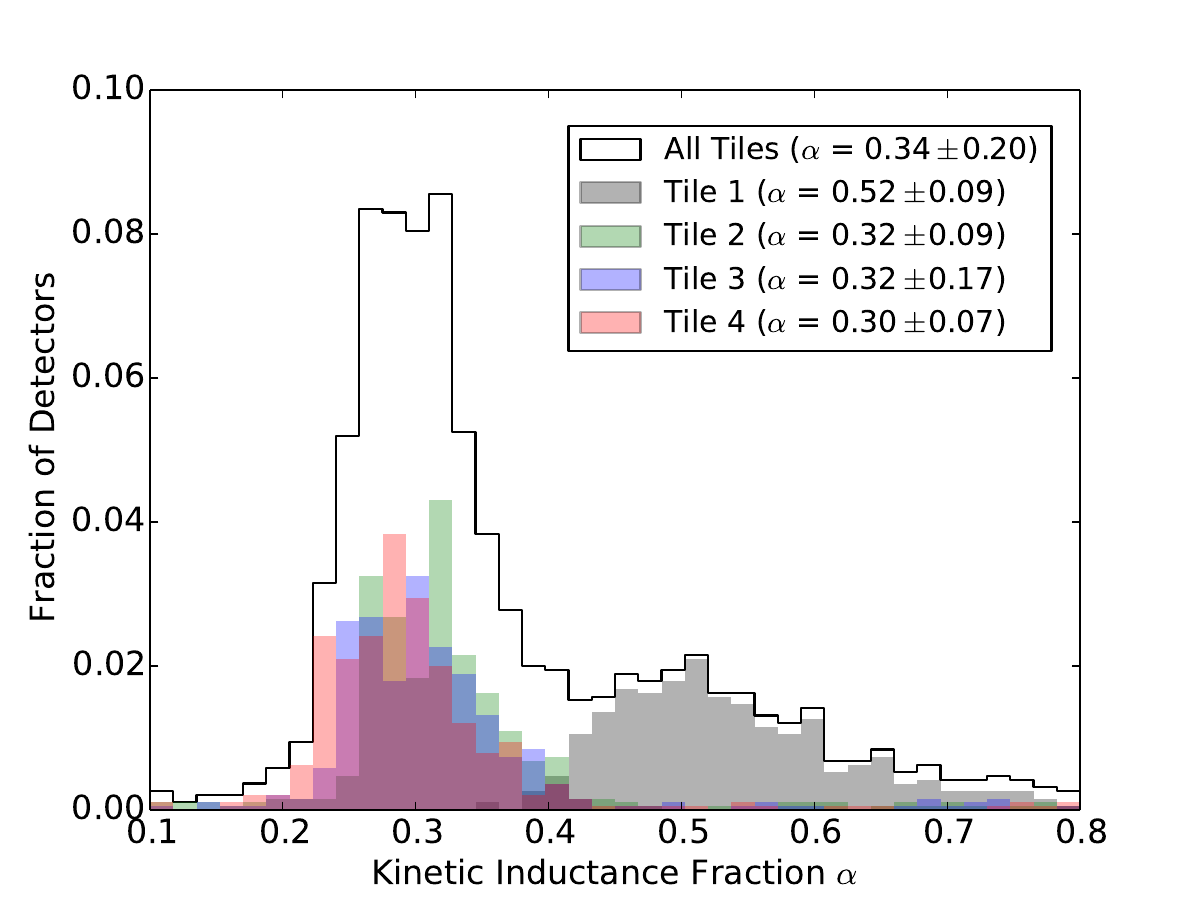}
    \hspace{0.005\textwidth}
    \includegraphics[width=0.49\textwidth]{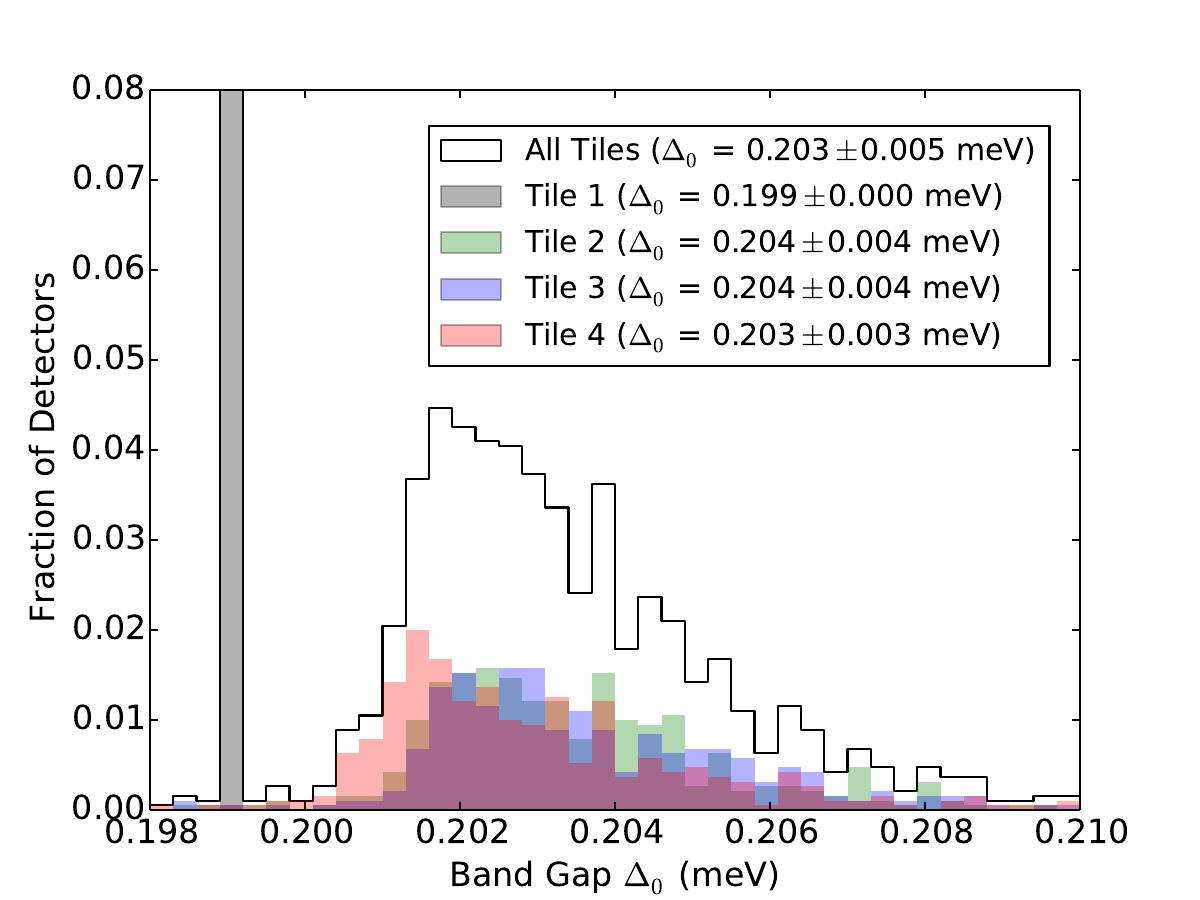}

    \centering 
    \includegraphics[width=0.49\textwidth]{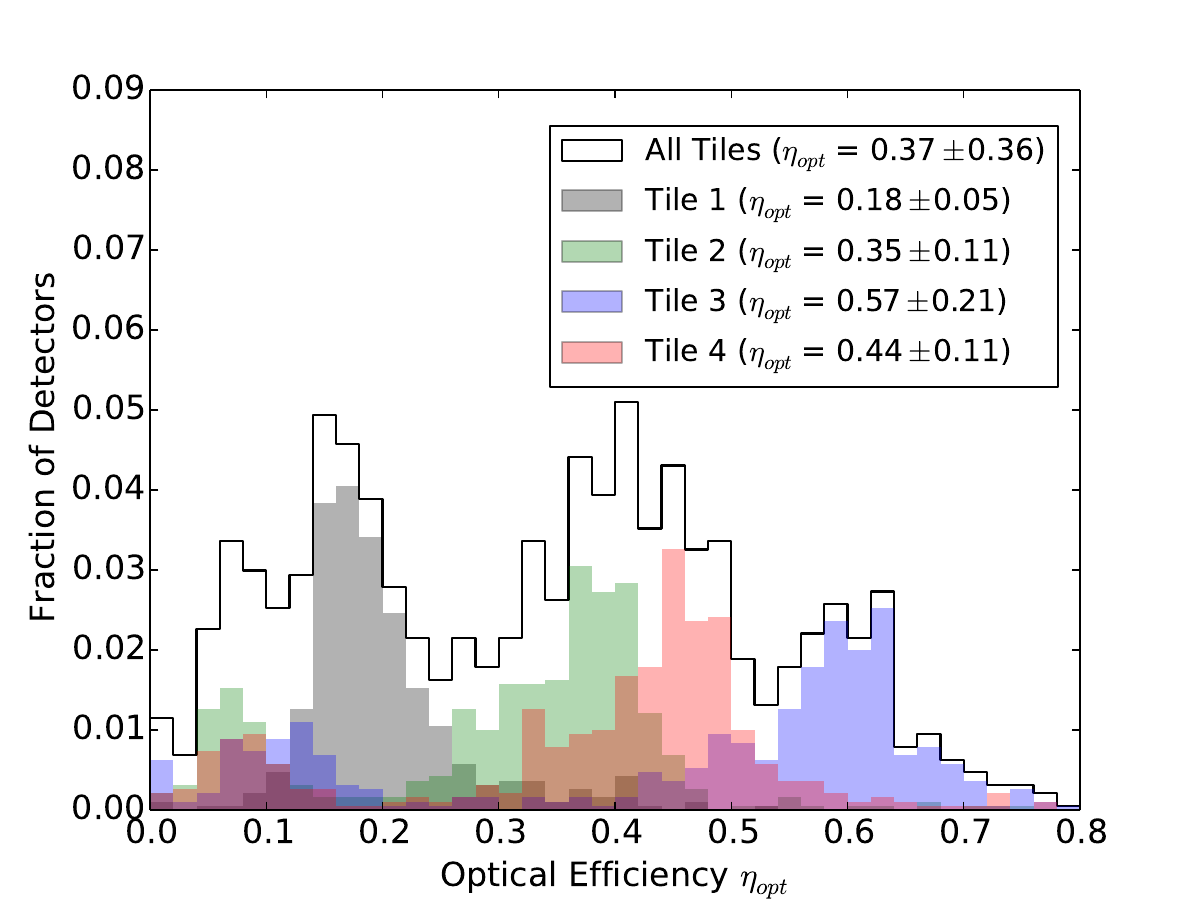}
    \hspace{0.005\textwidth}
    \includegraphics[width=0.49\textwidth]{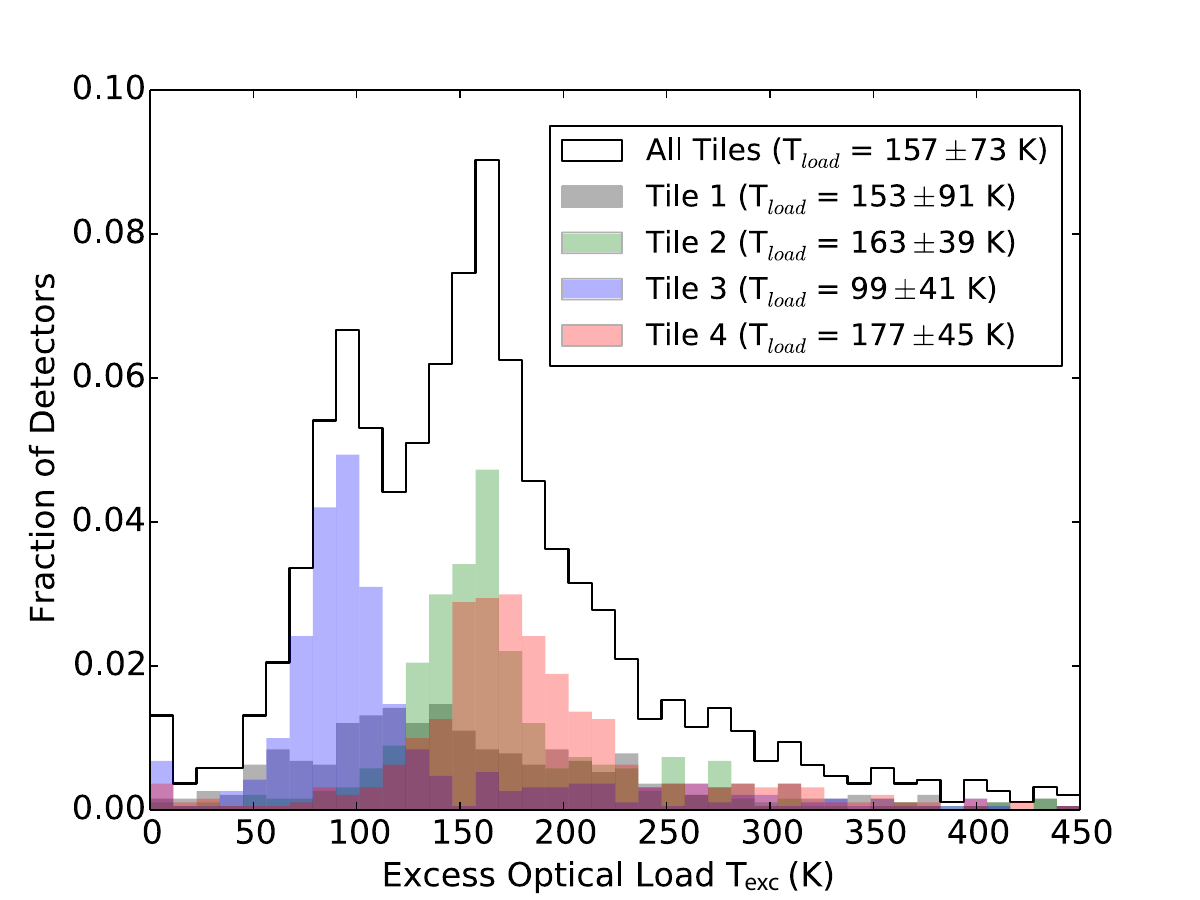}
    \caption{Distribution of fitted KID parameter values for each of the four tiles based on the fits shown in Figure~\ref{fig:resonator_fits} and described in the text. The superconducting properties of the KIDs, described by the kinetic inductance fraction and band gap shown in the tops plots, are consistent among tiles 2--4. As illustrated in Figure~\ref{fig:resonator_fits}, the KIDs on tile 1 are not well described by our resonator model, and the fitted parameters differ from the other tiles. The bottom plots illustrate the optical characteristics of the KIDs, in particular the end-to-end optical efficiency from the cryostat window to the KID along with the excess optical load due to stray light.}
    \label{fig:resonator_params}
\end{figure*}

Our FPA is populated with feedhorn-coupled aluminum KIDs on silicon wafers, following the general design presented in \citet{McCarrick2018} and \citet{Dober2016}. In brief, a monolithic aluminum block is used to create an array of smooth-walled feedhorns and circular waveguides via direct drilling \citep{Groppi2016}. The inductive portion of the KID is placed at the exit aperture of the waveguide, and serves as the absorbing element for the incoming radiation. Two KIDs, sensitive to orthogonal polarizations, are associated with each feedhorn. The KIDs are front-illuminated on the silicon wafer, which has a thickness of 1/4$~\lambda$ and a reflective aluminum layer on the backside. A single FPA unit consists of 480 feedhorns and 960 KIDs, layed out in a hexagonal close-pack configuration within a square region approximately 100~mm in size. The full focal plane consists of four such units, for a total of 3,840 KIDs. See Figure~\ref{fig:cold_stage}.

\subsection{Detector Characterization}
\label{sec:res_model}

 To characterize the performance of the detectors, we follow the general procedure described in detail in \citet{Sayers2020}. In brief, we first perform a sweep with a vector network analyzer (VNA) to search for resonances within the designed band of 200--600~MHz. As summarized in Table~\ref{tab:fpu_summary}, we typically identify $\sim 900$ non-overlapping resonances per tile based on the criteria of being separated from the nearest resonance by at least four line-widths. This corresponds to a per-tile detector yield of $\ge 90$ per cent. Next, we characterize each resonance as a function of tile temperature, in increments of 25~mK between 75--450~mK, with the cryostat in a dark configuration with aluminum covers over the optical windows in the cryostat shell and the 34~K and 3~K stages. See Figure~\ref{fig:resonator_fits}. Using the center frequency of the resonator at each temperature, we fit a four parameter model that includes the resonant frequency in the limit of zero temperature, the maximum quasiparticle lifetime $\tau_{max}$, the superconducting bandgap $\Delta_0$, and the kinetic inductance fraction $\alpha$. Among these parameters, $\Delta_0$ and $\alpha$ have the largest impact on performance. For tiles 2--4, $\Delta_0$ is constrained to a fractional precision of $\simeq 1$ per cent in each resonator, is consistent between tiles, and has a spread of approximately 2 per cent between resonators. Similarly, $\alpha$ is constrained to a fractional precision of $\simeq 6$ per cent in each resonator, is consistent between tiles, and has a spread of approximately 30 per cent between resonators. In contrast, the data for tile 1 are not well described by our resonator model, as illustrated in the top left panel of Figure~\ref{fig:resonator_fits}. This tile was fabricated separately from the others, and the aluminum detector layer appears to have some contamination causing it to deviate from the expected behavior. We have thus fixed $\Delta_0$ for that tile to the value obtained from a direct measurement of the superconducting transition temperature of the aluminum feedline on the tile, equal to 1.31~K or 0.199~meV.

\begin{figure}[t]
  \centering
  \includegraphics[width=\columnwidth]{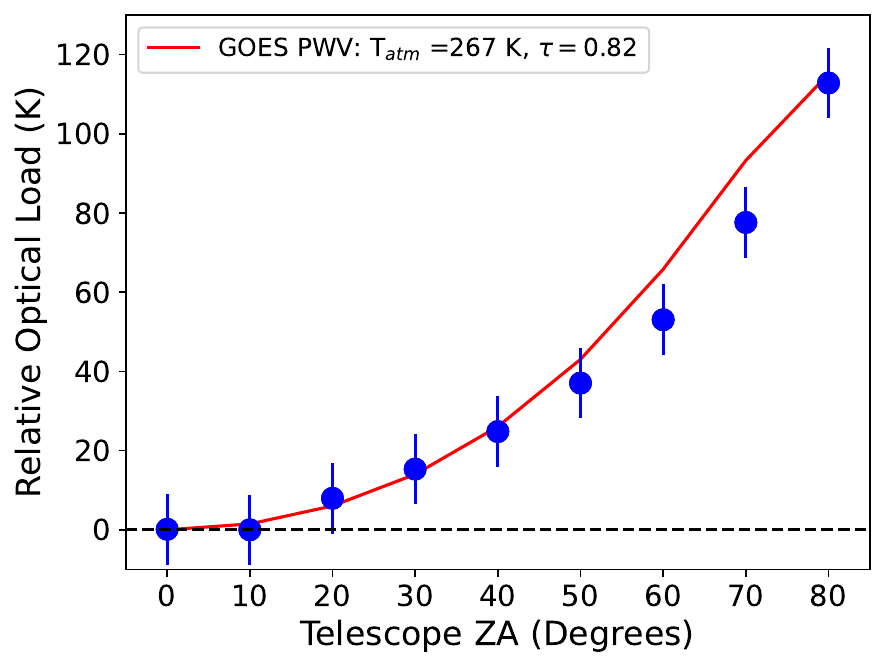}
  \caption{Top: Measured optical load relative to zenith as a function of telescope ZA. Overlaid in red is the expectation based on the measured atmospheric characteristics from the GOES data and the \texttt{fyodor} analysis package \cite{Meier2021}. The agreement is good, indicating that our lab-based calibration model illustrated in Figure~\ref{fig:resonator_fits} accurately describes the KIDs.}
  \label{fig:skydip}
\end{figure}

\begin{table}
     \begin{tabular}{c|c|c|c|c|c}
       Tile & resonances & $\Delta_0$ (meV) & $\alpha$ & $\eta_{opt}$ & $T_{exc}$ (K) \\ \hline
       1 & 866 & 0.199 & $0.52 \pm 0.09$ & $0.18 \pm 0.05$ & $153 \pm 91$ \\
       2 & 903 & $0.204 \pm 0.004$ & $0.32 \pm 0.09$ & $0.35 \pm 0.11$ & $163 \pm 39$ \\
       3 & 885 & $0.204 \pm 0.004$ & $0.32 \pm 0.17$ & $0.57 \pm 0.21$ & $\phantom{1}99 \pm 41$ \\
       4 & 935 & $0.203 \pm 0.003$ & $0.30 \pm 0.07$ & $0.44 \pm 0.11$ & $177 \pm 45$ \\
     \end{tabular}
     \caption{Summary of KID parameters obtained from the fits illustrated in Figure~\ref{fig:resonator_fits} for each of the four detector tiles. Also listed is the total number of resonances identified within the design band of 200--600~MHz. Each tile was fabricated with 960 total KIDs, and so the overall detector yield is $\gtrsim 90$ per cent. The range indicated for each parameter corresponds to the rms among detectors on a given tile, and is generally larger than the fitted precision per KID.}
     \label{tab:fpu_summary}
 \end{table}

The next step in our detector characterization also involves measuring the resonator frequency as a function of temperature, this time while completely filling the outside of the cryostat window with a blackbody absorber either at ambient temperature (approximately 300~K) or emersed in liquid nitrogen (approximately 77~K). Again following the procedure detailed in \citet{Sayers2020}, we difference the 300~K and 77~K resonance frequencies at each detector temperature spanning the range 125--450~mK, where the lowest temperatures used in the dark measurement are excluded because they cannot be readily achieved when the cryostat window is open. See Figure~\ref{fig:resonator_fits}. Using the functions defined in \citet{Sayers2020}, we fit an expanded model to the optical data with four additional free parameters: the detector response to incident optical power $C$ (in GHz~nm$^{-1}$) and the total excess optical load $T_{exc}$, along with two parameters related to possible heating of the quasiparticles relative to the superconductor, $n$ and $P_{opt}/g$. The excess optical load refers to the inferred optical power in the absence of signal from outside the cryostat window, and is due to emission from surfaces inside the cryostat. From these fits, we can then derive the optical efficiency of the imager $\eta_{opt}$, assuming fixed values for the efficiency of quasiparticle creation ($\eta_{qp} = 0.60$), the mm-wave bandwidth ($\Delta \nu = 30$~GHz), the area of the inductive portion of the KID ($A_L = 54475$~$\mu$m$^2$), and the recombination constant ($R = 9.4$~$\mu$m$^3$ s$^{-1}$) \cite{Golwala2012}. We also assume a thickness of the KID, allowing for variation in fabrication, of $t_L = 40$~nm with an rms of 10~nm.

The parameters most relevant to imager performance are the optical efficiency and excess optical load. Typically, $\eta_{opt}$ is constrained to a per-KID fractional precision of $\simeq 25$ per cent, limited entirely by the prior on the detector thickness, with a similar spread in values between resonators. The per-KID value of $T_{exc}$ is generally constrained to a precision of $\simeq 50$~K, corresponding to a fractional uncertainty of $\simeq 30$ per cent, again with a similar spread in values between resonators. See Figure~\ref{fig:resonator_params} and Table~\ref{tab:fpu_summary}. Aggregating all tiles, the median optical efficiency is $\eta_{opt} = 0.37$, which increases to $\eta_{opt} = 0.45$ when excluding Tile 1. For comparison, the predicted median optical efficiency is $\eta_{opt} = 0.39$, estimated based on: a total transmission of 0.68 for the optical filter stack described in Table~\ref{tab:filters}; a total optical coupling efficiency of 0.60 for the system comprised of the feedhorn, waveguide, and absorbing inductive portion of the KID; and a median efficiency of 0.96 as a result of vignetting. This vignetting is primarily due to the filter holder on the 34~K stage, and this median detector is located approximately 30~mm from the edge of the focal plane. It is important to note that, while the median efficiency due to vignetting is 0.96, a small number of detectors have much higher vignetting. As specific examples of the efficiency due to vignetting, 25th percentile detectors located 15~mm from the focal plane edge have an efficiency of 0.80 and 10th percentile detectors located 5~mm from the focal plane edge have an efficiency of 0.61. Returning to the overall median, our measured optical efficiency is in good agreement with the predicated median value of $\eta_{opt} = 0.39$, and it is also comparable to the typical value of $\eta_{opt} \simeq 0.4$ achieved in state-of-the-art imagers at these wavelengths \cite{Choi2018,Sobrin2022}. 

Again aggregating all tiles, the median excess optical load is $T_{exc} = 157$~K. This is approximately an order of magnitude higher than the best values obtained from ground-based imagers in this wavelength regime when utilizing a cryogenic aperture stop and baffling \cite{Grace2014,Sobrin2018}, both of which are absent from our system. This allows for direct coupling to 34~K and ambient temperature surfaces over a large range of potential incidence angles. While the exact mechanisms responsible for producing the excess load have not been determined, we do note that the transmission of the QMC shaders and LPE filters is modified at large incidence \cite{Ade2006}, and so the large-angle coupling to warm surfaces can result in both significant in-band optical power being absorbed by the detector, along with out-of-band power at higher frequencies. Regardless of the source of this loading, it is still smaller than the expected ambient optical load of $\simeq 300$~K for terrestrial imaging. Thus, the impact on sensitivity from this excess load is modest.

With fitted parameter values for the resonator model for each KID, it is then possible to determine the loading-dependent calibration factor to convert from a fractional frequency shift to a signal brightness in units of K. To verify this calibration model, we performed a skydip measurement to characterize the resonant frequency of each detector as a function of telescope ZA. See Figure~\ref{fig:skydip}. Based on this measurement, the resonator model suggests that the total optical load on the KIDs is approximately 120~K colder when the telescope is pointed to zenith ($\textrm{ZA} = 0$) compared to the horizon ($\textrm{ZA} = 90$). To assess this result, we determined the column depth of precipitable water vapor (PWV) using the the publicly available \texttt{fyodor} package based on time-synchronous data from the Geostationary Operational Environmental Satellites (GOES) \cite{Meier2021}. The PWV was approximately 28~mm, which suggests a zenith opacity of 0.82 in our 148~GHz observing band \cite{Pardo2001}. The GOES data further allow an estimate of the effective temperature of the atmosphere weighted by the distribution of water vapor, which was 267~K. As shown in Figure~\ref{fig:skydip}, the predicted optical loading from the atmosphere as a function of ZA based on the GOES data agrees very well with our measured values based on the fitted resonator model, indicating that the model parameters obtained from our lab-based calibration are accurate.

\subsection{Detector Noise}

\begin{figure*}[t]
  \centering
  \includegraphics[width=0.95\columnwidth]{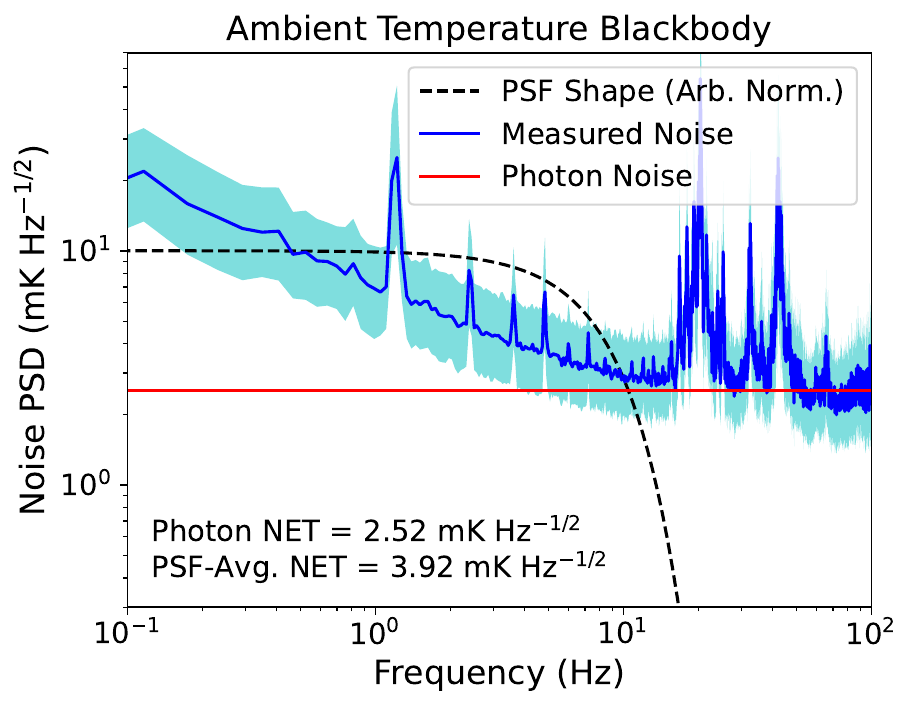}
  \hspace{0.08\columnwidth}
  \includegraphics[width=0.95\columnwidth]{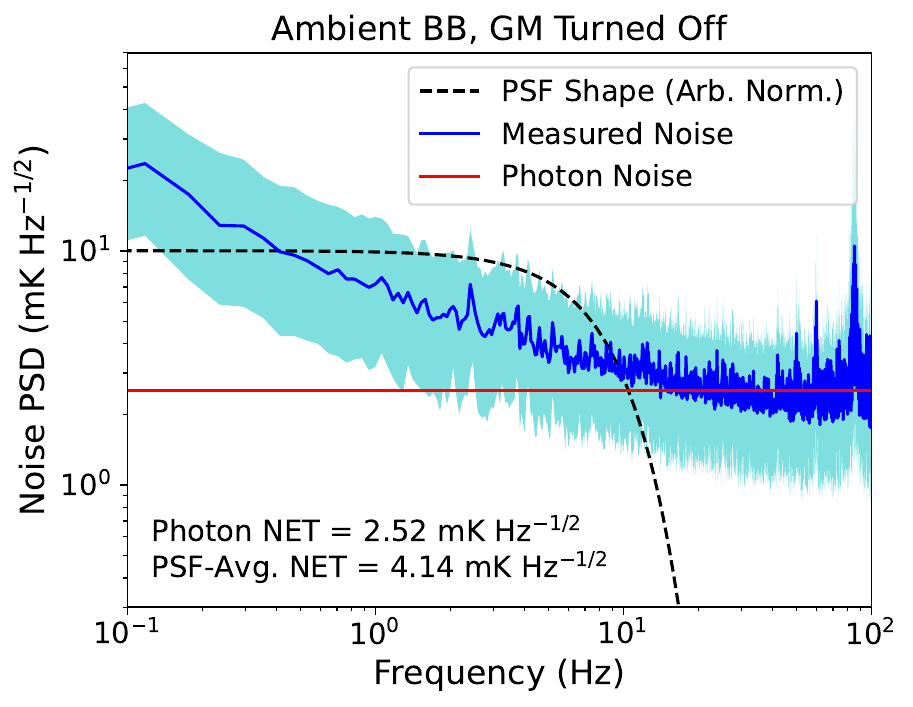}
  \includegraphics[width=0.95\columnwidth]{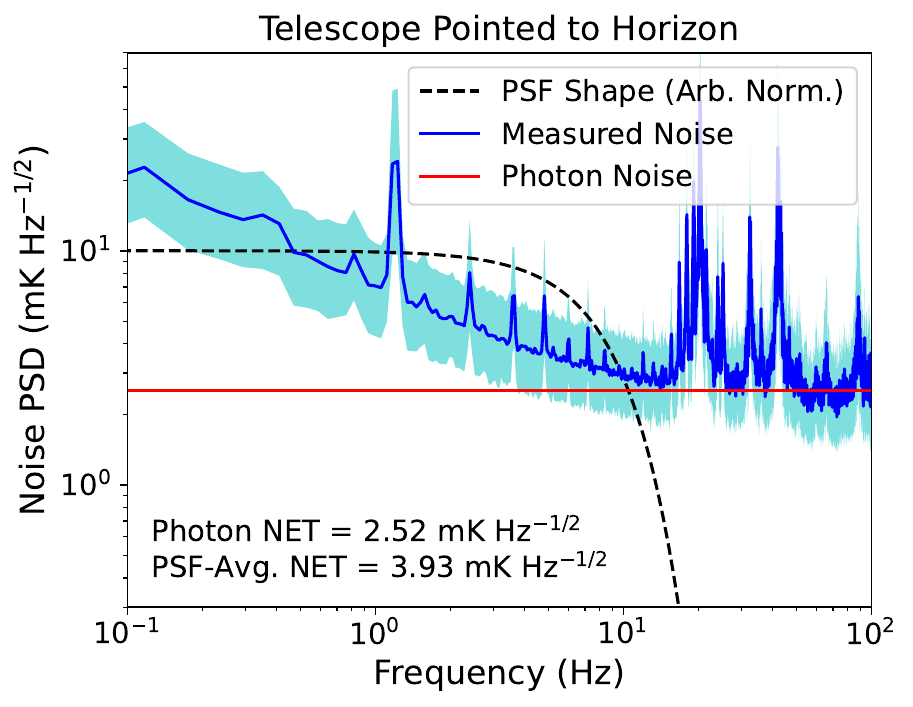}
  \hspace{0.08\columnwidth}
  \includegraphics[width=0.95\columnwidth]{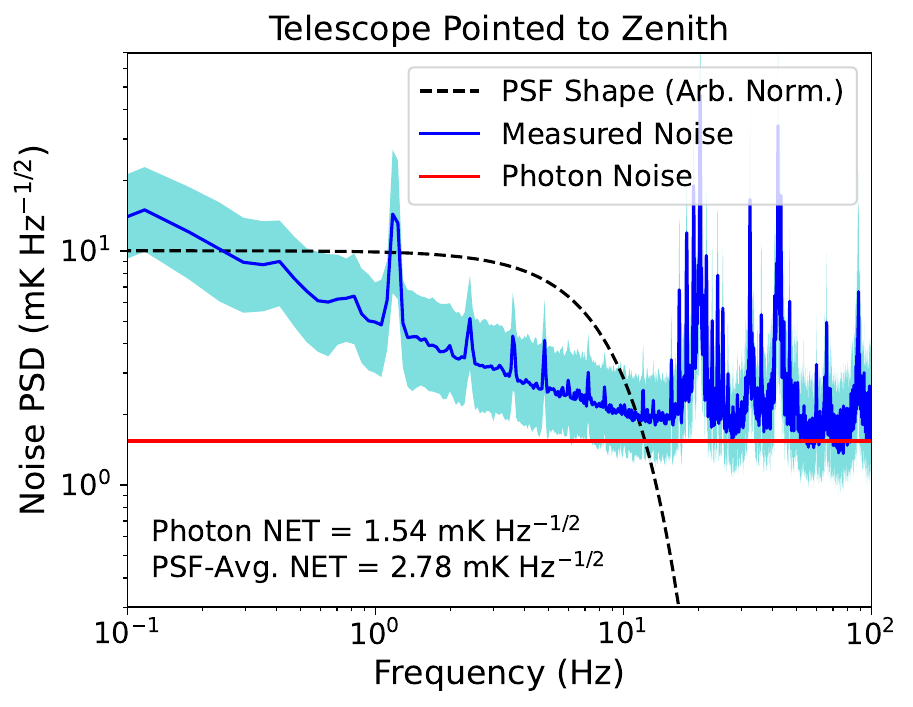}
  \caption{Top: Measured detector noise PSDs in four different configurations. Clockwise from the upper left the plots show noise PSDs with a blackbody absorber filling the instrument window; again with the blackbody absorber filling the window but with the GM cryocooler turned off; with the telescope pointed to zenith ($\textrm{ZA} = 0$), and with the telescope pointed to the horizon ($\textrm{ZA} = 90$). The photon noise limit in each case is denoted by a red line, with its value given in the lower left of each plot. the median per-KID noise spectrum is shown as a dark blue line, with the cyan region enclosing 68 per cent of the KIDs. Also displayed in the lower left of each plot is the effective NET weighted by the imager PSF, which has a HWHM of 7.4~Hz at our nominal dither speed of 1.5\degrees~s$^{-1}$. At frequencies $\gtrsim 10$~Hz the imager is photon noise limited away from the narrow pickup lines associated with the GM cryocooler. Excess noise is observed below $\simeq 10$~Hz with an amplitude scaling like $f^{-1/2}$. We have excluded TLS and microphonics or other pickup from the GM cryocooler as possible sources for this low-frequency noise, and it does not appear in lab-based measurements. The origin of this low-frequency noise is currently unclear, but we plan future testing to isolate its source.}
  \label{fig:noise_spectra}
\end{figure*}

To measure the {\it in situ} KID noise, we collect data streams from all of the detectors for a fixed length of time, generally a few minutes. Using the LO sweep measurement described in Section~\ref{sec:readout}, we then convert the complex-valued transmission of each detector to a fractional frequency shift. This frequency shift is then converted to units of brightness temperature based on the resonator model detailed in Section~\ref{sec:res_model}. The resulting data streams are generally correlated, due to, for example, temperature fluctuations common to all the detectors. To remove this correlated noise, we form an average template from all of the KID data streams, and then subtract this average template from each individual KID data stream. The data streams are then separated into non-overlapping 10-second blocks, and the noise power spectral density (PSD) is computed separately for each KID and each block. A single noise PSD is then estimated for each individual KID by averaging the PSDs obtained from all of the 10-second blocks. To determine the array-average performance, at each spectral frequency we determine the median noise PSD along with the range enclosing 68 per cent of the KIDs. The results from this analysis are shown in Figure~\ref{fig:noise_spectra} and discussed below.

The SKIPR instrument is nominally designed to image sources with an ambient temperature background near 300~K. Ideally, the detector noise should be limited by the random arrival of photons from this background, along with the $\simeq 150$~K of excess loading we measure, and this photon noise corresponds to a PSD with a noise equivalent temperature (NET) of $\simeq 2.5$~mK~Hz$^{-1/2}$. From our resonator model, we also predict a quasi-negligible contribution from generation-recombination of quasiparticles in the KIDs, with an expected noise PSD of $\simeq 0.6$~mK~Hz$^{-1/2}$. As shown in Figure~\ref{fig:readout_noise}, there is an additional quasi-negligible contribution from the readout electronics, with an amplitude of $\simeq 0.7$~mK~Hz$^{-1/2}$. Previous lab-based measurements of similar detectors in \citet{Sayers2020} indicate that noise from two-level-systems (TLS) is negligible. Thus, we expect the total KID noise PSD to be approximately equal to the value of $\simeq 2.5$~mK~Hz$^{-1/2}$ due to photon noise.

To assess the KID noise, we collected data in three different configurations with the expected ambient background of $\simeq 300$~K. Two of these configurations involve covering the imager window with a blackbody absorber, in one case with the GM cryocooler operating normally and in the other case with it turned off. The third configuration involved orienting the imager telescope to the horizon (i.e., $\textrm{ZA} = 90$). As illustrated in Figure~\ref{fig:noise_spectra}, the noise PSD from all three of these configurations is consistent, and equal to the expected NET value of 2.5~mK~Hz$^{-1/2}$ at frequencies $\gtrsim 10$~Hz with the exception of a number of narrow pickup lines, nearly all of which are associated with the GM cryocooler. We note that the cryocooler operates at approximately 1.2~Hz, and these vibrations produce the large noise pickup line at that frequency along with smaller amplitude lines at the associated harmonics. The lowest natural frequencies of the bipod supports for the focal plane are between 35--50~Hz, and so vibrations of those structures due to the GM cryocooler likely produce the two broad pickup features at those frequencies. The origin of the pickup features between 20--30~Hz is unclear.

At lower frequencies the smooth component of the noise PSD has a rising amplitude, scaling approximately like $f^{-1/2}$. Although the spectral shape differs from the typical expectation for TLS, which is $f^{-1/4}$, we performed a range of tests to more definitively assess TLS as a potential noise source. In particular, previous work has demonstrated that the amplitude of TLS noise scales with both resonator temperature $T_{res}$ and readout power $P_{res}$ \cite{zmuidzinas_2012,Gao2008}, and so we collected noise data varying $T_{res}$ between 150--300~mK and $P_{res}$ decreasing from its nominal value by as much as 18~dB. Higher readout powers are not possible because the resonators bifurcate. Over the range of temperatures we considered, the amplitude of the low-frequency noise scales like $T_{res}^{0.1}$, far from the expected scaling of $\simeq T_{res}^{-1.7}$ for TLS. The data with varying readout power also did not match expectations for TLS, with our measured scaling of $P_{res}^{-0.2}$ being much shallower than the predicted $P_{res}^{-0.5}$. We thus determine that the excess low frequency noise in our data is likely not due to TLS. We can also exclude microphonics or other pickup noise associated with the GM cryocooler, since the low-frequency noise spectrum is unchanged with the cryocooler turned off. Furthermore, as detailed in \citet{Sayers2020}, this low-frequency noise did not appear in lab-based measurements of the KIDs with this geometry. It is currently unclear why this additional noise is observed at the outdoor location, and future testing is planned to isolate the cause.

While additional characterization is needed to understand the source of the excess low-frequency noise, we have estimated its impact on the imager performance. Weighting by the PSF shape for the typical telescope dithering speed of 1.5\degrees\ s$^{-1}$, corresponding to a half-width at half maximum (HWHM) of 7.4~Hz for our median measured FWHM of 0.089\degrees, the effective noise PSD viewing a 300~K ambient background corresponds to an NET of $\simeq 3.9$~mK~Hz$^{-1/2}$, a factor of approximately $\times$1.5 higher than the photon noise limit. As seen in the noise spectra in Figure~\ref{fig:noise_spectra}, this degradation factor also includes the impact of a small number of narrow pickup lines associated with the GM cryocooler. In addition, we note that faster dithering of the telescope provides a general strategy to mitigate the low-frequency noise by modulating the PSF signal band to higher frequencies. However, as illustrated by the noise spectra in Figure~\ref{fig:noise_spectra}, there is a significant forest of pickup lines associated with the GM cryocooler at $\gtrsim 20$~Hz. Because dithering faster than $\simeq 1.5$\degrees\ s$^{-1}$ would move these lines into the PSF signal band, this is not a viable option for SKIPR.

Along with collecting noise data viewing a 300~K ambient background, we also collected noise data with the telescope oriented to zenith (i.e., $\textrm{ZA} = 0$), where the ambient background is approximately 180~K. Under these conditions, the expected photon noise NET decreases to $\simeq 1.5$~mK~Hz$^{-1/2}$. As shown in the bottom right panel of Figure~\ref{fig:noise_spectra}, we measure a noise spectrum close to this value at frequencies $\gtrsim 10$~Hz. The rising spectrum at low frequency seen with a 300~K ambient background is also apparent with a similar amplitude and spectral shape. At zenith, where the atmosphere is more transmissive, it is also possible to observe noise due turbulent fluctuations of atmospheric water vapor \cite{Sayers2010,Errard2015}, which have a noise spectrum that scales approximately like $f^{-1.5}$. Such atmospheric fluctuations are not evident in our measured noise spectra while looking at zenith. While the common-mode template removal applied to the data streams could remove the atmospheric fluctuations, we note that there is also no excess low-frequency noise when comparing the noise spectrum of the raw data streams prior to common-mode subtraction. Thus, we conclude that the excess low-frequency noise of unknown origin must have a larger amplitude than any noise due to atmospheric fluctuations, at least for $f \ge 0.1$~Hz. For the noise data collected while looking toward zenith, we again compute the PSF-weighted NET, finding a value of $\simeq 2.8$~mK~Hz$^{-1/2}$, corresponding to a factor of $\times$1.8 worse than the expected photon limit.

\section{\label{sec:summary}Summary\protect}

Motivated by the desire to obtain passive imaging through optical obscurants, we have developed SKIPR. It provides photometric polarization-sensitive imaging at 150~GHz via an FPA populated with 3,840 KIDs. The FPA is cooled to its nominal operating temperature by a GM cryocooler paired with a two-stage ADR, the latter of which provides approximately six hours of hold time at 200~mK via a cycle that requires a little over one hour to complete. The SKIPR readout electronics are based on an RFSoC module, with a single readout channel dedicated to the 960 KIDs on each of the four FPA tiles. The measured readout noise is well below the fundamental noise floor set by the random arrival of photons from the ambient background.

SKIPR utilizes a telescope with a crossed Dragone geometry and a 1.59~m diameter primary mirror on a controllable alt/az mount. The image quality is excellent over the entire 3.43\degrees\ square FOV, and the measured PSF shapes are in good agreement with the diffraction-limited expectation of a FWHM equal to 0.082\degrees. At a level of $\simeq -20$~dB relative to the peak, some stray-light artifacts are visible in the measured PSFs of detectors near the edge of the FOV. While imaging, the telescope is dithered at 1.5\degrees\ s$^{-1}$ in the AZ direction, with a positional accuracy of $\lesssim 0.002$\degrees\ and a typical velocity jitter of $\simeq 0.02$\degrees\ s$^{-1}$.

To calibrate the KIDs populating the SKIPR FPA, we performed a detailed lab-based characterization. From these measurements, we find good uniformity among the detectors. This includes an average optical efficiency of 0.37, in good agreement with the expected value of 0.39, and an excess optical loading of 157~K, well below the $\simeq 300$~K load due to the ambient background. To verify the accuracy of our calibration, we measured the detector response as a function of telescope ZA, which we found to be in good agreement with expectations based on the real-time properties of the atmosphere measured by the GOES satellites.

Finally, we measured the {\it in situ} noise performance of the KIDs, finding that it reaches a level of $\simeq 2.5$~mK~Hz$^{-1/2}$ limited by the random arrival of photons from the ambient $\simeq 300$~K background. However, we also observe pickup noise in narrow spectral lines due to the GM cryocooler, particularly above $\simeq 20$~Hz, along with a rising noise spectrum below $\simeq 10$~Hz. In aggregate, these non-idealities degrade the effective PSF-weighted NET of the SKIPR system by a factor of approximately $\times$1.5 relative to the fundamental photon noise limit. 

In aggregate, our characterization of SKIPR indicates that it performs close to expectations, and the system is thus ready for imaging tests.

\section*{Aknowledgements}

This work was supported by the Office of Naval Research, award numbers N000141812846 and N000142012430. This   research was carried out at the Jet Propulsion Laboratory, California Institute of Technology, and was sponsored by the Office of Naval Research through an agreement with the National Aeronautics and Space Administration. Reference  herein to any specific commercial  product, process, or service by trade name, trademark, manufacturer, or otherwise, does not constitute or imply its endorsement by the United States Government, the Office of Naval Research, or the Jet Propulsion Laboratory, California Institute of Technology. This document has been approved for public release with document number DCN\# 2024-11-1-293.

\section*{References}
\bibliography{SKIPR1}

\begin{thebibliography}{55}%
\makeatletter
\providecommand \@ifxundefined [1]{%
 \@ifx{#1\undefined}
}%
\providecommand \@ifnum [1]{%
 \ifnum #1\expandafter \@firstoftwo
 \else \expandafter \@secondoftwo
 \fi
}%
\providecommand \@ifx [1]{%
 \ifx #1\expandafter \@firstoftwo
 \else \expandafter \@secondoftwo
 \fi
}%
\providecommand \natexlab [1]{#1}%
\providecommand \enquote  [1]{``#1''}%
\providecommand \bibnamefont  [1]{#1}%
\providecommand \bibfnamefont [1]{#1}%
\providecommand \citenamefont [1]{#1}%
\providecommand \href@noop [0]{\@secondoftwo}%
\providecommand \href [0]{\begingroup \@sanitize@url \@href}%
\providecommand \@href[1]{\@@startlink{#1}\@@href}%
\providecommand \@@href[1]{\endgroup#1\@@endlink}%
\providecommand \@sanitize@url [0]{\catcode `\\12\catcode `\$12\catcode
  `\&12\catcode `\#12\catcode `\^12\catcode `\_12\catcode `\%12\relax}%
\providecommand \@@startlink[1]{}%
\providecommand \@@endlink[0]{}%
\providecommand \url  [0]{\begingroup\@sanitize@url \@url }%
\providecommand \@url [1]{\endgroup\@href {#1}{\urlprefix }}%
\providecommand \urlprefix  [0]{URL }%
\providecommand \Eprint [0]{\href }%
\providecommand \doibase [0]{http://dx.doi.org/}%
\providecommand \selectlanguage [0]{\@gobble}%
\providecommand \bibinfo  [0]{\@secondoftwo}%
\providecommand \bibfield  [0]{\@secondoftwo}%
\providecommand \translation [1]{[#1]}%
\providecommand \BibitemOpen [0]{}%
\providecommand \bibitemStop [0]{}%
\providecommand \bibitemNoStop [0]{.\EOS\space}%
\providecommand \EOS [0]{\spacefactor3000\relax}%
\providecommand \BibitemShut  [1]{\csname bibitem#1\endcsname}%
\let\auto@bib@innerbib\@empty
\bibitem [{\citenamefont {{Chen}}(1975)}]{Chen1975}%
  \BibitemOpen
  \bibfield  {author} {\bibinfo {author} {\bibfnamefont {C.~C.}\ \bibnamefont
  {{Chen}}},\ }\href@noop {} {\enquote {\bibinfo {title} {{Attenuation of
  electromagnetic radiation by haze, fog, clouds, and rain}},}\ } (\bibinfo
  {year} {1975})\BibitemShut {NoStop}%
\bibitem [{\citenamefont {{Kokhanovsky}}(2004)}]{Kokhanovsky2004}%
  \BibitemOpen
  \bibfield  {author} {\bibinfo {author} {\bibfnamefont {A.}~\bibnamefont
  {{Kokhanovsky}}},\ }\bibfield  {title} {\enquote {\bibinfo {title} {{Optical
  properties of terrestrial clouds}},}\ }\href {\doibase
  10.1016/S0012-8252(03)00042-4} {\bibfield  {journal} {\bibinfo  {journal}
  {Earth Science Reviews}\ }\textbf {\bibinfo {volume} {64}},\ \bibinfo {pages}
  {189--241} (\bibinfo {year} {2004})}\BibitemShut {NoStop}%
\bibitem [{\citenamefont {{Warren}}(2019)}]{Warren2019}%
  \BibitemOpen
  \bibfield  {author} {\bibinfo {author} {\bibfnamefont {S.~G.}\ \bibnamefont
  {{Warren}}},\ }\bibfield  {title} {\enquote {\bibinfo {title} {{Optical
  properties of ice and snow}},}\ }\href {\doibase 10.1098/rsta.2018.0161}
  {\bibfield  {journal} {\bibinfo  {journal} {Philosophical Transactions of the
  Royal Society of London Series A}\ }\textbf {\bibinfo {volume} {377}},\
  \bibinfo {pages} {20180161} (\bibinfo {year} {2019})}\BibitemShut {NoStop}%
\bibitem [{\citenamefont {{Liebe}}(1985)}]{Liebe1985}%
  \BibitemOpen
  \bibfield  {author} {\bibinfo {author} {\bibfnamefont {H.~J.}\ \bibnamefont
  {{Liebe}}},\ }\bibfield  {title} {\enquote {\bibinfo {title} {{An updated
  model for millimeter wave propagation in moist air}},}\ }\href {\doibase
  10.1029/RS020i005p01069} {\bibfield  {journal} {\bibinfo  {journal} {Radio
  Science}\ }\textbf {\bibinfo {volume} {20}},\ \bibinfo {pages} {1069--1089}
  (\bibinfo {year} {1985})}\BibitemShut {NoStop}%
\bibitem [{\citenamefont {{Wentz}}\ and\ \citenamefont
  {{Meissner}}(2016)}]{Wentz2016}%
  \BibitemOpen
  \bibfield  {author} {\bibinfo {author} {\bibfnamefont {F.~J.}\ \bibnamefont
  {{Wentz}}}\ and\ \bibinfo {author} {\bibfnamefont {T.}~\bibnamefont
  {{Meissner}}},\ }\bibfield  {title} {\enquote {\bibinfo {title} {{Atmospheric
  absorption model for dry air and water vapor at microwave frequencies below
  100 GHz derived from spaceborne radiometer observations}},}\ }\href {\doibase
  10.1002/2015RS005858} {\bibfield  {journal} {\bibinfo  {journal} {Radio
  Science}\ }\textbf {\bibinfo {volume} {51}},\ \bibinfo {pages} {381--391}
  (\bibinfo {year} {2016})}\BibitemShut {NoStop}%
\bibitem [{\citenamefont {{Pardo}}, \citenamefont {{Cernicharo}},\ and\
  \citenamefont {{Serabyn}}(2001)}]{Pardo2001}%
  \BibitemOpen
  \bibfield  {author} {\bibinfo {author} {\bibfnamefont {J.~R.}\ \bibnamefont
  {{Pardo}}}, \bibinfo {author} {\bibfnamefont {J.}~\bibnamefont
  {{Cernicharo}}}, \ and\ \bibinfo {author} {\bibfnamefont {E.}~\bibnamefont
  {{Serabyn}}},\ }\bibfield  {title} {\enquote {\bibinfo {title} {{Atmospheric
  transmission at microwaves (ATM): an improved model for
  millimeter/submillimeter applications}},}\ }\href {\doibase 10.1109/8.982447}
  {\bibfield  {journal} {\bibinfo  {journal} {IEEE Transactions on Antennas and
  Propagation}\ }\textbf {\bibinfo {volume} {49}},\ \bibinfo {pages}
  {1683--1694} (\bibinfo {year} {2001})}\BibitemShut {NoStop}%
\bibitem [{\citenamefont {{Paine}}(2019)}]{Paine2019}%
  \BibitemOpen
  \bibfield  {author} {\bibinfo {author} {\bibfnamefont {S.}~\bibnamefont
  {{Paine}}},\ }\href {\doibase 10.5281/zenodo.3406483} {\enquote {\bibinfo
  {title} {{The am atmospheric model}},}\ } (\bibinfo {year}
  {2019})\BibitemShut {NoStop}%
\bibitem [{\citenamefont {{Samluk}}\ \emph {et~al.}(2012)\citenamefont
  {{Samluk}}, \citenamefont {{Schuetz}}, \citenamefont {{Dillon}},
  \citenamefont {{Stein}}, \citenamefont {{Robbins}}, \citenamefont
  {{Mackrides}}, \citenamefont {{Martin}}, \citenamefont {{Wilson}},
  \citenamefont {{Chen}},\ and\ \citenamefont {{Prather}}}]{Samluk2012}%
  \BibitemOpen
  \bibfield  {author} {\bibinfo {author} {\bibfnamefont {J.~P.}\ \bibnamefont
  {{Samluk}}}, \bibinfo {author} {\bibfnamefont {C.~A.}\ \bibnamefont
  {{Schuetz}}}, \bibinfo {author} {\bibfnamefont {T.}~\bibnamefont {{Dillon}}},
  \bibinfo {author} {\bibfnamefont {E.~L.}\ \bibnamefont {{Stein}}}, \bibinfo
  {author} {\bibfnamefont {A.}~\bibnamefont {{Robbins}}}, \bibinfo {author}
  {\bibfnamefont {D.~G.}\ \bibnamefont {{Mackrides}}}, \bibinfo {author}
  {\bibfnamefont {R.~D.}\ \bibnamefont {{Martin}}}, \bibinfo {author}
  {\bibfnamefont {J.}~\bibnamefont {{Wilson}}}, \bibinfo {author}
  {\bibfnamefont {C.}~\bibnamefont {{Chen}}}, \ and\ \bibinfo {author}
  {\bibfnamefont {D.~W.}\ \bibnamefont {{Prather}}},\ }\bibfield  {title}
  {\enquote {\bibinfo {title} {{Q-Band Millimeter Wave Imaging in the Far-Field
  Enabled by Optical Upconversion Methodology}},}\ }\href {\doibase
  10.1007/s10762-011-9850-1} {\bibfield  {journal} {\bibinfo  {journal}
  {Journal of Infrared and Millimeter Waves}\ }\textbf {\bibinfo {volume}
  {33}},\ \bibinfo {pages} {54--66} (\bibinfo {year} {2012})}\BibitemShut
  {NoStop}%
\bibitem [{\citenamefont {{Jarosik}}\ \emph {et~al.}(2003)\citenamefont
  {{Jarosik}}, \citenamefont {{Bennett}}, \citenamefont {{Halpern}},
  \citenamefont {{Hinshaw}}, \citenamefont {{Kogut}}, \citenamefont {{Limon}},
  \citenamefont {{Meyer}}, \citenamefont {{Page}}, \citenamefont
  {{Pospieszalski}}, \citenamefont {{Spergel}}, \citenamefont {{Tucker}},
  \citenamefont {{Wilkinson}}, \citenamefont {{Wollack}}, \citenamefont
  {{Wright}},\ and\ \citenamefont {{Zhang}}}]{Jarosik2003}%
  \BibitemOpen
  \bibfield  {author} {\bibinfo {author} {\bibfnamefont {N.}~\bibnamefont
  {{Jarosik}}}, \bibinfo {author} {\bibfnamefont {C.~L.}\ \bibnamefont
  {{Bennett}}}, \bibinfo {author} {\bibfnamefont {M.}~\bibnamefont
  {{Halpern}}}, \bibinfo {author} {\bibfnamefont {G.}~\bibnamefont
  {{Hinshaw}}}, \bibinfo {author} {\bibfnamefont {A.}~\bibnamefont {{Kogut}}},
  \bibinfo {author} {\bibfnamefont {M.}~\bibnamefont {{Limon}}}, \bibinfo
  {author} {\bibfnamefont {S.~S.}\ \bibnamefont {{Meyer}}}, \bibinfo {author}
  {\bibfnamefont {L.}~\bibnamefont {{Page}}}, \bibinfo {author} {\bibfnamefont
  {M.}~\bibnamefont {{Pospieszalski}}}, \bibinfo {author} {\bibfnamefont
  {D.~N.}\ \bibnamefont {{Spergel}}}, \bibinfo {author} {\bibfnamefont {G.~S.}\
  \bibnamefont {{Tucker}}}, \bibinfo {author} {\bibfnamefont {D.~T.}\
  \bibnamefont {{Wilkinson}}}, \bibinfo {author} {\bibfnamefont
  {E.}~\bibnamefont {{Wollack}}}, \bibinfo {author} {\bibfnamefont {E.~L.}\
  \bibnamefont {{Wright}}}, \ and\ \bibinfo {author} {\bibfnamefont
  {Z.}~\bibnamefont {{Zhang}}},\ }\bibfield  {title} {\enquote {\bibinfo
  {title} {{Design, Implementation, and Testing of the Microwave Anisotropy
  Probe Radiometers}},}\ }\href {\doibase 10.1086/346080} {\bibfield  {journal}
  {\bibinfo  {journal} {\apjs}\ }\textbf {\bibinfo {volume} {145}},\ \bibinfo
  {pages} {413--436} (\bibinfo {year} {2003})},\ \Eprint
  {http://arxiv.org/abs/astro-ph/0301164} {arXiv:astro-ph/0301164 [astro-ph]}
  \BibitemShut {NoStop}%
\bibitem [{\citenamefont {{Crill}}\ \emph {et~al.}(2003)\citenamefont
  {{Crill}}, \citenamefont {{Ade}}, \citenamefont {{Artusa}}, \citenamefont
  {{Bhatia}}, \citenamefont {{Bock}}, \citenamefont {{Boscaleri}},
  \citenamefont {{Cardoni}}, \citenamefont {{Church}}, \citenamefont {{Coble}},
  \citenamefont {{de Bernardis}}, \citenamefont {{de Troia}}, \citenamefont
  {{Farese}}, \citenamefont {{Ganga}}, \citenamefont {{Giacometti}},
  \citenamefont {{Haynes}}, \citenamefont {{Hivon}}, \citenamefont {{Hristov}},
  \citenamefont {{Iacoangeli}}, \citenamefont {{Jones}}, \citenamefont
  {{Lange}}, \citenamefont {{Martinis}}, \citenamefont {{Masi}}, \citenamefont
  {{Mason}}, \citenamefont {{Mauskopf}}, \citenamefont {{Miglio}},
  \citenamefont {{Montroy}}, \citenamefont {{Netterfield}}, \citenamefont
  {{Paine}}, \citenamefont {{Pascale}}, \citenamefont {{Piacentini}},
  \citenamefont {{Polenta}}, \citenamefont {{Pongetti}}, \citenamefont
  {{Romeo}}, \citenamefont {{Ruhl}}, \citenamefont {{Scaramuzzi}},
  \citenamefont {{Sforna}},\ and\ \citenamefont {{Turner}}}]{Crill2003}%
  \BibitemOpen
  \bibfield  {author} {\bibinfo {author} {\bibfnamefont {B.~P.}\ \bibnamefont
  {{Crill}}}, \bibinfo {author} {\bibfnamefont {P.~A.~R.}\ \bibnamefont
  {{Ade}}}, \bibinfo {author} {\bibfnamefont {D.~R.}\ \bibnamefont {{Artusa}}},
  \bibinfo {author} {\bibfnamefont {R.~S.}\ \bibnamefont {{Bhatia}}}, \bibinfo
  {author} {\bibfnamefont {J.~J.}\ \bibnamefont {{Bock}}}, \bibinfo {author}
  {\bibfnamefont {A.}~\bibnamefont {{Boscaleri}}}, \bibinfo {author}
  {\bibfnamefont {P.}~\bibnamefont {{Cardoni}}}, \bibinfo {author}
  {\bibfnamefont {S.~E.}\ \bibnamefont {{Church}}}, \bibinfo {author}
  {\bibfnamefont {K.}~\bibnamefont {{Coble}}}, \bibinfo {author} {\bibfnamefont
  {P.}~\bibnamefont {{de Bernardis}}}, \bibinfo {author} {\bibfnamefont
  {G.}~\bibnamefont {{de Troia}}}, \bibinfo {author} {\bibfnamefont
  {P.}~\bibnamefont {{Farese}}}, \bibinfo {author} {\bibfnamefont {K.~M.}\
  \bibnamefont {{Ganga}}}, \bibinfo {author} {\bibfnamefont {M.}~\bibnamefont
  {{Giacometti}}}, \bibinfo {author} {\bibfnamefont {C.~V.}\ \bibnamefont
  {{Haynes}}}, \bibinfo {author} {\bibfnamefont {E.}~\bibnamefont {{Hivon}}},
  \bibinfo {author} {\bibfnamefont {V.~V.}\ \bibnamefont {{Hristov}}}, \bibinfo
  {author} {\bibfnamefont {A.}~\bibnamefont {{Iacoangeli}}}, \bibinfo {author}
  {\bibfnamefont {W.~C.}\ \bibnamefont {{Jones}}}, \bibinfo {author}
  {\bibfnamefont {A.~E.}\ \bibnamefont {{Lange}}}, \bibinfo {author}
  {\bibfnamefont {L.}~\bibnamefont {{Martinis}}}, \bibinfo {author}
  {\bibfnamefont {S.}~\bibnamefont {{Masi}}}, \bibinfo {author} {\bibfnamefont
  {P.~V.}\ \bibnamefont {{Mason}}}, \bibinfo {author} {\bibfnamefont {P.~D.}\
  \bibnamefont {{Mauskopf}}}, \bibinfo {author} {\bibfnamefont
  {L.}~\bibnamefont {{Miglio}}}, \bibinfo {author} {\bibfnamefont
  {T.}~\bibnamefont {{Montroy}}}, \bibinfo {author} {\bibfnamefont {C.~B.}\
  \bibnamefont {{Netterfield}}}, \bibinfo {author} {\bibfnamefont {C.~G.}\
  \bibnamefont {{Paine}}}, \bibinfo {author} {\bibfnamefont {E.}~\bibnamefont
  {{Pascale}}}, \bibinfo {author} {\bibfnamefont {F.}~\bibnamefont
  {{Piacentini}}}, \bibinfo {author} {\bibfnamefont {G.}~\bibnamefont
  {{Polenta}}}, \bibinfo {author} {\bibfnamefont {F.}~\bibnamefont
  {{Pongetti}}}, \bibinfo {author} {\bibfnamefont {G.}~\bibnamefont {{Romeo}}},
  \bibinfo {author} {\bibfnamefont {J.~E.}\ \bibnamefont {{Ruhl}}}, \bibinfo
  {author} {\bibfnamefont {F.}~\bibnamefont {{Scaramuzzi}}}, \bibinfo {author}
  {\bibfnamefont {D.}~\bibnamefont {{Sforna}}}, \ and\ \bibinfo {author}
  {\bibfnamefont {A.~D.}\ \bibnamefont {{Turner}}},\ }\bibfield  {title}
  {\enquote {\bibinfo {title} {{BOOMERANG: A Balloon-borne Millimeter-Wave
  Telescope and Total Power Receiver for Mapping Anisotropy in the Cosmic
  Microwave Background}},}\ }\href {\doibase 10.1086/376894} {\bibfield
  {journal} {\bibinfo  {journal} {\apjs}\ }\textbf {\bibinfo {volume} {148}},\
  \bibinfo {pages} {527--541} (\bibinfo {year} {2003})},\ \Eprint
  {http://arxiv.org/abs/astro-ph/0206254} {arXiv:astro-ph/0206254 [astro-ph]}
  \BibitemShut {NoStop}%
\bibitem [{\citenamefont {{Bischoff}}\ \emph {et~al.}(2013)\citenamefont
  {{Bischoff}}, \citenamefont {{Brizius}}, \citenamefont {{Buder}},
  \citenamefont {{Chinone}}, \citenamefont {{Cleary}}, \citenamefont
  {{Dumoulin}}, \citenamefont {{Kusaka}}, \citenamefont {{Monsalve}},
  \citenamefont {{N{\ae}ss}}, \citenamefont {{Newburgh}}, \citenamefont
  {{Nixon}}, \citenamefont {{Reeves}}, \citenamefont {{Smith}}, \citenamefont
  {{Vanderlinde}}, \citenamefont {{Wehus}}, \citenamefont {{Bogdan}},
  \citenamefont {{Bustos}}, \citenamefont {{Church}}, \citenamefont {{Davis}},
  \citenamefont {{Dickinson}}, \citenamefont {{Eriksen}}, \citenamefont
  {{Gaier}}, \citenamefont {{Gundersen}}, \citenamefont {{Hasegawa}},
  \citenamefont {{Hazumi}}, \citenamefont {{Holler}}, \citenamefont
  {{Huffenberger}}, \citenamefont {{Imbriale}}, \citenamefont {{Ishidoshiro}},
  \citenamefont {{Jones}}, \citenamefont {{Kangaslahti}}, \citenamefont
  {{Kapner}}, \citenamefont {{Lawrence}}, \citenamefont {{Leitch}},
  \citenamefont {{Limon}}, \citenamefont {{McMahon}}, \citenamefont {{Miller}},
  \citenamefont {{Nagai}}, \citenamefont {{Nguyen}}, \citenamefont {{Pearson}},
  \citenamefont {{Piccirillo}}, \citenamefont {{Radford}}, \citenamefont
  {{Readhead}}, \citenamefont {{Richards}}, \citenamefont {{Samtleben}},
  \citenamefont {{Seiffert}}, \citenamefont {{Shepherd}}, \citenamefont
  {{Staggs}}, \citenamefont {{Tajima}}, \citenamefont {{Thompson}},
  \citenamefont {{Williamson}}, \citenamefont {{Winstein}}, \citenamefont
  {{Wollack}},\ and\ \citenamefont {{Zwart}}}]{Bischoff2013}%
  \BibitemOpen
  \bibfield  {author} {\bibinfo {author} {\bibfnamefont {C.}~\bibnamefont
  {{Bischoff}}}, \bibinfo {author} {\bibfnamefont {A.}~\bibnamefont
  {{Brizius}}}, \bibinfo {author} {\bibfnamefont {I.}~\bibnamefont {{Buder}}},
  \bibinfo {author} {\bibfnamefont {Y.}~\bibnamefont {{Chinone}}}, \bibinfo
  {author} {\bibfnamefont {K.}~\bibnamefont {{Cleary}}}, \bibinfo {author}
  {\bibfnamefont {R.~N.}\ \bibnamefont {{Dumoulin}}}, \bibinfo {author}
  {\bibfnamefont {A.}~\bibnamefont {{Kusaka}}}, \bibinfo {author}
  {\bibfnamefont {R.}~\bibnamefont {{Monsalve}}}, \bibinfo {author}
  {\bibfnamefont {S.~K.}\ \bibnamefont {{N{\ae}ss}}}, \bibinfo {author}
  {\bibfnamefont {L.~B.}\ \bibnamefont {{Newburgh}}}, \bibinfo {author}
  {\bibfnamefont {G.}~\bibnamefont {{Nixon}}}, \bibinfo {author} {\bibfnamefont
  {R.}~\bibnamefont {{Reeves}}}, \bibinfo {author} {\bibfnamefont {K.~M.}\
  \bibnamefont {{Smith}}}, \bibinfo {author} {\bibfnamefont {K.}~\bibnamefont
  {{Vanderlinde}}}, \bibinfo {author} {\bibfnamefont {I.~K.}\ \bibnamefont
  {{Wehus}}}, \bibinfo {author} {\bibfnamefont {M.}~\bibnamefont {{Bogdan}}},
  \bibinfo {author} {\bibfnamefont {R.}~\bibnamefont {{Bustos}}}, \bibinfo
  {author} {\bibfnamefont {S.~E.}\ \bibnamefont {{Church}}}, \bibinfo {author}
  {\bibfnamefont {R.}~\bibnamefont {{Davis}}}, \bibinfo {author} {\bibfnamefont
  {C.}~\bibnamefont {{Dickinson}}}, \bibinfo {author} {\bibfnamefont {H.~K.}\
  \bibnamefont {{Eriksen}}}, \bibinfo {author} {\bibfnamefont {T.}~\bibnamefont
  {{Gaier}}}, \bibinfo {author} {\bibfnamefont {J.~O.}\ \bibnamefont
  {{Gundersen}}}, \bibinfo {author} {\bibfnamefont {M.}~\bibnamefont
  {{Hasegawa}}}, \bibinfo {author} {\bibfnamefont {M.}~\bibnamefont
  {{Hazumi}}}, \bibinfo {author} {\bibfnamefont {C.}~\bibnamefont {{Holler}}},
  \bibinfo {author} {\bibfnamefont {K.~M.}\ \bibnamefont {{Huffenberger}}},
  \bibinfo {author} {\bibfnamefont {W.~A.}\ \bibnamefont {{Imbriale}}},
  \bibinfo {author} {\bibfnamefont {K.}~\bibnamefont {{Ishidoshiro}}}, \bibinfo
  {author} {\bibfnamefont {M.~E.}\ \bibnamefont {{Jones}}}, \bibinfo {author}
  {\bibfnamefont {P.}~\bibnamefont {{Kangaslahti}}}, \bibinfo {author}
  {\bibfnamefont {D.~J.}\ \bibnamefont {{Kapner}}}, \bibinfo {author}
  {\bibfnamefont {C.~R.}\ \bibnamefont {{Lawrence}}}, \bibinfo {author}
  {\bibfnamefont {E.~M.}\ \bibnamefont {{Leitch}}}, \bibinfo {author}
  {\bibfnamefont {M.}~\bibnamefont {{Limon}}}, \bibinfo {author} {\bibfnamefont
  {J.~J.}\ \bibnamefont {{McMahon}}}, \bibinfo {author} {\bibfnamefont {A.~D.}\
  \bibnamefont {{Miller}}}, \bibinfo {author} {\bibfnamefont {M.}~\bibnamefont
  {{Nagai}}}, \bibinfo {author} {\bibfnamefont {H.}~\bibnamefont {{Nguyen}}},
  \bibinfo {author} {\bibfnamefont {T.~J.}\ \bibnamefont {{Pearson}}}, \bibinfo
  {author} {\bibfnamefont {L.}~\bibnamefont {{Piccirillo}}}, \bibinfo {author}
  {\bibfnamefont {S.~J.~E.}\ \bibnamefont {{Radford}}}, \bibinfo {author}
  {\bibfnamefont {A.~C.~S.}\ \bibnamefont {{Readhead}}}, \bibinfo {author}
  {\bibfnamefont {J.~L.}\ \bibnamefont {{Richards}}}, \bibinfo {author}
  {\bibfnamefont {D.}~\bibnamefont {{Samtleben}}}, \bibinfo {author}
  {\bibfnamefont {M.}~\bibnamefont {{Seiffert}}}, \bibinfo {author}
  {\bibfnamefont {M.~C.}\ \bibnamefont {{Shepherd}}}, \bibinfo {author}
  {\bibfnamefont {S.~T.}\ \bibnamefont {{Staggs}}}, \bibinfo {author}
  {\bibfnamefont {O.}~\bibnamefont {{Tajima}}}, \bibinfo {author}
  {\bibfnamefont {K.~L.}\ \bibnamefont {{Thompson}}}, \bibinfo {author}
  {\bibfnamefont {R.}~\bibnamefont {{Williamson}}}, \bibinfo {author}
  {\bibfnamefont {B.}~\bibnamefont {{Winstein}}}, \bibinfo {author}
  {\bibfnamefont {E.~J.}\ \bibnamefont {{Wollack}}}, \ and\ \bibinfo {author}
  {\bibfnamefont {J.~T.~L.}\ \bibnamefont {{Zwart}}},\ }\bibfield  {title}
  {\enquote {\bibinfo {title} {{The Q/U Imaging ExperimenT Instrument}},}\
  }\href {\doibase 10.1088/0004-637X/768/1/9} {\bibfield  {journal} {\bibinfo
  {journal} {\apj}\ }\textbf {\bibinfo {volume} {768}},\ \bibinfo {eid} {9}
  (\bibinfo {year} {2013})},\ \Eprint {http://arxiv.org/abs/1207.5562}
  {arXiv:1207.5562 [astro-ph.IM]} \BibitemShut {NoStop}%
\bibitem [{\citenamefont {{Yoon}}\ \emph {et~al.}(2006)\citenamefont {{Yoon}},
  \citenamefont {{Ade}}, \citenamefont {{Barkats}}, \citenamefont {{Battle}},
  \citenamefont {{Bierman}}, \citenamefont {{Bock}}, \citenamefont {{Brevik}},
  \citenamefont {{Chiang}}, \citenamefont {{Crites}}, \citenamefont {{Dowell}},
  \citenamefont {{Duband}}, \citenamefont {{Griffin}}, \citenamefont {{Hivon}},
  \citenamefont {{Holzapfel}}, \citenamefont {{Hristov}}, \citenamefont
  {{Keating}}, \citenamefont {{Kovac}}, \citenamefont {{Kuo}}, \citenamefont
  {{Lange}}, \citenamefont {{Leitch}}, \citenamefont {{Mason}}, \citenamefont
  {{Nguyen}}, \citenamefont {{Ponthieu}}, \citenamefont {{Takahashi}},
  \citenamefont {{Renbarger}}, \citenamefont {{Weintraub}},\ and\ \citenamefont
  {{Woolsey}}}]{Yoon2006}%
  \BibitemOpen
  \bibfield  {author} {\bibinfo {author} {\bibfnamefont {K.~W.}\ \bibnamefont
  {{Yoon}}}, \bibinfo {author} {\bibfnamefont {P.~A.~R.}\ \bibnamefont
  {{Ade}}}, \bibinfo {author} {\bibfnamefont {D.}~\bibnamefont {{Barkats}}},
  \bibinfo {author} {\bibfnamefont {J.~O.}\ \bibnamefont {{Battle}}}, \bibinfo
  {author} {\bibfnamefont {E.~M.}\ \bibnamefont {{Bierman}}}, \bibinfo {author}
  {\bibfnamefont {J.~J.}\ \bibnamefont {{Bock}}}, \bibinfo {author}
  {\bibfnamefont {J.~A.}\ \bibnamefont {{Brevik}}}, \bibinfo {author}
  {\bibfnamefont {H.~C.}\ \bibnamefont {{Chiang}}}, \bibinfo {author}
  {\bibfnamefont {A.}~\bibnamefont {{Crites}}}, \bibinfo {author}
  {\bibfnamefont {C.~D.}\ \bibnamefont {{Dowell}}}, \bibinfo {author}
  {\bibfnamefont {L.}~\bibnamefont {{Duband}}}, \bibinfo {author}
  {\bibfnamefont {G.~S.}\ \bibnamefont {{Griffin}}}, \bibinfo {author}
  {\bibfnamefont {E.~F.}\ \bibnamefont {{Hivon}}}, \bibinfo {author}
  {\bibfnamefont {W.~L.}\ \bibnamefont {{Holzapfel}}}, \bibinfo {author}
  {\bibfnamefont {V.~V.}\ \bibnamefont {{Hristov}}}, \bibinfo {author}
  {\bibfnamefont {B.~G.}\ \bibnamefont {{Keating}}}, \bibinfo {author}
  {\bibfnamefont {J.~M.}\ \bibnamefont {{Kovac}}}, \bibinfo {author}
  {\bibfnamefont {C.~L.}\ \bibnamefont {{Kuo}}}, \bibinfo {author}
  {\bibfnamefont {A.~E.}\ \bibnamefont {{Lange}}}, \bibinfo {author}
  {\bibfnamefont {E.~M.}\ \bibnamefont {{Leitch}}}, \bibinfo {author}
  {\bibfnamefont {P.~V.}\ \bibnamefont {{Mason}}}, \bibinfo {author}
  {\bibfnamefont {H.~T.}\ \bibnamefont {{Nguyen}}}, \bibinfo {author}
  {\bibfnamefont {N.}~\bibnamefont {{Ponthieu}}}, \bibinfo {author}
  {\bibfnamefont {Y.~D.}\ \bibnamefont {{Takahashi}}}, \bibinfo {author}
  {\bibfnamefont {T.}~\bibnamefont {{Renbarger}}}, \bibinfo {author}
  {\bibfnamefont {L.~C.}\ \bibnamefont {{Weintraub}}}, \ and\ \bibinfo {author}
  {\bibfnamefont {D.}~\bibnamefont {{Woolsey}}},\ }\bibfield  {title} {\enquote
  {\bibinfo {title} {{The Robinson Gravitational Wave Background Telescope
  (BICEP): a bolometric large angular scale CMB polarimeter}},}\ }in\ \href
  {\doibase 10.1117/12.672652} {\emph {\bibinfo {booktitle} {Millimeter and
  Submillimeter Detectors and Instrumentation for Astronomy III}}},\ \bibinfo
  {series} {Society of Photo-Optical Instrumentation Engineers (SPIE)
  Conference Series}, Vol.\ \bibinfo {volume} {6275},\ \bibinfo {editor}
  {edited by\ \bibinfo {editor} {\bibfnamefont {J.}~\bibnamefont
  {{Zmuidzinas}}}, \bibinfo {editor} {\bibfnamefont {W.~S.}\ \bibnamefont
  {{Holland}}}, \bibinfo {editor} {\bibfnamefont {S.}~\bibnamefont
  {{Withington}}}, \ and\ \bibinfo {editor} {\bibfnamefont {W.~D.}\
  \bibnamefont {{Duncan}}}}\ (\bibinfo {year} {2006})\ p.\ \bibinfo {pages}
  {62751K},\ \Eprint {http://arxiv.org/abs/astro-ph/0606278}
  {arXiv:astro-ph/0606278 [astro-ph]} \BibitemShut {NoStop}%
\bibitem [{\citenamefont {{Everett}}\ \emph {et~al.}(2018)\citenamefont
  {{Everett}}, \citenamefont {{Ade}}, \citenamefont {{Ahmed}}, \citenamefont
  {{Anderson}}, \citenamefont {{Austermann}}, \citenamefont {{Avva}},
  \citenamefont {{Thakur}}, \citenamefont {{Bender}}, \citenamefont {{Benson}},
  \citenamefont {{Carlstrom}}, \citenamefont {{Carter}}, \citenamefont
  {{Cecil}}, \citenamefont {{Chang}}, \citenamefont {{Cliche}}, \citenamefont
  {{Cukierman}}, \citenamefont {{Denison}}, \citenamefont {{de Haan}},
  \citenamefont {{Ding}}, \citenamefont {{Dobbs}}, \citenamefont {{Dutcher}},
  \citenamefont {{Foster}}, \citenamefont {{Gannon}}, \citenamefont
  {{Gilbert}}, \citenamefont {{Groh}}, \citenamefont {{Halverson}},
  \citenamefont {{Harke-Hosemann}}, \citenamefont {{Harrington}}, \citenamefont
  {{Henning}}, \citenamefont {{Hilton}}, \citenamefont {{Holzapfel}},
  \citenamefont {{Huang}}, \citenamefont {{Irwin}}, \citenamefont {{Jeong}},
  \citenamefont {{Jonas}}, \citenamefont {{Khaire}}, \citenamefont {{Kofman}},
  \citenamefont {{Korman}}, \citenamefont {{Kubik}}, \citenamefont
  {{Kuhlmann}}, \citenamefont {{Kuo}}, \citenamefont {{Lee}}, \citenamefont
  {{Lowitz}}, \citenamefont {{Meyer}}, \citenamefont {{Michalik}},
  \citenamefont {{Montgomery}}, \citenamefont {{Nadolski}}, \citenamefont
  {{Natoli}}, \citenamefont {{Nguyen}}, \citenamefont {{Noble}}, \citenamefont
  {{Novosad}}, \citenamefont {{Padin}}, \citenamefont {{Pan}}, \citenamefont
  {{Pearson}}, \citenamefont {{Posada}}, \citenamefont {{Rahlin}},
  \citenamefont {{Ruhl}}, \citenamefont {{Saunders}}, \citenamefont {{Sayre}},
  \citenamefont {{Shirley}}, \citenamefont {{Shirokoff}}, \citenamefont
  {{Smecher}}, \citenamefont {{Sobrin}}, \citenamefont {{Stark}}, \citenamefont
  {{Story}}, \citenamefont {{Suzuki}}, \citenamefont {{Tang}}, \citenamefont
  {{Thompson}}, \citenamefont {{Tucker}}, \citenamefont {{Vale}}, \citenamefont
  {{Vanderlinde}}, \citenamefont {{Vieira}}, \citenamefont {{Wang}},
  \citenamefont {{Whitehorn}}, \citenamefont {{Yefremenko}}, \citenamefont
  {{Yoon}},\ and\ \citenamefont {{Young}}}]{Everett2018}%
  \BibitemOpen
  \bibfield  {author} {\bibinfo {author} {\bibfnamefont {W.}~\bibnamefont
  {{Everett}}}, \bibinfo {author} {\bibfnamefont {P.~A.~R.}\ \bibnamefont
  {{Ade}}}, \bibinfo {author} {\bibfnamefont {Z.}~\bibnamefont {{Ahmed}}},
  \bibinfo {author} {\bibfnamefont {A.~J.}\ \bibnamefont {{Anderson}}},
  \bibinfo {author} {\bibfnamefont {J.~E.}\ \bibnamefont {{Austermann}}},
  \bibinfo {author} {\bibfnamefont {J.~S.}\ \bibnamefont {{Avva}}}, \bibinfo
  {author} {\bibfnamefont {R.~B.}\ \bibnamefont {{Thakur}}}, \bibinfo {author}
  {\bibfnamefont {A.~N.}\ \bibnamefont {{Bender}}}, \bibinfo {author}
  {\bibfnamefont {B.~A.}\ \bibnamefont {{Benson}}}, \bibinfo {author}
  {\bibfnamefont {J.~E.}\ \bibnamefont {{Carlstrom}}}, \bibinfo {author}
  {\bibfnamefont {F.~W.}\ \bibnamefont {{Carter}}}, \bibinfo {author}
  {\bibfnamefont {T.}~\bibnamefont {{Cecil}}}, \bibinfo {author} {\bibfnamefont
  {C.~L.}\ \bibnamefont {{Chang}}}, \bibinfo {author} {\bibfnamefont {J.~F.}\
  \bibnamefont {{Cliche}}}, \bibinfo {author} {\bibfnamefont {A.}~\bibnamefont
  {{Cukierman}}}, \bibinfo {author} {\bibfnamefont {E.~V.}\ \bibnamefont
  {{Denison}}}, \bibinfo {author} {\bibfnamefont {T.}~\bibnamefont {{de
  Haan}}}, \bibinfo {author} {\bibfnamefont {J.}~\bibnamefont {{Ding}}},
  \bibinfo {author} {\bibfnamefont {M.~A.}\ \bibnamefont {{Dobbs}}}, \bibinfo
  {author} {\bibfnamefont {D.}~\bibnamefont {{Dutcher}}}, \bibinfo {author}
  {\bibfnamefont {A.}~\bibnamefont {{Foster}}}, \bibinfo {author}
  {\bibfnamefont {R.~N.}\ \bibnamefont {{Gannon}}}, \bibinfo {author}
  {\bibfnamefont {A.}~\bibnamefont {{Gilbert}}}, \bibinfo {author}
  {\bibfnamefont {J.~C.}\ \bibnamefont {{Groh}}}, \bibinfo {author}
  {\bibfnamefont {N.~W.}\ \bibnamefont {{Halverson}}}, \bibinfo {author}
  {\bibfnamefont {A.~H.}\ \bibnamefont {{Harke-Hosemann}}}, \bibinfo {author}
  {\bibfnamefont {N.~L.}\ \bibnamefont {{Harrington}}}, \bibinfo {author}
  {\bibfnamefont {J.~W.}\ \bibnamefont {{Henning}}}, \bibinfo {author}
  {\bibfnamefont {G.~C.}\ \bibnamefont {{Hilton}}}, \bibinfo {author}
  {\bibfnamefont {W.~L.}\ \bibnamefont {{Holzapfel}}}, \bibinfo {author}
  {\bibfnamefont {N.}~\bibnamefont {{Huang}}}, \bibinfo {author} {\bibfnamefont
  {K.~D.}\ \bibnamefont {{Irwin}}}, \bibinfo {author} {\bibfnamefont {O.~B.}\
  \bibnamefont {{Jeong}}}, \bibinfo {author} {\bibfnamefont {M.}~\bibnamefont
  {{Jonas}}}, \bibinfo {author} {\bibfnamefont {T.}~\bibnamefont {{Khaire}}},
  \bibinfo {author} {\bibfnamefont {A.~M.}\ \bibnamefont {{Kofman}}}, \bibinfo
  {author} {\bibfnamefont {M.}~\bibnamefont {{Korman}}}, \bibinfo {author}
  {\bibfnamefont {D.}~\bibnamefont {{Kubik}}}, \bibinfo {author} {\bibfnamefont
  {S.}~\bibnamefont {{Kuhlmann}}}, \bibinfo {author} {\bibfnamefont {C.~L.}\
  \bibnamefont {{Kuo}}}, \bibinfo {author} {\bibfnamefont {A.~T.}\ \bibnamefont
  {{Lee}}}, \bibinfo {author} {\bibfnamefont {A.~E.}\ \bibnamefont {{Lowitz}}},
  \bibinfo {author} {\bibfnamefont {S.~S.}\ \bibnamefont {{Meyer}}}, \bibinfo
  {author} {\bibfnamefont {D.}~\bibnamefont {{Michalik}}}, \bibinfo {author}
  {\bibfnamefont {J.}~\bibnamefont {{Montgomery}}}, \bibinfo {author}
  {\bibfnamefont {A.}~\bibnamefont {{Nadolski}}}, \bibinfo {author}
  {\bibfnamefont {T.}~\bibnamefont {{Natoli}}}, \bibinfo {author}
  {\bibfnamefont {H.}~\bibnamefont {{Nguyen}}}, \bibinfo {author}
  {\bibfnamefont {G.~I.}\ \bibnamefont {{Noble}}}, \bibinfo {author}
  {\bibfnamefont {V.}~\bibnamefont {{Novosad}}}, \bibinfo {author}
  {\bibfnamefont {S.}~\bibnamefont {{Padin}}}, \bibinfo {author} {\bibfnamefont
  {Z.}~\bibnamefont {{Pan}}}, \bibinfo {author} {\bibfnamefont
  {J.}~\bibnamefont {{Pearson}}}, \bibinfo {author} {\bibfnamefont {C.~M.}\
  \bibnamefont {{Posada}}}, \bibinfo {author} {\bibfnamefont {A.}~\bibnamefont
  {{Rahlin}}}, \bibinfo {author} {\bibfnamefont {J.~E.}\ \bibnamefont
  {{Ruhl}}}, \bibinfo {author} {\bibfnamefont {L.~J.}\ \bibnamefont
  {{Saunders}}}, \bibinfo {author} {\bibfnamefont {J.~T.}\ \bibnamefont
  {{Sayre}}}, \bibinfo {author} {\bibfnamefont {I.}~\bibnamefont {{Shirley}}},
  \bibinfo {author} {\bibfnamefont {E.}~\bibnamefont {{Shirokoff}}}, \bibinfo
  {author} {\bibfnamefont {G.}~\bibnamefont {{Smecher}}}, \bibinfo {author}
  {\bibfnamefont {J.~A.}\ \bibnamefont {{Sobrin}}}, \bibinfo {author}
  {\bibfnamefont {A.~A.}\ \bibnamefont {{Stark}}}, \bibinfo {author}
  {\bibfnamefont {K.~T.}\ \bibnamefont {{Story}}}, \bibinfo {author}
  {\bibfnamefont {A.}~\bibnamefont {{Suzuki}}}, \bibinfo {author}
  {\bibfnamefont {Q.~Y.}\ \bibnamefont {{Tang}}}, \bibinfo {author}
  {\bibfnamefont {K.~L.}\ \bibnamefont {{Thompson}}}, \bibinfo {author}
  {\bibfnamefont {C.}~\bibnamefont {{Tucker}}}, \bibinfo {author}
  {\bibfnamefont {L.~R.}\ \bibnamefont {{Vale}}}, \bibinfo {author}
  {\bibfnamefont {K.}~\bibnamefont {{Vanderlinde}}}, \bibinfo {author}
  {\bibfnamefont {J.~D.}\ \bibnamefont {{Vieira}}}, \bibinfo {author}
  {\bibfnamefont {G.}~\bibnamefont {{Wang}}}, \bibinfo {author} {\bibfnamefont
  {N.}~\bibnamefont {{Whitehorn}}}, \bibinfo {author} {\bibfnamefont
  {V.}~\bibnamefont {{Yefremenko}}}, \bibinfo {author} {\bibfnamefont {K.~W.}\
  \bibnamefont {{Yoon}}}, \ and\ \bibinfo {author} {\bibfnamefont {M.~R.}\
  \bibnamefont {{Young}}},\ }\bibfield  {title} {\enquote {\bibinfo {title}
  {{Design and Bolometer Characterization of the SPT-3G First-Year Focal
  Plane}},}\ }\href {\doibase 10.1007/s10909-018-2057-2} {\bibfield  {journal}
  {\bibinfo  {journal} {Journal of Low Temperature Physics}\ }\textbf {\bibinfo
  {volume} {193}},\ \bibinfo {pages} {1085--1093} (\bibinfo {year} {2018})},\
  \Eprint {http://arxiv.org/abs/1902.09640} {arXiv:1902.09640 [astro-ph.IM]}
  \BibitemShut {NoStop}%
\bibitem [{\citenamefont {{Westbrook}}\ \emph {et~al.}(2018)\citenamefont
  {{Westbrook}}, \citenamefont {{Ade}}, \citenamefont {{Aguilar}},
  \citenamefont {{Akiba}}, \citenamefont {{Arnold}}, \citenamefont
  {{Baccigalupi}}, \citenamefont {{Barron}}, \citenamefont {{Beck}},
  \citenamefont {{Beckman}}, \citenamefont {{Bender}}, \citenamefont
  {{Bianchini}}, \citenamefont {{Boettger}}, \citenamefont {{Borrill}},
  \citenamefont {{Chapman}}, \citenamefont {{Chinone}}, \citenamefont
  {{Coppi}}, \citenamefont {{Crowley}}, \citenamefont {{Cukierman}},
  \citenamefont {{de Haan}}, \citenamefont {{D{\"u}nner}}, \citenamefont
  {{Dobbs}}, \citenamefont {{Elleflot}}, \citenamefont {{Errard}},
  \citenamefont {{Fabbian}}, \citenamefont {{Feeney}}, \citenamefont {{Feng}},
  \citenamefont {{Fuller}}, \citenamefont {{Galitzki}}, \citenamefont
  {{Gilbert}}, \citenamefont {{Goeckner-Wald}}, \citenamefont {{Groh}},
  \citenamefont {{Halverson}}, \citenamefont {{Hamada}}, \citenamefont
  {{Hasegawa}}, \citenamefont {{Hazumi}}, \citenamefont {{Hill}}, \citenamefont
  {{Holzapfel}}, \citenamefont {{Howe}}, \citenamefont {{Inoue}}, \citenamefont
  {{Jaehnig}}, \citenamefont {{Jaffe}}, \citenamefont {{Jeong}}, \citenamefont
  {{Kaneko}}, \citenamefont {{Katayama}}, \citenamefont {{Keating}},
  \citenamefont {{Keskitalo}}, \citenamefont {{Kisner}}, \citenamefont
  {{Krachmalnicoff}}, \citenamefont {{Kusaka}}, \citenamefont {{Le Jeune}},
  \citenamefont {{Lee}}, \citenamefont {{Leon}}, \citenamefont {{Linder}},
  \citenamefont {{Lowry}}, \citenamefont {{Madurowicz}}, \citenamefont {{Mak}},
  \citenamefont {{Matsuda}}, \citenamefont {{May}}, \citenamefont {{Miller}},
  \citenamefont {{Minami}}, \citenamefont {{Montgomery}}, \citenamefont
  {{Navaroli}}, \citenamefont {{Nishino}}, \citenamefont {{Peloton}},
  \citenamefont {{Pham}}, \citenamefont {{Piccirillo}}, \citenamefont
  {{Plambeck}}, \citenamefont {{Poletti}}, \citenamefont {{Puglisi}},
  \citenamefont {{Raum}}, \citenamefont {{Rebeiz}}, \citenamefont
  {{Reichardt}}, \citenamefont {{Richards}}, \citenamefont {{Roberts}},
  \citenamefont {{Ross}}, \citenamefont {{Rotermund}}, \citenamefont
  {{Segawa}}, \citenamefont {{Sherwin}}, \citenamefont {{Silva-Feaver}},
  \citenamefont {{Siritanasak}}, \citenamefont {{Stompor}}, \citenamefont
  {{Suzuki}}, \citenamefont {{Tajima}}, \citenamefont {{Takakura}},
  \citenamefont {{Takatori}}, \citenamefont {{Tanabe}}, \citenamefont {{Tat}},
  \citenamefont {{Teply}}, \citenamefont {{Tikhomirov}}, \citenamefont
  {{Tomaru}}, \citenamefont {{Tsai}}, \citenamefont {{Whitehorn}},\ and\
  \citenamefont {{Zahn}}}]{Westbrook2018}%
  \BibitemOpen
  \bibfield  {author} {\bibinfo {author} {\bibfnamefont {B.}~\bibnamefont
  {{Westbrook}}}, \bibinfo {author} {\bibfnamefont {P.~A.~R.}\ \bibnamefont
  {{Ade}}}, \bibinfo {author} {\bibfnamefont {M.}~\bibnamefont {{Aguilar}}},
  \bibinfo {author} {\bibfnamefont {Y.}~\bibnamefont {{Akiba}}}, \bibinfo
  {author} {\bibfnamefont {K.}~\bibnamefont {{Arnold}}}, \bibinfo {author}
  {\bibfnamefont {C.}~\bibnamefont {{Baccigalupi}}}, \bibinfo {author}
  {\bibfnamefont {D.}~\bibnamefont {{Barron}}}, \bibinfo {author}
  {\bibfnamefont {D.}~\bibnamefont {{Beck}}}, \bibinfo {author} {\bibfnamefont
  {S.}~\bibnamefont {{Beckman}}}, \bibinfo {author} {\bibfnamefont {A.~N.}\
  \bibnamefont {{Bender}}}, \bibinfo {author} {\bibfnamefont {F.}~\bibnamefont
  {{Bianchini}}}, \bibinfo {author} {\bibfnamefont {D.}~\bibnamefont
  {{Boettger}}}, \bibinfo {author} {\bibfnamefont {J.}~\bibnamefont
  {{Borrill}}}, \bibinfo {author} {\bibfnamefont {S.}~\bibnamefont
  {{Chapman}}}, \bibinfo {author} {\bibfnamefont {Y.}~\bibnamefont
  {{Chinone}}}, \bibinfo {author} {\bibfnamefont {G.}~\bibnamefont {{Coppi}}},
  \bibinfo {author} {\bibfnamefont {K.}~\bibnamefont {{Crowley}}}, \bibinfo
  {author} {\bibfnamefont {A.}~\bibnamefont {{Cukierman}}}, \bibinfo {author}
  {\bibfnamefont {T.}~\bibnamefont {{de Haan}}}, \bibinfo {author}
  {\bibfnamefont {R.}~\bibnamefont {{D{\"u}nner}}}, \bibinfo {author}
  {\bibfnamefont {M.}~\bibnamefont {{Dobbs}}}, \bibinfo {author} {\bibfnamefont
  {T.}~\bibnamefont {{Elleflot}}}, \bibinfo {author} {\bibfnamefont
  {J.}~\bibnamefont {{Errard}}}, \bibinfo {author} {\bibfnamefont
  {G.}~\bibnamefont {{Fabbian}}}, \bibinfo {author} {\bibfnamefont {S.~M.}\
  \bibnamefont {{Feeney}}}, \bibinfo {author} {\bibfnamefont {C.}~\bibnamefont
  {{Feng}}}, \bibinfo {author} {\bibfnamefont {G.}~\bibnamefont {{Fuller}}},
  \bibinfo {author} {\bibfnamefont {N.}~\bibnamefont {{Galitzki}}}, \bibinfo
  {author} {\bibfnamefont {A.}~\bibnamefont {{Gilbert}}}, \bibinfo {author}
  {\bibfnamefont {N.}~\bibnamefont {{Goeckner-Wald}}}, \bibinfo {author}
  {\bibfnamefont {J.}~\bibnamefont {{Groh}}}, \bibinfo {author} {\bibfnamefont
  {N.~W.}\ \bibnamefont {{Halverson}}}, \bibinfo {author} {\bibfnamefont
  {T.}~\bibnamefont {{Hamada}}}, \bibinfo {author} {\bibfnamefont
  {M.}~\bibnamefont {{Hasegawa}}}, \bibinfo {author} {\bibfnamefont
  {M.}~\bibnamefont {{Hazumi}}}, \bibinfo {author} {\bibfnamefont {C.~A.}\
  \bibnamefont {{Hill}}}, \bibinfo {author} {\bibfnamefont {W.}~\bibnamefont
  {{Holzapfel}}}, \bibinfo {author} {\bibfnamefont {L.}~\bibnamefont {{Howe}}},
  \bibinfo {author} {\bibfnamefont {Y.}~\bibnamefont {{Inoue}}}, \bibinfo
  {author} {\bibfnamefont {G.}~\bibnamefont {{Jaehnig}}}, \bibinfo {author}
  {\bibfnamefont {A.}~\bibnamefont {{Jaffe}}}, \bibinfo {author} {\bibfnamefont
  {O.}~\bibnamefont {{Jeong}}}, \bibinfo {author} {\bibfnamefont
  {D.}~\bibnamefont {{Kaneko}}}, \bibinfo {author} {\bibfnamefont
  {N.}~\bibnamefont {{Katayama}}}, \bibinfo {author} {\bibfnamefont
  {B.}~\bibnamefont {{Keating}}}, \bibinfo {author} {\bibfnamefont
  {R.}~\bibnamefont {{Keskitalo}}}, \bibinfo {author} {\bibfnamefont
  {T.}~\bibnamefont {{Kisner}}}, \bibinfo {author} {\bibfnamefont
  {N.}~\bibnamefont {{Krachmalnicoff}}}, \bibinfo {author} {\bibfnamefont
  {A.}~\bibnamefont {{Kusaka}}}, \bibinfo {author} {\bibfnamefont
  {M.}~\bibnamefont {{Le Jeune}}}, \bibinfo {author} {\bibfnamefont {A.~T.}\
  \bibnamefont {{Lee}}}, \bibinfo {author} {\bibfnamefont {D.}~\bibnamefont
  {{Leon}}}, \bibinfo {author} {\bibfnamefont {E.}~\bibnamefont {{Linder}}},
  \bibinfo {author} {\bibfnamefont {L.}~\bibnamefont {{Lowry}}}, \bibinfo
  {author} {\bibfnamefont {A.}~\bibnamefont {{Madurowicz}}}, \bibinfo {author}
  {\bibfnamefont {D.}~\bibnamefont {{Mak}}}, \bibinfo {author} {\bibfnamefont
  {F.}~\bibnamefont {{Matsuda}}}, \bibinfo {author} {\bibfnamefont
  {A.}~\bibnamefont {{May}}}, \bibinfo {author} {\bibfnamefont {N.~J.}\
  \bibnamefont {{Miller}}}, \bibinfo {author} {\bibfnamefont {Y.}~\bibnamefont
  {{Minami}}}, \bibinfo {author} {\bibfnamefont {J.}~\bibnamefont
  {{Montgomery}}}, \bibinfo {author} {\bibfnamefont {M.}~\bibnamefont
  {{Navaroli}}}, \bibinfo {author} {\bibfnamefont {H.}~\bibnamefont
  {{Nishino}}}, \bibinfo {author} {\bibfnamefont {J.}~\bibnamefont
  {{Peloton}}}, \bibinfo {author} {\bibfnamefont {A.}~\bibnamefont {{Pham}}},
  \bibinfo {author} {\bibfnamefont {L.}~\bibnamefont {{Piccirillo}}}, \bibinfo
  {author} {\bibfnamefont {D.}~\bibnamefont {{Plambeck}}}, \bibinfo {author}
  {\bibfnamefont {D.}~\bibnamefont {{Poletti}}}, \bibinfo {author}
  {\bibfnamefont {G.}~\bibnamefont {{Puglisi}}}, \bibinfo {author}
  {\bibfnamefont {C.}~\bibnamefont {{Raum}}}, \bibinfo {author} {\bibfnamefont
  {G.}~\bibnamefont {{Rebeiz}}}, \bibinfo {author} {\bibfnamefont {C.~L.}\
  \bibnamefont {{Reichardt}}}, \bibinfo {author} {\bibfnamefont {P.~L.}\
  \bibnamefont {{Richards}}}, \bibinfo {author} {\bibfnamefont
  {H.}~\bibnamefont {{Roberts}}}, \bibinfo {author} {\bibfnamefont
  {C.}~\bibnamefont {{Ross}}}, \bibinfo {author} {\bibfnamefont {K.~M.}\
  \bibnamefont {{Rotermund}}}, \bibinfo {author} {\bibfnamefont
  {Y.}~\bibnamefont {{Segawa}}}, \bibinfo {author} {\bibfnamefont
  {B.}~\bibnamefont {{Sherwin}}}, \bibinfo {author} {\bibfnamefont
  {M.}~\bibnamefont {{Silva-Feaver}}}, \bibinfo {author} {\bibfnamefont
  {P.}~\bibnamefont {{Siritanasak}}}, \bibinfo {author} {\bibfnamefont
  {R.}~\bibnamefont {{Stompor}}}, \bibinfo {author} {\bibfnamefont
  {A.}~\bibnamefont {{Suzuki}}}, \bibinfo {author} {\bibfnamefont
  {O.}~\bibnamefont {{Tajima}}}, \bibinfo {author} {\bibfnamefont
  {S.}~\bibnamefont {{Takakura}}}, \bibinfo {author} {\bibfnamefont
  {S.}~\bibnamefont {{Takatori}}}, \bibinfo {author} {\bibfnamefont
  {D.}~\bibnamefont {{Tanabe}}}, \bibinfo {author} {\bibfnamefont
  {R.}~\bibnamefont {{Tat}}}, \bibinfo {author} {\bibfnamefont {G.~P.}\
  \bibnamefont {{Teply}}}, \bibinfo {author} {\bibfnamefont {A.}~\bibnamefont
  {{Tikhomirov}}}, \bibinfo {author} {\bibfnamefont {T.}~\bibnamefont
  {{Tomaru}}}, \bibinfo {author} {\bibfnamefont {C.}~\bibnamefont {{Tsai}}},
  \bibinfo {author} {\bibfnamefont {N.}~\bibnamefont {{Whitehorn}}}, \ and\
  \bibinfo {author} {\bibfnamefont {A.}~\bibnamefont {{Zahn}}},\ }\bibfield
  {title} {\enquote {\bibinfo {title} {{The POLARBEAR-2 and Simons Array Focal
  Plane Fabrication Status}},}\ }\href {\doibase 10.1007/s10909-018-2059-0}
  {\bibfield  {journal} {\bibinfo  {journal} {Journal of Low Temperature
  Physics}\ }\textbf {\bibinfo {volume} {193}},\ \bibinfo {pages} {758--770}
  (\bibinfo {year} {2018})},\ \Eprint {http://arxiv.org/abs/2210.04117}
  {arXiv:2210.04117 [astro-ph.IM]} \BibitemShut {NoStop}%
\bibitem [{\citenamefont {{Gordon}}\ \emph {et~al.}(2020)\citenamefont
  {{Gordon}}, \citenamefont {{Sinclair}}, \citenamefont {{Mauskopf}},
  \citenamefont {{Coppi}}, \citenamefont {{Devlin}}, \citenamefont {{Dober}},
  \citenamefont {{Fissel}}, \citenamefont {{Galitzki}}, \citenamefont {{Gao}},
  \citenamefont {{Hubmayr}}, \citenamefont {{Lourie}}, \citenamefont {{Lowe}},
  \citenamefont {{McKenney}}, \citenamefont {{Nati}},\ and\ \citenamefont
  {{Romualdez}}}]{Gordon2020}%
  \BibitemOpen
  \bibfield  {author} {\bibinfo {author} {\bibfnamefont {S.}~\bibnamefont
  {{Gordon}}}, \bibinfo {author} {\bibfnamefont {A.}~\bibnamefont
  {{Sinclair}}}, \bibinfo {author} {\bibfnamefont {P.}~\bibnamefont
  {{Mauskopf}}}, \bibinfo {author} {\bibfnamefont {G.}~\bibnamefont {{Coppi}}},
  \bibinfo {author} {\bibfnamefont {M.}~\bibnamefont {{Devlin}}}, \bibinfo
  {author} {\bibfnamefont {B.}~\bibnamefont {{Dober}}}, \bibinfo {author}
  {\bibfnamefont {L.}~\bibnamefont {{Fissel}}}, \bibinfo {author}
  {\bibfnamefont {N.}~\bibnamefont {{Galitzki}}}, \bibinfo {author}
  {\bibfnamefont {J.}~\bibnamefont {{Gao}}}, \bibinfo {author} {\bibfnamefont
  {J.}~\bibnamefont {{Hubmayr}}}, \bibinfo {author} {\bibfnamefont
  {N.}~\bibnamefont {{Lourie}}}, \bibinfo {author} {\bibfnamefont
  {I.}~\bibnamefont {{Lowe}}}, \bibinfo {author} {\bibfnamefont
  {C.}~\bibnamefont {{McKenney}}}, \bibinfo {author} {\bibfnamefont
  {F.}~\bibnamefont {{Nati}}}, \ and\ \bibinfo {author} {\bibfnamefont
  {J.}~\bibnamefont {{Romualdez}}},\ }\bibfield  {title} {\enquote {\bibinfo
  {title} {{Preflight Detector Characterization of BLAST-TNG}},}\ }\href
  {\doibase 10.1007/s10909-020-02459-6} {\bibfield  {journal} {\bibinfo
  {journal} {Journal of Low Temperature Physics}\ }\textbf {\bibinfo {volume}
  {200}},\ \bibinfo {pages} {400--406} (\bibinfo {year} {2020})}\BibitemShut
  {NoStop}%
\bibitem [{\citenamefont {{Sayers}}\ \emph {et~al.}(2020)\citenamefont
  {{Sayers}}, \citenamefont {{Day}}, \citenamefont {{Cunnane}}, \citenamefont
  {{Eom}}, \citenamefont {{LeDuc}}, \citenamefont {{O'Brient}}, \citenamefont
  {{Runyan}}, \citenamefont {{Bryan}}, \citenamefont {{Gordon}}, \citenamefont
  {{Mauskopf}}, \citenamefont {{Johnson}}, \citenamefont {{McCarrick}},\ and\
  \citenamefont {{Bhandarkar}}}]{Sayers2020}%
  \BibitemOpen
  \bibfield  {author} {\bibinfo {author} {\bibfnamefont {J.}~\bibnamefont
  {{Sayers}}}, \bibinfo {author} {\bibfnamefont {P.~K.}\ \bibnamefont {{Day}}},
  \bibinfo {author} {\bibfnamefont {D.~P.}\ \bibnamefont {{Cunnane}}}, \bibinfo
  {author} {\bibfnamefont {B.~H.}\ \bibnamefont {{Eom}}}, \bibinfo {author}
  {\bibfnamefont {H.~G.}\ \bibnamefont {{LeDuc}}}, \bibinfo {author}
  {\bibfnamefont {R.~C.}\ \bibnamefont {{O'Brient}}}, \bibinfo {author}
  {\bibfnamefont {M.~C.}\ \bibnamefont {{Runyan}}}, \bibinfo {author}
  {\bibfnamefont {S.~A.}\ \bibnamefont {{Bryan}}}, \bibinfo {author}
  {\bibfnamefont {S.~B.}\ \bibnamefont {{Gordon}}}, \bibinfo {author}
  {\bibfnamefont {P.~D.}\ \bibnamefont {{Mauskopf}}}, \bibinfo {author}
  {\bibfnamefont {B.~R.}\ \bibnamefont {{Johnson}}}, \bibinfo {author}
  {\bibfnamefont {H.}~\bibnamefont {{McCarrick}}}, \ and\ \bibinfo {author}
  {\bibfnamefont {T.~A.}\ \bibnamefont {{Bhandarkar}}},\ }\bibfield  {title}
  {\enquote {\bibinfo {title} {{A millimeter-wave kinetic inductance detector
  camera for long-range imaging through optical obscurants}},}\ }in\ \href
  {\doibase 10.1117/12.2557428} {\emph {\bibinfo {booktitle} {Passive and
  Active Millimeter-Wave Imaging XXIII}}},\ \bibinfo {series} {Society of
  Photo-Optical Instrumentation Engineers (SPIE) Conference Series}, Vol.\
  \bibinfo {volume} {11411},\ \bibinfo {editor} {edited by\ \bibinfo {editor}
  {\bibfnamefont {D.~A.}\ \bibnamefont {{Wikner}}}\ and\ \bibinfo {editor}
  {\bibfnamefont {D.~A.}\ \bibnamefont {{Robertson}}}}\ (\bibinfo {year}
  {2020})\ p.\ \bibinfo {pages} {114110H}\BibitemShut {NoStop}%
\bibitem [{Note1()}]{Note1}%
  \BibitemOpen
  \bibinfo {note} {Https://starcryo.com}\BibitemShut {NoStop}%
\bibitem [{Note2()}]{Note2}%
  \BibitemOpen
  \bibinfo {note} {Https://www.shicryogenics.com}\BibitemShut {NoStop}%
\bibitem [{\citenamefont {{Fickett}}, \citenamefont {{Reed}},\ and\
  \citenamefont {{Dalder}}(1979)}]{Fickett1979}%
  \BibitemOpen
  \bibfield  {author} {\bibinfo {author} {\bibfnamefont {F.~R.}\ \bibnamefont
  {{Fickett}}}, \bibinfo {author} {\bibfnamefont {R.~P.}\ \bibnamefont
  {{Reed}}}, \ and\ \bibinfo {author} {\bibfnamefont {E.~N.~C.}\ \bibnamefont
  {{Dalder}}},\ }\bibfield  {title} {\enquote {\bibinfo {title} {{Structures,
  insulators and conductors for large superconducting magnets*1}},}\ }\href
  {\doibase 10.1016/0022-3115(79)90515-4} {\bibfield  {journal} {\bibinfo
  {journal} {Journal of Nuclear Materials}\ }\textbf {\bibinfo {volume}
  {85-86}},\ \bibinfo {pages} {353--360} (\bibinfo {year} {1979})}\BibitemShut
  {NoStop}%
\bibitem [{Note3()}]{Note3}%
  \BibitemOpen
  \bibinfo {note} {Https://www.carlisleit.com/brands/micro-coax/}\BibitemShut
  {NoStop}%
\bibitem [{\citenamefont {Marquardt}, \citenamefont {Le},\ and\ \citenamefont
  {Radebaugh}(2002)}]{Marquardt2002}%
  \BibitemOpen
  \bibfield  {author} {\bibinfo {author} {\bibfnamefont {E.~D.}\ \bibnamefont
  {Marquardt}}, \bibinfo {author} {\bibfnamefont {J.~P.}\ \bibnamefont {Le}}, \
  and\ \bibinfo {author} {\bibfnamefont {R.}~\bibnamefont {Radebaugh}},\
  }\enquote {\bibinfo {title} {Cryogenic material properties database},}\ in\
  \href {\doibase 10.1007/0-306-47112-4_84} {\emph {\bibinfo {booktitle}
  {Cryocoolers 11}}},\ \bibinfo {editor} {edited by\ \bibinfo {editor}
  {\bibfnamefont {R.~G.}\ \bibnamefont {Ross}}}\ (\bibinfo  {publisher}
  {Springer US},\ \bibinfo {address} {Boston, MA},\ \bibinfo {year} {2002})\
  pp.\ \bibinfo {pages} {681--687}\BibitemShut {NoStop}%
\bibitem [{Note4()}]{Note4}%
  \BibitemOpen
  \bibinfo {note} {Http://www.qmcinstruments.co.uk/}\BibitemShut {NoStop}%
\bibitem [{\citenamefont {{Gudmundsson}}\ \emph {et~al.}(2015)\citenamefont
  {{Gudmundsson}}, \citenamefont {{Ade}}, \citenamefont {{Amiri}},
  \citenamefont {{Benton}}, \citenamefont {{Bock}}, \citenamefont {{Bond}},
  \citenamefont {{Bryan}}, \citenamefont {{Chiang}}, \citenamefont
  {{Contaldi}}, \citenamefont {{Crill}}, \citenamefont {{Dore}}, \citenamefont
  {{Filippini}}, \citenamefont {{Fraisse}}, \citenamefont {{Gambrel}},
  \citenamefont {{Gandilo}}, \citenamefont {{Hasselfield}}, \citenamefont
  {{Halpern}}, \citenamefont {{Hilton}}, \citenamefont {{Holmes}},
  \citenamefont {{Hristov}}, \citenamefont {{Irwin}}, \citenamefont {{Jones}},
  \citenamefont {{Kermish}}, \citenamefont {{MacTavish}}, \citenamefont
  {{Mason}}, \citenamefont {{Megerian}}, \citenamefont {{Moncelsi}},
  \citenamefont {{Montroy}}, \citenamefont {{Morford}}, \citenamefont {{Nagy}},
  \citenamefont {{Netterfield}}, \citenamefont {{Rahlin}}, \citenamefont
  {{Reintsema}}, \citenamefont {{Ruhl}}, \citenamefont {{Runyan}},
  \citenamefont {{Shariff}}, \citenamefont {{Soler}}, \citenamefont
  {{Trangsrud}}, \citenamefont {{Tucker}}, \citenamefont {{Tucker}},
  \citenamefont {{Turner}}, \citenamefont {{Wiebe}}, \citenamefont {{Young}},
  \citenamefont {{Spider Collaboration}}, \citenamefont {{Abe}}, \citenamefont
  {{Amiri}}, \citenamefont {{Benton}}, \citenamefont {{Bock}}, \citenamefont
  {{Bond}}, \citenamefont {{Bryan}}, \citenamefont {{Chiang}}, \citenamefont
  {{Contaldi}}, \citenamefont {{Crill}}, \citenamefont {{Dore}}, \citenamefont
  {{Filippini}}, \citenamefont {{Fraisse}}, \citenamefont {{Gambrel}},
  \citenamefont {{Gandilo}}, \citenamefont {{Hasselfield}}, \citenamefont
  {{Halpern}}, \citenamefont {{Hilton}}, \citenamefont {{Holmes}},
  \citenamefont {{Hristov}}, \citenamefont {{Irwin}}, \citenamefont {{Jones}},
  \citenamefont {{Kermish}}, \citenamefont {{MacTavish}}, \citenamefont
  {{Mason}}, \citenamefont {{Megerian}}, \citenamefont {{Moncelsi}},
  \citenamefont {{Montroy}}, \citenamefont {{Morford}}, \citenamefont {{Nagy}},
  \citenamefont {{Netterfield}}, \citenamefont {{Rahlin}}, \citenamefont
  {{Reintsema}}, \citenamefont {{Ruhl}}, \citenamefont {{Runyan}},
  \citenamefont {{Shariff}}, \citenamefont {{Soler}}, \citenamefont
  {{Trangsrud}}, \citenamefont {{Tucker}}, \citenamefont {{Tucker}},
  \citenamefont {{Turner}}, \citenamefont {{Wiebe}},\ and\ \citenamefont
  {{Young}}}]{Gudmundsson2015}%
  \BibitemOpen
  \bibfield  {author} {\bibinfo {author} {\bibfnamefont {J.~E.}\ \bibnamefont
  {{Gudmundsson}}}, \bibinfo {author} {\bibfnamefont {P.~A.~R.}\ \bibnamefont
  {{Ade}}}, \bibinfo {author} {\bibfnamefont {M.}~\bibnamefont {{Amiri}}},
  \bibinfo {author} {\bibfnamefont {S.~J.}\ \bibnamefont {{Benton}}}, \bibinfo
  {author} {\bibfnamefont {J.~J.}\ \bibnamefont {{Bock}}}, \bibinfo {author}
  {\bibfnamefont {J.~R.}\ \bibnamefont {{Bond}}}, \bibinfo {author}
  {\bibfnamefont {S.~A.}\ \bibnamefont {{Bryan}}}, \bibinfo {author}
  {\bibfnamefont {H.~C.}\ \bibnamefont {{Chiang}}}, \bibinfo {author}
  {\bibfnamefont {C.~R.}\ \bibnamefont {{Contaldi}}}, \bibinfo {author}
  {\bibfnamefont {B.~P.}\ \bibnamefont {{Crill}}}, \bibinfo {author}
  {\bibfnamefont {O.}~\bibnamefont {{Dore}}}, \bibinfo {author} {\bibfnamefont
  {J.~P.}\ \bibnamefont {{Filippini}}}, \bibinfo {author} {\bibfnamefont
  {A.~A.}\ \bibnamefont {{Fraisse}}}, \bibinfo {author} {\bibfnamefont
  {A.}~\bibnamefont {{Gambrel}}}, \bibinfo {author} {\bibfnamefont {N.~N.}\
  \bibnamefont {{Gandilo}}}, \bibinfo {author} {\bibfnamefont {M.}~\bibnamefont
  {{Hasselfield}}}, \bibinfo {author} {\bibfnamefont {M.}~\bibnamefont
  {{Halpern}}}, \bibinfo {author} {\bibfnamefont {G.}~\bibnamefont {{Hilton}}},
  \bibinfo {author} {\bibfnamefont {W.}~\bibnamefont {{Holmes}}}, \bibinfo
  {author} {\bibfnamefont {V.~V.}\ \bibnamefont {{Hristov}}}, \bibinfo {author}
  {\bibfnamefont {K.~D.}\ \bibnamefont {{Irwin}}}, \bibinfo {author}
  {\bibfnamefont {W.~C.}\ \bibnamefont {{Jones}}}, \bibinfo {author}
  {\bibfnamefont {Z.}~\bibnamefont {{Kermish}}}, \bibinfo {author}
  {\bibfnamefont {C.~J.}\ \bibnamefont {{MacTavish}}}, \bibinfo {author}
  {\bibfnamefont {P.~V.}\ \bibnamefont {{Mason}}}, \bibinfo {author}
  {\bibfnamefont {K.}~\bibnamefont {{Megerian}}}, \bibinfo {author}
  {\bibfnamefont {L.}~\bibnamefont {{Moncelsi}}}, \bibinfo {author}
  {\bibfnamefont {T.~E.}\ \bibnamefont {{Montroy}}}, \bibinfo {author}
  {\bibfnamefont {T.~A.}\ \bibnamefont {{Morford}}}, \bibinfo {author}
  {\bibfnamefont {J.~M.}\ \bibnamefont {{Nagy}}}, \bibinfo {author}
  {\bibfnamefont {C.~B.}\ \bibnamefont {{Netterfield}}}, \bibinfo {author}
  {\bibfnamefont {A.~S.}\ \bibnamefont {{Rahlin}}}, \bibinfo {author}
  {\bibfnamefont {C.~D.}\ \bibnamefont {{Reintsema}}}, \bibinfo {author}
  {\bibfnamefont {J.~E.}\ \bibnamefont {{Ruhl}}}, \bibinfo {author}
  {\bibfnamefont {M.~C.}\ \bibnamefont {{Runyan}}}, \bibinfo {author}
  {\bibfnamefont {J.~A.}\ \bibnamefont {{Shariff}}}, \bibinfo {author}
  {\bibfnamefont {J.~D.}\ \bibnamefont {{Soler}}}, \bibinfo {author}
  {\bibfnamefont {A.}~\bibnamefont {{Trangsrud}}}, \bibinfo {author}
  {\bibfnamefont {C.}~\bibnamefont {{Tucker}}}, \bibinfo {author}
  {\bibfnamefont {R.~S.}\ \bibnamefont {{Tucker}}}, \bibinfo {author}
  {\bibfnamefont {A.~D.}\ \bibnamefont {{Turner}}}, \bibinfo {author}
  {\bibfnamefont {D.~V.}\ \bibnamefont {{Wiebe}}}, \bibinfo {author}
  {\bibfnamefont {E.}~\bibnamefont {{Young}}}, \bibinfo {author} {\bibnamefont
  {{Spider Collaboration}}}, \bibinfo {author} {\bibfnamefont {P.~A.~R.}\
  \bibnamefont {{Abe}}}, \bibinfo {author} {\bibfnamefont {M.}~\bibnamefont
  {{Amiri}}}, \bibinfo {author} {\bibfnamefont {S.~J.}\ \bibnamefont
  {{Benton}}}, \bibinfo {author} {\bibfnamefont {J.~J.}\ \bibnamefont
  {{Bock}}}, \bibinfo {author} {\bibfnamefont {J.~R.}\ \bibnamefont {{Bond}}},
  \bibinfo {author} {\bibfnamefont {S.~A.}\ \bibnamefont {{Bryan}}}, \bibinfo
  {author} {\bibfnamefont {H.~C.}\ \bibnamefont {{Chiang}}}, \bibinfo {author}
  {\bibfnamefont {C.~R.}\ \bibnamefont {{Contaldi}}}, \bibinfo {author}
  {\bibfnamefont {B.~P.}\ \bibnamefont {{Crill}}}, \bibinfo {author}
  {\bibfnamefont {O.}~\bibnamefont {{Dore}}}, \bibinfo {author} {\bibfnamefont
  {J.~P.}\ \bibnamefont {{Filippini}}}, \bibinfo {author} {\bibfnamefont
  {A.~A.}\ \bibnamefont {{Fraisse}}}, \bibinfo {author} {\bibfnamefont
  {A.}~\bibnamefont {{Gambrel}}}, \bibinfo {author} {\bibfnamefont {N.~N.}\
  \bibnamefont {{Gandilo}}}, \bibinfo {author} {\bibfnamefont {M.}~\bibnamefont
  {{Hasselfield}}}, \bibinfo {author} {\bibfnamefont {M.}~\bibnamefont
  {{Halpern}}}, \bibinfo {author} {\bibfnamefont {G.}~\bibnamefont {{Hilton}}},
  \bibinfo {author} {\bibfnamefont {W.}~\bibnamefont {{Holmes}}}, \bibinfo
  {author} {\bibfnamefont {V.~V.}\ \bibnamefont {{Hristov}}}, \bibinfo {author}
  {\bibfnamefont {K.~D.}\ \bibnamefont {{Irwin}}}, \bibinfo {author}
  {\bibfnamefont {W.~C.}\ \bibnamefont {{Jones}}}, \bibinfo {author}
  {\bibfnamefont {Z.}~\bibnamefont {{Kermish}}}, \bibinfo {author}
  {\bibfnamefont {C.~J.}\ \bibnamefont {{MacTavish}}}, \bibinfo {author}
  {\bibfnamefont {P.~V.}\ \bibnamefont {{Mason}}}, \bibinfo {author}
  {\bibfnamefont {K.}~\bibnamefont {{Megerian}}}, \bibinfo {author}
  {\bibfnamefont {L.}~\bibnamefont {{Moncelsi}}}, \bibinfo {author}
  {\bibfnamefont {T.~E.}\ \bibnamefont {{Montroy}}}, \bibinfo {author}
  {\bibfnamefont {T.~A.}\ \bibnamefont {{Morford}}}, \bibinfo {author}
  {\bibfnamefont {J.~M.}\ \bibnamefont {{Nagy}}}, \bibinfo {author}
  {\bibfnamefont {C.~B.}\ \bibnamefont {{Netterfield}}}, \bibinfo {author}
  {\bibfnamefont {A.~S.}\ \bibnamefont {{Rahlin}}}, \bibinfo {author}
  {\bibfnamefont {C.~D.}\ \bibnamefont {{Reintsema}}}, \bibinfo {author}
  {\bibfnamefont {J.~E.}\ \bibnamefont {{Ruhl}}}, \bibinfo {author}
  {\bibfnamefont {M.~C.}\ \bibnamefont {{Runyan}}}, \bibinfo {author}
  {\bibfnamefont {J.~A.}\ \bibnamefont {{Shariff}}}, \bibinfo {author}
  {\bibfnamefont {J.~D.}\ \bibnamefont {{Soler}}}, \bibinfo {author}
  {\bibfnamefont {A.}~\bibnamefont {{Trangsrud}}}, \bibinfo {author}
  {\bibfnamefont {C.}~\bibnamefont {{Tucker}}}, \bibinfo {author}
  {\bibfnamefont {R.~S.}\ \bibnamefont {{Tucker}}}, \bibinfo {author}
  {\bibfnamefont {A.~D.}\ \bibnamefont {{Turner}}}, \bibinfo {author}
  {\bibfnamefont {D.~V.}\ \bibnamefont {{Wiebe}}}, \ and\ \bibinfo {author}
  {\bibfnamefont {E.}~\bibnamefont {{Young}}},\ }\bibfield  {title} {\enquote
  {\bibinfo {title} {{The thermal design, characterization, and performance of
  the SPIDER long-duration balloon cryostat}},}\ }\href {\doibase
  10.1016/j.cryogenics.2015.09.002} {\bibfield  {journal} {\bibinfo  {journal}
  {Cryogenics}\ }\textbf {\bibinfo {volume} {72}},\ \bibinfo {pages} {65--76}
  (\bibinfo {year} {2015})},\ \Eprint {http://arxiv.org/abs/1506.06953}
  {arXiv:1506.06953 [astro-ph.IM]} \BibitemShut {NoStop}%
\bibitem [{\citenamefont {{Miyakita}}\ \emph {et~al.}(2014)\citenamefont
  {{Miyakita}}, \citenamefont {{Hatakenaka}}, \citenamefont {{Sugita}},
  \citenamefont {{Saitoh}},\ and\ \citenamefont {{Hirai}}}]{Miyakita2014}%
  \BibitemOpen
  \bibfield  {author} {\bibinfo {author} {\bibfnamefont {T.}~\bibnamefont
  {{Miyakita}}}, \bibinfo {author} {\bibfnamefont {R.}~\bibnamefont
  {{Hatakenaka}}}, \bibinfo {author} {\bibfnamefont {H.}~\bibnamefont
  {{Sugita}}}, \bibinfo {author} {\bibfnamefont {M.}~\bibnamefont {{Saitoh}}},
  \ and\ \bibinfo {author} {\bibfnamefont {T.}~\bibnamefont {{Hirai}}},\
  }\bibfield  {title} {\enquote {\bibinfo {title} {{Development of a new
  multi-layer insulation blanket with non-interlayer-contact spacer for space
  cryogenic mission}},}\ }\href {\doibase 10.1016/j.cryogenics.2014.04.008}
  {\bibfield  {journal} {\bibinfo  {journal} {Cryogenics}\ }\textbf {\bibinfo
  {volume} {64}},\ \bibinfo {pages} {112--120} (\bibinfo {year}
  {2014})}\BibitemShut {NoStop}%
\bibitem [{Note5()}]{Note5}%
  \BibitemOpen
  \bibinfo {note} {Https://www.cosmicmicrowavetechnology.com/}\BibitemShut
  {NoStop}%
\bibitem [{\citenamefont {{Kellaris}}\ \emph {et~al.}(2014)\citenamefont
  {{Kellaris}}, \citenamefont {{Daal}}, \citenamefont {{Epland}}, \citenamefont
  {{Pepin}}, \citenamefont {{Kamaev}}, \citenamefont {{Cushman}}, \citenamefont
  {{Kramer}}, \citenamefont {{Sadoulet}}, \citenamefont {{Mirabolfathi}},
  \citenamefont {{Golwala}},\ and\ \citenamefont {{Runyan}}}]{Kelleris2014}%
  \BibitemOpen
  \bibfield  {author} {\bibinfo {author} {\bibfnamefont {N.}~\bibnamefont
  {{Kellaris}}}, \bibinfo {author} {\bibfnamefont {M.}~\bibnamefont {{Daal}}},
  \bibinfo {author} {\bibfnamefont {M.}~\bibnamefont {{Epland}}}, \bibinfo
  {author} {\bibfnamefont {M.}~\bibnamefont {{Pepin}}}, \bibinfo {author}
  {\bibfnamefont {O.}~\bibnamefont {{Kamaev}}}, \bibinfo {author}
  {\bibfnamefont {P.}~\bibnamefont {{Cushman}}}, \bibinfo {author}
  {\bibfnamefont {E.}~\bibnamefont {{Kramer}}}, \bibinfo {author}
  {\bibfnamefont {B.}~\bibnamefont {{Sadoulet}}}, \bibinfo {author}
  {\bibfnamefont {N.}~\bibnamefont {{Mirabolfathi}}}, \bibinfo {author}
  {\bibfnamefont {S.}~\bibnamefont {{Golwala}}}, \ and\ \bibinfo {author}
  {\bibfnamefont {M.}~\bibnamefont {{Runyan}}},\ }\bibfield  {title} {\enquote
  {\bibinfo {title} {{Sub-Kelvin Thermal Conductivity and Radioactivity of Some
  Useful Materials in Low Background Cryogenic Experiments}},}\ }\href
  {\doibase 10.1007/s10909-013-1048-6} {\bibfield  {journal} {\bibinfo
  {journal} {Journal of Low Temperature Physics}\ }\textbf {\bibinfo {volume}
  {176}},\ \bibinfo {pages} {201--208} (\bibinfo {year} {2014})}\BibitemShut
  {NoStop}%
\bibitem [{Note6()}]{Note6}%
  \BibitemOpen
  \bibinfo {note} {Https://keycom.co.jp/}\BibitemShut {NoStop}%
\bibitem [{\citenamefont {{Daal}}\ \emph {et~al.}(2019)\citenamefont {{Daal}},
  \citenamefont {{Zobrist}}, \citenamefont {{Kellaris}}, \citenamefont
  {{Sadoulet}},\ and\ \citenamefont {{Robertson}}}]{Daal2019}%
  \BibitemOpen
  \bibfield  {author} {\bibinfo {author} {\bibfnamefont {M.}~\bibnamefont
  {{Daal}}}, \bibinfo {author} {\bibfnamefont {N.}~\bibnamefont {{Zobrist}}},
  \bibinfo {author} {\bibfnamefont {N.}~\bibnamefont {{Kellaris}}}, \bibinfo
  {author} {\bibfnamefont {B.}~\bibnamefont {{Sadoulet}}}, \ and\ \bibinfo
  {author} {\bibfnamefont {M.}~\bibnamefont {{Robertson}}},\ }\bibfield
  {title} {\enquote {\bibinfo {title} {{Properties of selected structural and
  flat flexible cabling materials for low temperature applications}},}\ }\href
  {\doibase 10.1016/j.cryogenics.2018.10.019} {\bibfield  {journal} {\bibinfo
  {journal} {Cryogenics}\ }\textbf {\bibinfo {volume} {98}},\ \bibinfo {pages}
  {47--59} (\bibinfo {year} {2019})},\ \Eprint
  {http://arxiv.org/abs/1810.10187} {arXiv:1810.10187 [physics.ins-det]}
  \BibitemShut {NoStop}%
\bibitem [{\citenamefont {{Drobizhev}}\ \emph {et~al.}(2017)\citenamefont
  {{Drobizhev}}, \citenamefont {{Reiten}}, \citenamefont {{Singh}},\ and\
  \citenamefont {{Kolomensky}}}]{Drobizhev2017}%
  \BibitemOpen
  \bibfield  {author} {\bibinfo {author} {\bibfnamefont {A.}~\bibnamefont
  {{Drobizhev}}}, \bibinfo {author} {\bibfnamefont {J.}~\bibnamefont
  {{Reiten}}}, \bibinfo {author} {\bibfnamefont {V.}~\bibnamefont {{Singh}}}, \
  and\ \bibinfo {author} {\bibfnamefont {Y.~G.}\ \bibnamefont {{Kolomensky}}},\
  }\bibfield  {title} {\enquote {\bibinfo {title} {{Thermal conductivity
  measurements of PTFE and Al$_{2}$O$_{3}$ ceramic at sub-Kelvin
  temperatures}},}\ }\href {\doibase 10.1016/j.cryogenics.2017.05.008}
  {\bibfield  {journal} {\bibinfo  {journal} {Cryogenics}\ }\textbf {\bibinfo
  {volume} {85}},\ \bibinfo {pages} {63--70} (\bibinfo {year} {2017})},\
  \Eprint {http://arxiv.org/abs/1612.02514} {arXiv:1612.02514
  [cond-mat.mtrl-sci]} \BibitemShut {NoStop}%
\bibitem [{\citenamefont {{Peroni}}\ \emph {et~al.}(1999)\citenamefont
  {{Peroni}}, \citenamefont {{Gottardi}}, \citenamefont {{Peruzzi}},
  \citenamefont {{Ponti}},\ and\ \citenamefont {{Ventura}}}]{Peroni1999}%
  \BibitemOpen
  \bibfield  {author} {\bibinfo {author} {\bibfnamefont {I.}~\bibnamefont
  {{Peroni}}}, \bibinfo {author} {\bibfnamefont {E.}~\bibnamefont
  {{Gottardi}}}, \bibinfo {author} {\bibfnamefont {A.}~\bibnamefont
  {{Peruzzi}}}, \bibinfo {author} {\bibfnamefont {G.}~\bibnamefont {{Ponti}}},
  \ and\ \bibinfo {author} {\bibfnamefont {G.}~\bibnamefont {{Ventura}}},\
  }\bibfield  {title} {\enquote {\bibinfo {title} {{Thermal conductivity of
  manganin below 1 K}},}\ }\href {\doibase 10.1016/S0920-5632(99)00606-4}
  {\bibfield  {journal} {\bibinfo  {journal} {Nuclear Physics B Proceedings
  Supplements}\ }\textbf {\bibinfo {volume} {78}},\ \bibinfo {pages} {573--575}
  (\bibinfo {year} {1999})}\BibitemShut {NoStop}%
\bibitem [{\citenamefont {{Bartlett}}, \citenamefont {{Hardy}},\ and\
  \citenamefont {{Hepburn}}(2015)}]{Bartlett2015}%
  \BibitemOpen
  \bibfield  {author} {\bibinfo {author} {\bibfnamefont {J.}~\bibnamefont
  {{Bartlett}}}, \bibinfo {author} {\bibfnamefont {G.}~\bibnamefont {{Hardy}}},
  \ and\ \bibinfo {author} {\bibfnamefont {I.~D.}\ \bibnamefont {{Hepburn}}},\
  }\bibfield  {title} {\enquote {\bibinfo {title} {{Performance of a fast
  response miniature Adiabatic Demagnetisation Refrigerator using a single
  crystal tungsten magnetoresistive heat switch}},}\ }\href {\doibase
  10.1016/j.cryogenics.2015.10.004} {\bibfield  {journal} {\bibinfo  {journal}
  {Cryogenics}\ }\textbf {\bibinfo {volume} {72}},\ \bibinfo {pages} {111--121}
  (\bibinfo {year} {2015})}\BibitemShut {NoStop}%
\bibitem [{Note7()}]{Note7}%
  \BibitemOpen
  \bibinfo {note} {Https://www.timet.com/}\BibitemShut {NoStop}%
\bibitem [{\citenamefont {{Gordon}}\ \emph {et~al.}(2016)\citenamefont
  {{Gordon}}, \citenamefont {{Dober}}, \citenamefont {{Sinclair}},
  \citenamefont {{Rowe}}, \citenamefont {{Bryan}}, \citenamefont {{Mauskopf}},
  \citenamefont {{Austermann}}, \citenamefont {{Devlin}}, \citenamefont
  {{Dicker}}, \citenamefont {{Gao}}, \citenamefont {{Hilton}}, \citenamefont
  {{Hubmayr}}, \citenamefont {{Jones}}, \citenamefont {{Klein}}, \citenamefont
  {{Lourie}}, \citenamefont {{McKenney}}, \citenamefont {{Nati}}, \citenamefont
  {{Soler}}, \citenamefont {{Strader}},\ and\ \citenamefont
  {{Vissers}}}]{Gordon2016}%
  \BibitemOpen
  \bibfield  {author} {\bibinfo {author} {\bibfnamefont {S.}~\bibnamefont
  {{Gordon}}}, \bibinfo {author} {\bibfnamefont {B.}~\bibnamefont {{Dober}}},
  \bibinfo {author} {\bibfnamefont {A.}~\bibnamefont {{Sinclair}}}, \bibinfo
  {author} {\bibfnamefont {S.}~\bibnamefont {{Rowe}}}, \bibinfo {author}
  {\bibfnamefont {S.}~\bibnamefont {{Bryan}}}, \bibinfo {author} {\bibfnamefont
  {P.}~\bibnamefont {{Mauskopf}}}, \bibinfo {author} {\bibfnamefont
  {J.}~\bibnamefont {{Austermann}}}, \bibinfo {author} {\bibfnamefont
  {M.}~\bibnamefont {{Devlin}}}, \bibinfo {author} {\bibfnamefont
  {S.}~\bibnamefont {{Dicker}}}, \bibinfo {author} {\bibfnamefont
  {J.}~\bibnamefont {{Gao}}}, \bibinfo {author} {\bibfnamefont {G.~C.}\
  \bibnamefont {{Hilton}}}, \bibinfo {author} {\bibfnamefont {J.}~\bibnamefont
  {{Hubmayr}}}, \bibinfo {author} {\bibfnamefont {G.}~\bibnamefont {{Jones}}},
  \bibinfo {author} {\bibfnamefont {J.}~\bibnamefont {{Klein}}}, \bibinfo
  {author} {\bibfnamefont {N.~P.}\ \bibnamefont {{Lourie}}}, \bibinfo {author}
  {\bibfnamefont {C.}~\bibnamefont {{McKenney}}}, \bibinfo {author}
  {\bibfnamefont {F.}~\bibnamefont {{Nati}}}, \bibinfo {author} {\bibfnamefont
  {J.~D.}\ \bibnamefont {{Soler}}}, \bibinfo {author} {\bibfnamefont
  {M.}~\bibnamefont {{Strader}}}, \ and\ \bibinfo {author} {\bibfnamefont
  {M.}~\bibnamefont {{Vissers}}},\ }\bibfield  {title} {\enquote {\bibinfo
  {title} {{An Open Source, FPGA-Based LeKID Readout for BLAST-TNG: Pre-Flight
  Results}},}\ }\href {\doibase 10.1142/S2251171716410038} {\bibfield
  {journal} {\bibinfo  {journal} {Journal of Astronomical Instrumentation}\
  }\textbf {\bibinfo {volume} {5}},\ \bibinfo {eid} {1641003} (\bibinfo {year}
  {2016})},\ \Eprint {http://arxiv.org/abs/1611.05400} {arXiv:1611.05400
  [astro-ph.IM]} \BibitemShut {NoStop}%
\bibitem [{\citenamefont {{Duan}}\ \emph {et~al.}(2010)\citenamefont {{Duan}},
  \citenamefont {{McHugh}}, \citenamefont {{Serfass}}, \citenamefont {{Mazin}},
  \citenamefont {{Merrill}}, \citenamefont {{Golwala}}, \citenamefont
  {{Downes}}, \citenamefont {{Czakon}}, \citenamefont {{Day}}, \citenamefont
  {{Gao}}, \citenamefont {{Glenn}}, \citenamefont {{Hollister}}, \citenamefont
  {{Leduc}}, \citenamefont {{Maloney}}, \citenamefont {{Noroozian}},
  \citenamefont {{Nguyen}}, \citenamefont {{Sayers}}, \citenamefont
  {{Schlaerth}}, \citenamefont {{Siegel}}, \citenamefont {{Vaillancourt}},
  \citenamefont {{Vayonakis}}, \citenamefont {{Wilson}},\ and\ \citenamefont
  {{Zmuidzinas}}}]{Duan2010}%
  \BibitemOpen
  \bibfield  {author} {\bibinfo {author} {\bibfnamefont {R.}~\bibnamefont
  {{Duan}}}, \bibinfo {author} {\bibfnamefont {S.}~\bibnamefont {{McHugh}}},
  \bibinfo {author} {\bibfnamefont {B.}~\bibnamefont {{Serfass}}}, \bibinfo
  {author} {\bibfnamefont {B.~A.}\ \bibnamefont {{Mazin}}}, \bibinfo {author}
  {\bibfnamefont {A.}~\bibnamefont {{Merrill}}}, \bibinfo {author}
  {\bibfnamefont {S.~R.}\ \bibnamefont {{Golwala}}}, \bibinfo {author}
  {\bibfnamefont {T.~P.}\ \bibnamefont {{Downes}}}, \bibinfo {author}
  {\bibfnamefont {N.~G.}\ \bibnamefont {{Czakon}}}, \bibinfo {author}
  {\bibfnamefont {P.~K.}\ \bibnamefont {{Day}}}, \bibinfo {author}
  {\bibfnamefont {J.}~\bibnamefont {{Gao}}}, \bibinfo {author} {\bibfnamefont
  {J.}~\bibnamefont {{Glenn}}}, \bibinfo {author} {\bibfnamefont {M.~I.}\
  \bibnamefont {{Hollister}}}, \bibinfo {author} {\bibfnamefont {H.~G.}\
  \bibnamefont {{Leduc}}}, \bibinfo {author} {\bibfnamefont {P.~R.}\
  \bibnamefont {{Maloney}}}, \bibinfo {author} {\bibfnamefont {O.}~\bibnamefont
  {{Noroozian}}}, \bibinfo {author} {\bibfnamefont {H.~T.}\ \bibnamefont
  {{Nguyen}}}, \bibinfo {author} {\bibfnamefont {J.}~\bibnamefont {{Sayers}}},
  \bibinfo {author} {\bibfnamefont {J.~A.}\ \bibnamefont {{Schlaerth}}},
  \bibinfo {author} {\bibfnamefont {S.}~\bibnamefont {{Siegel}}}, \bibinfo
  {author} {\bibfnamefont {J.~E.}\ \bibnamefont {{Vaillancourt}}}, \bibinfo
  {author} {\bibfnamefont {A.}~\bibnamefont {{Vayonakis}}}, \bibinfo {author}
  {\bibfnamefont {P.~R.}\ \bibnamefont {{Wilson}}}, \ and\ \bibinfo {author}
  {\bibfnamefont {J.}~\bibnamefont {{Zmuidzinas}}},\ }\bibfield  {title}
  {\enquote {\bibinfo {title} {{An open-source readout for MKIDs}},}\ }in\
  \href {\doibase 10.1117/12.856832} {\emph {\bibinfo {booktitle} {Millimeter,
  Submillimeter, and Far-Infrared Detectors and Instrumentation for Astronomy
  V}}},\ \bibinfo {series} {Society of Photo-Optical Instrumentation Engineers
  (SPIE) Conference Series}, Vol.\ \bibinfo {volume} {7741},\ \bibinfo {editor}
  {edited by\ \bibinfo {editor} {\bibfnamefont {W.~S.}\ \bibnamefont
  {{Holland}}}\ and\ \bibinfo {editor} {\bibfnamefont {J.}~\bibnamefont
  {{Zmuidzinas}}}}\ (\bibinfo {year} {2010})\ p.\ \bibinfo {pages}
  {77411V}\BibitemShut {NoStop}%
\bibitem [{Note8()}]{Note8}%
  \BibitemOpen
  \bibinfo {note}
  {Https://www.xilinx.com/products/boards-and-kits/zcu111.html}\BibitemShut
  {NoStop}%
\bibitem [{\citenamefont {{Sinclair}}\ \emph {et~al.}(2022)\citenamefont
  {{Sinclair}}, \citenamefont {{Stephenson}}, \citenamefont {{Roberson}},
  \citenamefont {{Weeks}}, \citenamefont {{Burgoyne}}, \citenamefont {{Huber}},
  \citenamefont {{Mauskopf}}, \citenamefont {{Chapman}}, \citenamefont
  {{Austermann}}, \citenamefont {{Choi}}, \citenamefont {{Duell}},
  \citenamefont {{Fich}}, \citenamefont {{Groppi}}, \citenamefont {{Huber}},
  \citenamefont {{Niemack}}, \citenamefont {{Nikola}}, \citenamefont {{Rossi}},
  \citenamefont {{Sriram}}, \citenamefont {{Stacey}}, \citenamefont
  {{Szakiel}}, \citenamefont {{Tsuchitori}}, \citenamefont {{Vavagiakis}},\
  and\ \citenamefont {{Wheeler}}}]{Sinclair2022}%
  \BibitemOpen
  \bibfield  {author} {\bibinfo {author} {\bibfnamefont {A.~K.}\ \bibnamefont
  {{Sinclair}}}, \bibinfo {author} {\bibfnamefont {R.~C.}\ \bibnamefont
  {{Stephenson}}}, \bibinfo {author} {\bibfnamefont {C.~A.}\ \bibnamefont
  {{Roberson}}}, \bibinfo {author} {\bibfnamefont {E.~L.}\ \bibnamefont
  {{Weeks}}}, \bibinfo {author} {\bibfnamefont {J.}~\bibnamefont {{Burgoyne}}},
  \bibinfo {author} {\bibfnamefont {A.~I.}\ \bibnamefont {{Huber}}}, \bibinfo
  {author} {\bibfnamefont {P.~M.}\ \bibnamefont {{Mauskopf}}}, \bibinfo
  {author} {\bibfnamefont {S.~C.}\ \bibnamefont {{Chapman}}}, \bibinfo {author}
  {\bibfnamefont {J.~E.}\ \bibnamefont {{Austermann}}}, \bibinfo {author}
  {\bibfnamefont {S.~K.}\ \bibnamefont {{Choi}}}, \bibinfo {author}
  {\bibfnamefont {C.~J.}\ \bibnamefont {{Duell}}}, \bibinfo {author}
  {\bibfnamefont {M.}~\bibnamefont {{Fich}}}, \bibinfo {author} {\bibfnamefont
  {C.~E.}\ \bibnamefont {{Groppi}}}, \bibinfo {author} {\bibfnamefont
  {Z.}~\bibnamefont {{Huber}}}, \bibinfo {author} {\bibfnamefont {M.~D.}\
  \bibnamefont {{Niemack}}}, \bibinfo {author} {\bibfnamefont {T.}~\bibnamefont
  {{Nikola}}}, \bibinfo {author} {\bibfnamefont {K.~M.}\ \bibnamefont
  {{Rossi}}}, \bibinfo {author} {\bibfnamefont {A.}~\bibnamefont {{Sriram}}},
  \bibinfo {author} {\bibfnamefont {G.~J.}\ \bibnamefont {{Stacey}}}, \bibinfo
  {author} {\bibfnamefont {E.}~\bibnamefont {{Szakiel}}}, \bibinfo {author}
  {\bibfnamefont {J.}~\bibnamefont {{Tsuchitori}}}, \bibinfo {author}
  {\bibfnamefont {E.~M.}\ \bibnamefont {{Vavagiakis}}}, \ and\ \bibinfo
  {author} {\bibfnamefont {J.~D.}\ \bibnamefont {{Wheeler}}},\ }\bibfield
  {title} {\enquote {\bibinfo {title} {{CCAT-prime: RFSoC based readout for
  frequency multiplexed kinetic inductance detectors}},}\ }in\ \href {\doibase
  10.1117/12.2629722} {\emph {\bibinfo {booktitle} {Millimeter, Submillimeter,
  and Far-Infrared Detectors and Instrumentation for Astronomy XI}}},\ \bibinfo
  {series} {Society of Photo-Optical Instrumentation Engineers (SPIE)
  Conference Series}, Vol.\ \bibinfo {volume} {12190},\ \bibinfo {editor}
  {edited by\ \bibinfo {editor} {\bibfnamefont {J.}~\bibnamefont
  {{Zmuidzinas}}}\ and\ \bibinfo {editor} {\bibfnamefont {J.-R.}\ \bibnamefont
  {{Gao}}}}\ (\bibinfo {year} {2022})\ p.\ \bibinfo {pages} {121900W},\ \Eprint
  {http://arxiv.org/abs/2208.07465} {arXiv:2208.07465 [astro-ph.IM]}
  \BibitemShut {NoStop}%
\bibitem [{\citenamefont {{Sinclair}}\ \emph {et~al.}(2024)\citenamefont
  {{Sinclair}}, \citenamefont {{Burgoyne}}, \citenamefont {{Huber}},
  \citenamefont {{Murphy}}, \citenamefont {{Choi}}, \citenamefont {{Duell}},
  \citenamefont {{Huber}}, \citenamefont {{Li}}, \citenamefont {{Chapman}},
  \citenamefont {{Niemack}}, \citenamefont {{Nikola}}, \citenamefont
  {{Vavagiakis}}, \citenamefont {{Walker}}, \citenamefont {{Wheeler}},
  \citenamefont {{Austermann}}, \citenamefont {{Lin}}, \citenamefont {{Xie}},
  \citenamefont {{Zou}},\ and\ \citenamefont {{Mauskopf}}}]{Sinclair2024}%
  \BibitemOpen
  \bibfield  {author} {\bibinfo {author} {\bibfnamefont {A.~K.}\ \bibnamefont
  {{Sinclair}}}, \bibinfo {author} {\bibfnamefont {J.}~\bibnamefont
  {{Burgoyne}}}, \bibinfo {author} {\bibfnamefont {A.~I.}\ \bibnamefont
  {{Huber}}}, \bibinfo {author} {\bibfnamefont {C.}~\bibnamefont {{Murphy}}},
  \bibinfo {author} {\bibfnamefont {S.~K.}\ \bibnamefont {{Choi}}}, \bibinfo
  {author} {\bibfnamefont {C.~J.}\ \bibnamefont {{Duell}}}, \bibinfo {author}
  {\bibfnamefont {Z.~B.}\ \bibnamefont {{Huber}}}, \bibinfo {author}
  {\bibfnamefont {Y.}~\bibnamefont {{Li}}}, \bibinfo {author} {\bibfnamefont
  {S.~C.}\ \bibnamefont {{Chapman}}}, \bibinfo {author} {\bibfnamefont {M.~D.}\
  \bibnamefont {{Niemack}}}, \bibinfo {author} {\bibfnamefont {T.}~\bibnamefont
  {{Nikola}}}, \bibinfo {author} {\bibfnamefont {E.~M.}\ \bibnamefont
  {{Vavagiakis}}}, \bibinfo {author} {\bibfnamefont {S.}~\bibnamefont
  {{Walker}}}, \bibinfo {author} {\bibfnamefont {J.~D.}\ \bibnamefont
  {{Wheeler}}}, \bibinfo {author} {\bibfnamefont {J.}~\bibnamefont
  {{Austermann}}}, \bibinfo {author} {\bibfnamefont {L.}~\bibnamefont {{Lin}}},
  \bibinfo {author} {\bibfnamefont {R.}~\bibnamefont {{Xie}}}, \bibinfo
  {author} {\bibfnamefont {B.}~\bibnamefont {{Zou}}}, \ and\ \bibinfo {author}
  {\bibfnamefont {P.~D.}\ \bibnamefont {{Mauskopf}}},\ }\bibfield  {title}
  {\enquote {\bibinfo {title} {{CCAT: Detector Noise Limited Performance of the
  RFSoC-based Readout Electronics for mm/sub-mm/far-IR KIDs}},}\ }\href
  {\doibase 10.48550/arXiv.2406.14892} {\bibfield  {journal} {\bibinfo
  {journal} {arXiv e-prints}\ ,\ \bibinfo {eid} {arXiv:2406.14892}} (\bibinfo
  {year} {2024})},\ \Eprint {http://arxiv.org/abs/2406.14892} {arXiv:2406.14892
  [astro-ph.IM]} \BibitemShut {NoStop}%
\bibitem [{\citenamefont {Dragone}(1978)}]{Dragone1978}%
  \BibitemOpen
  \bibfield  {author} {\bibinfo {author} {\bibfnamefont {C.}~\bibnamefont
  {Dragone}},\ }\bibfield  {title} {\enquote {\bibinfo {title} {Offset
  multireflector antennas with perfect pattern symmetry and polarization
  discrimination},}\ }\href {\doibase 10.1002/j.1538-7305.1978.tb02171.x}
  {\bibfield  {journal} {\bibinfo  {journal} {The Bell System Technical
  Journal}\ }\textbf {\bibinfo {volume} {57}},\ \bibinfo {pages} {2663--2684}
  (\bibinfo {year} {1978})}\BibitemShut {NoStop}%
\bibitem [{\citenamefont {{Ruze}}(1966)}]{Ruze1966}%
  \BibitemOpen
  \bibfield  {author} {\bibinfo {author} {\bibfnamefont {J.}~\bibnamefont
  {{Ruze}}},\ }\bibfield  {title} {\enquote {\bibinfo {title} {{Antenna
  Tolerance Theory -- A Review}},}\ }\href@noop {} {\bibfield  {journal}
  {\bibinfo  {journal} {IEEE Proceedings}\ }\textbf {\bibinfo {volume} {54}},\
  \bibinfo {pages} {633--642} (\bibinfo {year} {1966})}\BibitemShut {NoStop}%
\bibitem [{\citenamefont {{Crumrine}}\ \emph {et~al.}(2018)\citenamefont
  {{Crumrine}}, \citenamefont {{Ade}}, \citenamefont {{Ahmed}}, \citenamefont
  {{Aikin}}, \citenamefont {{Alexander}}, \citenamefont {{Barkats}},
  \citenamefont {{Benton}}, \citenamefont {{Bischoff}}, \citenamefont {{Bock}},
  \citenamefont {{Bowens-Rubin}}, \citenamefont {{Brevik}}, \citenamefont
  {{Buder}}, \citenamefont {{Bullock}}, \citenamefont {{Buza}}, \citenamefont
  {{Connors}}, \citenamefont {{Cornelison}}, \citenamefont {{Crill}},
  \citenamefont {{Dierickx}}, \citenamefont {{Duband}}, \citenamefont
  {{Dvorkin}}, \citenamefont {{Filippini}}, \citenamefont {{Fliescher}},
  \citenamefont {{Grayson}}, \citenamefont {{Hall}}, \citenamefont {{Halpern}},
  \citenamefont {{Harrison}}, \citenamefont {{Hildebrandt}}, \citenamefont
  {{Hilton}}, \citenamefont {{Hui}}, \citenamefont {{Irwin}}, \citenamefont
  {{Kang}}, \citenamefont {{Karkare}}, \citenamefont {{Karpel}}, \citenamefont
  {{Kaufman}}, \citenamefont {{Keating}}, \citenamefont {{Kefeli}},
  \citenamefont {{Kernasovskiy}}, \citenamefont {{Kovac}}, \citenamefont
  {{Kuo}}, \citenamefont {{Larsen}}, \citenamefont {{Lau}}, \citenamefont
  {{Leitch}}, \citenamefont {{Lueker}}, \citenamefont {{Megerian}},
  \citenamefont {{Moncelsi}}, \citenamefont {{Namikawa}}, \citenamefont
  {{Netterfield}}, \citenamefont {{Nguyen}}, \citenamefont {{O'Brient}},
  \citenamefont {{Ogburn}}, \citenamefont {{Palladino}}, \citenamefont
  {{Pryke}}, \citenamefont {{Racine}}, \citenamefont {{Richter}}, \citenamefont
  {{Schwarz}}, \citenamefont {{Schillaci}}, \citenamefont {{Sheehy}},
  \citenamefont {{Soliman}}, \citenamefont {{St. Germaine}}, \citenamefont
  {{Staniszewski}}, \citenamefont {{Steinbach}}, \citenamefont {{Sudiwala}},
  \citenamefont {{Teply}}, \citenamefont {{Thompson}}, \citenamefont {{Tolan}},
  \citenamefont {{Tucker}}, \citenamefont {{Turner}}, \citenamefont
  {{Umilt{\`a}}}, \citenamefont {{Vieregg}}, \citenamefont {{Wandui}},
  \citenamefont {{Weber}}, \citenamefont {{Wiebe}}, \citenamefont {{Willmert}},
  \citenamefont {{Wong}}, \citenamefont {{Wu}}, \citenamefont {{Yang}},
  \citenamefont {{Yoon}},\ and\ \citenamefont {{Zhang}}}]{Crumrine2018}%
  \BibitemOpen
  \bibfield  {author} {\bibinfo {author} {\bibfnamefont {M.}~\bibnamefont
  {{Crumrine}}}, \bibinfo {author} {\bibfnamefont {P.~A.~R.}\ \bibnamefont
  {{Ade}}}, \bibinfo {author} {\bibfnamefont {Z.}~\bibnamefont {{Ahmed}}},
  \bibinfo {author} {\bibfnamefont {R.~W.}\ \bibnamefont {{Aikin}}}, \bibinfo
  {author} {\bibfnamefont {K.~D.}\ \bibnamefont {{Alexander}}}, \bibinfo
  {author} {\bibfnamefont {D.}~\bibnamefont {{Barkats}}}, \bibinfo {author}
  {\bibfnamefont {S.~J.}\ \bibnamefont {{Benton}}}, \bibinfo {author}
  {\bibfnamefont {C.~A.}\ \bibnamefont {{Bischoff}}}, \bibinfo {author}
  {\bibfnamefont {J.~J.}\ \bibnamefont {{Bock}}}, \bibinfo {author}
  {\bibfnamefont {R.}~\bibnamefont {{Bowens-Rubin}}}, \bibinfo {author}
  {\bibfnamefont {J.~A.}\ \bibnamefont {{Brevik}}}, \bibinfo {author}
  {\bibfnamefont {I.}~\bibnamefont {{Buder}}}, \bibinfo {author} {\bibfnamefont
  {E.}~\bibnamefont {{Bullock}}}, \bibinfo {author} {\bibfnamefont
  {V.}~\bibnamefont {{Buza}}}, \bibinfo {author} {\bibfnamefont
  {J.}~\bibnamefont {{Connors}}}, \bibinfo {author} {\bibfnamefont
  {J.}~\bibnamefont {{Cornelison}}}, \bibinfo {author} {\bibfnamefont {B.~P.}\
  \bibnamefont {{Crill}}}, \bibinfo {author} {\bibfnamefont {M.}~\bibnamefont
  {{Dierickx}}}, \bibinfo {author} {\bibfnamefont {L.}~\bibnamefont
  {{Duband}}}, \bibinfo {author} {\bibfnamefont {C.}~\bibnamefont {{Dvorkin}}},
  \bibinfo {author} {\bibfnamefont {J.~P.}\ \bibnamefont {{Filippini}}},
  \bibinfo {author} {\bibfnamefont {S.}~\bibnamefont {{Fliescher}}}, \bibinfo
  {author} {\bibfnamefont {J.~A.}\ \bibnamefont {{Grayson}}}, \bibinfo {author}
  {\bibfnamefont {G.}~\bibnamefont {{Hall}}}, \bibinfo {author} {\bibfnamefont
  {M.}~\bibnamefont {{Halpern}}}, \bibinfo {author} {\bibfnamefont {S.~A.}\
  \bibnamefont {{Harrison}}}, \bibinfo {author} {\bibfnamefont {S.~R.}\
  \bibnamefont {{Hildebrandt}}}, \bibinfo {author} {\bibfnamefont {G.~C.}\
  \bibnamefont {{Hilton}}}, \bibinfo {author} {\bibfnamefont {H.}~\bibnamefont
  {{Hui}}}, \bibinfo {author} {\bibfnamefont {K.~D.}\ \bibnamefont {{Irwin}}},
  \bibinfo {author} {\bibfnamefont {J.~H.}\ \bibnamefont {{Kang}}}, \bibinfo
  {author} {\bibfnamefont {K.~S.}\ \bibnamefont {{Karkare}}}, \bibinfo {author}
  {\bibfnamefont {E.}~\bibnamefont {{Karpel}}}, \bibinfo {author}
  {\bibfnamefont {J.~P.}\ \bibnamefont {{Kaufman}}}, \bibinfo {author}
  {\bibfnamefont {B.~G.}\ \bibnamefont {{Keating}}}, \bibinfo {author}
  {\bibfnamefont {S.}~\bibnamefont {{Kefeli}}}, \bibinfo {author}
  {\bibfnamefont {S.~A.}\ \bibnamefont {{Kernasovskiy}}}, \bibinfo {author}
  {\bibfnamefont {J.~M.}\ \bibnamefont {{Kovac}}}, \bibinfo {author}
  {\bibfnamefont {C.~L.}\ \bibnamefont {{Kuo}}}, \bibinfo {author}
  {\bibfnamefont {N.~A.}\ \bibnamefont {{Larsen}}}, \bibinfo {author}
  {\bibfnamefont {K.}~\bibnamefont {{Lau}}}, \bibinfo {author} {\bibfnamefont
  {E.~M.}\ \bibnamefont {{Leitch}}}, \bibinfo {author} {\bibfnamefont {M.~V.}\
  \bibnamefont {{Lueker}}}, \bibinfo {author} {\bibfnamefont {K.~G.}\
  \bibnamefont {{Megerian}}}, \bibinfo {author} {\bibfnamefont
  {L.}~\bibnamefont {{Moncelsi}}}, \bibinfo {author} {\bibfnamefont
  {T.}~\bibnamefont {{Namikawa}}}, \bibinfo {author} {\bibfnamefont {C.~B.}\
  \bibnamefont {{Netterfield}}}, \bibinfo {author} {\bibfnamefont {H.~T.}\
  \bibnamefont {{Nguyen}}}, \bibinfo {author} {\bibfnamefont {R.}~\bibnamefont
  {{O'Brient}}}, \bibinfo {author} {\bibfnamefont {R.~W.}\ \bibnamefont
  {{Ogburn}}}, \bibinfo {author} {\bibfnamefont {S.}~\bibnamefont
  {{Palladino}}}, \bibinfo {author} {\bibfnamefont {C.}~\bibnamefont
  {{Pryke}}}, \bibinfo {author} {\bibfnamefont {B.}~\bibnamefont {{Racine}}},
  \bibinfo {author} {\bibfnamefont {S.}~\bibnamefont {{Richter}}}, \bibinfo
  {author} {\bibfnamefont {R.}~\bibnamefont {{Schwarz}}}, \bibinfo {author}
  {\bibfnamefont {A.}~\bibnamefont {{Schillaci}}}, \bibinfo {author}
  {\bibfnamefont {C.~D.}\ \bibnamefont {{Sheehy}}}, \bibinfo {author}
  {\bibfnamefont {A.}~\bibnamefont {{Soliman}}}, \bibinfo {author}
  {\bibfnamefont {T.}~\bibnamefont {{St. Germaine}}}, \bibinfo {author}
  {\bibfnamefont {Z.~K.}\ \bibnamefont {{Staniszewski}}}, \bibinfo {author}
  {\bibfnamefont {B.}~\bibnamefont {{Steinbach}}}, \bibinfo {author}
  {\bibfnamefont {R.~V.}\ \bibnamefont {{Sudiwala}}}, \bibinfo {author}
  {\bibfnamefont {G.~P.}\ \bibnamefont {{Teply}}}, \bibinfo {author}
  {\bibfnamefont {K.~L.}\ \bibnamefont {{Thompson}}}, \bibinfo {author}
  {\bibfnamefont {J.~E.}\ \bibnamefont {{Tolan}}}, \bibinfo {author}
  {\bibfnamefont {C.~E.}\ \bibnamefont {{Tucker}}}, \bibinfo {author}
  {\bibfnamefont {A.~D.}\ \bibnamefont {{Turner}}}, \bibinfo {author}
  {\bibfnamefont {C.}~\bibnamefont {{Umilt{\`a}}}}, \bibinfo {author}
  {\bibfnamefont {A.~G.}\ \bibnamefont {{Vieregg}}}, \bibinfo {author}
  {\bibfnamefont {A.}~\bibnamefont {{Wandui}}}, \bibinfo {author}
  {\bibfnamefont {A.~C.}\ \bibnamefont {{Weber}}}, \bibinfo {author}
  {\bibfnamefont {D.~V.}\ \bibnamefont {{Wiebe}}}, \bibinfo {author}
  {\bibfnamefont {J.}~\bibnamefont {{Willmert}}}, \bibinfo {author}
  {\bibfnamefont {C.~L.}\ \bibnamefont {{Wong}}}, \bibinfo {author}
  {\bibfnamefont {W.~L.~K.}\ \bibnamefont {{Wu}}}, \bibinfo {author}
  {\bibfnamefont {E.}~\bibnamefont {{Yang}}}, \bibinfo {author} {\bibfnamefont
  {K.~W.}\ \bibnamefont {{Yoon}}}, \ and\ \bibinfo {author} {\bibfnamefont
  {C.}~\bibnamefont {{Zhang}}},\ }\bibfield  {title} {\enquote {\bibinfo
  {title} {{BICEP Array cryostat and mount design}},}\ }in\ \href {\doibase
  10.1117/12.2312829} {\emph {\bibinfo {booktitle} {Millimeter, Submillimeter,
  and Far-Infrared Detectors and Instrumentation for Astronomy IX}}},\ \bibinfo
  {series} {Society of Photo-Optical Instrumentation Engineers (SPIE)
  Conference Series}, Vol.\ \bibinfo {volume} {10708},\ \bibinfo {editor}
  {edited by\ \bibinfo {editor} {\bibfnamefont {J.}~\bibnamefont
  {{Zmuidzinas}}}\ and\ \bibinfo {editor} {\bibfnamefont {J.-R.}\ \bibnamefont
  {{Gao}}}}\ (\bibinfo {year} {2018})\ p.\ \bibinfo {pages} {107082D},\ \Eprint
  {http://arxiv.org/abs/1808.00569} {arXiv:1808.00569 [astro-ph.IM]}
  \BibitemShut {NoStop}%
\bibitem [{\citenamefont {{Lee}}\ \emph {et~al.}(2020)\citenamefont {{Lee}},
  \citenamefont {{Choi}}, \citenamefont {{G{\'e}nova-Santos}}, \citenamefont
  {{Hattori}}, \citenamefont {{Hazumi}}, \citenamefont {{Honda}}, \citenamefont
  {{Ikemitsu}}, \citenamefont {{Ishida}}, \citenamefont {{Ishitsuka}},
  \citenamefont {{Jo}}, \citenamefont {{Karatsu}}, \citenamefont {{Kiuchi}},
  \citenamefont {{Komine}}, \citenamefont {{Koyano}}, \citenamefont
  {{Kutsuma}}, \citenamefont {{Mima}}, \citenamefont {{Minowa}}, \citenamefont
  {{Moon}}, \citenamefont {{Nagai}}, \citenamefont {{Nagasaki}}, \citenamefont
  {{Naruse}}, \citenamefont {{Oguri}}, \citenamefont {{Otani}}, \citenamefont
  {{Peel}}, \citenamefont {{Rebolo}}, \citenamefont {{Rubi{\~n}o-Mart{\'\i}n}},
  \citenamefont {{Sekimoto}}, \citenamefont {{Suzuki}}, \citenamefont
  {{Taino}}, \citenamefont {{Tajima}}, \citenamefont {{Tomita}}, \citenamefont
  {{Uchida}}, \citenamefont {{Won}},\ and\ \citenamefont
  {{Yoshida}}}]{Lee2020}%
  \BibitemOpen
  \bibfield  {author} {\bibinfo {author} {\bibfnamefont {K.}~\bibnamefont
  {{Lee}}}, \bibinfo {author} {\bibfnamefont {J.}~\bibnamefont {{Choi}}},
  \bibinfo {author} {\bibfnamefont {R.~T.}\ \bibnamefont
  {{G{\'e}nova-Santos}}}, \bibinfo {author} {\bibfnamefont {M.}~\bibnamefont
  {{Hattori}}}, \bibinfo {author} {\bibfnamefont {M.}~\bibnamefont {{Hazumi}}},
  \bibinfo {author} {\bibfnamefont {S.}~\bibnamefont {{Honda}}}, \bibinfo
  {author} {\bibfnamefont {T.}~\bibnamefont {{Ikemitsu}}}, \bibinfo {author}
  {\bibfnamefont {H.}~\bibnamefont {{Ishida}}}, \bibinfo {author}
  {\bibfnamefont {H.}~\bibnamefont {{Ishitsuka}}}, \bibinfo {author}
  {\bibfnamefont {Y.}~\bibnamefont {{Jo}}}, \bibinfo {author} {\bibfnamefont
  {K.}~\bibnamefont {{Karatsu}}}, \bibinfo {author} {\bibfnamefont
  {K.}~\bibnamefont {{Kiuchi}}}, \bibinfo {author} {\bibfnamefont
  {J.}~\bibnamefont {{Komine}}}, \bibinfo {author} {\bibfnamefont
  {R.}~\bibnamefont {{Koyano}}}, \bibinfo {author} {\bibfnamefont
  {H.}~\bibnamefont {{Kutsuma}}}, \bibinfo {author} {\bibfnamefont
  {S.}~\bibnamefont {{Mima}}}, \bibinfo {author} {\bibfnamefont
  {M.}~\bibnamefont {{Minowa}}}, \bibinfo {author} {\bibfnamefont
  {J.}~\bibnamefont {{Moon}}}, \bibinfo {author} {\bibfnamefont
  {M.}~\bibnamefont {{Nagai}}}, \bibinfo {author} {\bibfnamefont
  {T.}~\bibnamefont {{Nagasaki}}}, \bibinfo {author} {\bibfnamefont
  {M.}~\bibnamefont {{Naruse}}}, \bibinfo {author} {\bibfnamefont
  {S.}~\bibnamefont {{Oguri}}}, \bibinfo {author} {\bibfnamefont
  {C.}~\bibnamefont {{Otani}}}, \bibinfo {author} {\bibfnamefont
  {M.}~\bibnamefont {{Peel}}}, \bibinfo {author} {\bibfnamefont
  {R.}~\bibnamefont {{Rebolo}}}, \bibinfo {author} {\bibfnamefont {J.~A.}\
  \bibnamefont {{Rubi{\~n}o-Mart{\'\i}n}}}, \bibinfo {author} {\bibfnamefont
  {Y.}~\bibnamefont {{Sekimoto}}}, \bibinfo {author} {\bibfnamefont
  {J.}~\bibnamefont {{Suzuki}}}, \bibinfo {author} {\bibfnamefont
  {T.}~\bibnamefont {{Taino}}}, \bibinfo {author} {\bibfnamefont
  {O.}~\bibnamefont {{Tajima}}}, \bibinfo {author} {\bibfnamefont
  {N.}~\bibnamefont {{Tomita}}}, \bibinfo {author} {\bibfnamefont
  {T.}~\bibnamefont {{Uchida}}}, \bibinfo {author} {\bibfnamefont
  {E.}~\bibnamefont {{Won}}}, \ and\ \bibinfo {author} {\bibfnamefont
  {M.}~\bibnamefont {{Yoshida}}},\ }\bibfield  {title} {\enquote {\bibinfo
  {title} {{GroundBIRD: A CMB Polarization Experiment with MKID Arrays}},}\
  }\href {\doibase 10.1007/s10909-020-02511-5} {\bibfield  {journal} {\bibinfo
  {journal} {Journal of Low Temperature Physics}\ }\textbf {\bibinfo {volume}
  {200}},\ \bibinfo {pages} {384--391} (\bibinfo {year} {2020})},\ \Eprint
  {http://arxiv.org/abs/2011.07705} {arXiv:2011.07705 [astro-ph.IM]}
  \BibitemShut {NoStop}%
\bibitem [{\citenamefont {{McCarrick}}\ \emph {et~al.}(2018)\citenamefont
  {{McCarrick}}, \citenamefont {{Jones}}, \citenamefont {{Johnson}},
  \citenamefont {{Abitbol}}, \citenamefont {{Ade}}, \citenamefont {{Bryan}},
  \citenamefont {{Day}}, \citenamefont {{Essinger-Hileman}}, \citenamefont
  {{Flanigan}}, \citenamefont {{Leduc}}, \citenamefont {{Limon}}, \citenamefont
  {{Mauskopf}}, \citenamefont {{Miller}},\ and\ \citenamefont
  {{Tucker}}}]{McCarrick2018}%
  \BibitemOpen
  \bibfield  {author} {\bibinfo {author} {\bibfnamefont {H.}~\bibnamefont
  {{McCarrick}}}, \bibinfo {author} {\bibfnamefont {G.}~\bibnamefont
  {{Jones}}}, \bibinfo {author} {\bibfnamefont {B.~R.}\ \bibnamefont
  {{Johnson}}}, \bibinfo {author} {\bibfnamefont {M.~H.}\ \bibnamefont
  {{Abitbol}}}, \bibinfo {author} {\bibfnamefont {P.~A.~R.}\ \bibnamefont
  {{Ade}}}, \bibinfo {author} {\bibfnamefont {S.}~\bibnamefont {{Bryan}}},
  \bibinfo {author} {\bibfnamefont {P.}~\bibnamefont {{Day}}}, \bibinfo
  {author} {\bibfnamefont {T.}~\bibnamefont {{Essinger-Hileman}}}, \bibinfo
  {author} {\bibfnamefont {D.}~\bibnamefont {{Flanigan}}}, \bibinfo {author}
  {\bibfnamefont {H.~G.}\ \bibnamefont {{Leduc}}}, \bibinfo {author}
  {\bibfnamefont {M.}~\bibnamefont {{Limon}}}, \bibinfo {author} {\bibfnamefont
  {P.}~\bibnamefont {{Mauskopf}}}, \bibinfo {author} {\bibfnamefont
  {A.}~\bibnamefont {{Miller}}}, \ and\ \bibinfo {author} {\bibfnamefont
  {C.}~\bibnamefont {{Tucker}}},\ }\bibfield  {title} {\enquote {\bibinfo
  {title} {{Design and performance of dual-polarization lumped-element kinetic
  inductance detectors for millimeter-wave polarimetry}},}\ }\href {\doibase
  10.1051/0004-6361/201732044} {\bibfield  {journal} {\bibinfo  {journal}
  {Astronomy \& Astrophysics}\ }\textbf {\bibinfo {volume} {610}},\ \bibinfo
  {eid} {A45} (\bibinfo {year} {2018})},\ \Eprint
  {http://arxiv.org/abs/1710.02239} {arXiv:1710.02239 [astro-ph.IM]}
  \BibitemShut {NoStop}%
\bibitem [{\citenamefont {{Dober}}\ \emph {et~al.}(2016)\citenamefont
  {{Dober}}, \citenamefont {{Austermann}}, \citenamefont {{Beall}},
  \citenamefont {{Becker}}, \citenamefont {{Che}}, \citenamefont {{Cho}},
  \citenamefont {{Devlin}}, \citenamefont {{Duff}}, \citenamefont {{Galitzki}},
  \citenamefont {{Gao}}, \citenamefont {{Groppi}}, \citenamefont {{Hilton}},
  \citenamefont {{Hubmayr}}, \citenamefont {{Irwin}}, \citenamefont
  {{McKenney}}, \citenamefont {{Li}}, \citenamefont {{Lourie}}, \citenamefont
  {{Mauskopf}}, \citenamefont {{Vissers}},\ and\ \citenamefont
  {{Wang}}}]{Dober2016}%
  \BibitemOpen
  \bibfield  {author} {\bibinfo {author} {\bibfnamefont {B.}~\bibnamefont
  {{Dober}}}, \bibinfo {author} {\bibfnamefont {J.~A.}\ \bibnamefont
  {{Austermann}}}, \bibinfo {author} {\bibfnamefont {J.~A.}\ \bibnamefont
  {{Beall}}}, \bibinfo {author} {\bibfnamefont {D.}~\bibnamefont {{Becker}}},
  \bibinfo {author} {\bibfnamefont {G.}~\bibnamefont {{Che}}}, \bibinfo
  {author} {\bibfnamefont {H.~M.}\ \bibnamefont {{Cho}}}, \bibinfo {author}
  {\bibfnamefont {M.}~\bibnamefont {{Devlin}}}, \bibinfo {author}
  {\bibfnamefont {S.~M.}\ \bibnamefont {{Duff}}}, \bibinfo {author}
  {\bibfnamefont {N.}~\bibnamefont {{Galitzki}}}, \bibinfo {author}
  {\bibfnamefont {J.}~\bibnamefont {{Gao}}}, \bibinfo {author} {\bibfnamefont
  {C.}~\bibnamefont {{Groppi}}}, \bibinfo {author} {\bibfnamefont {G.~C.}\
  \bibnamefont {{Hilton}}}, \bibinfo {author} {\bibfnamefont {J.}~\bibnamefont
  {{Hubmayr}}}, \bibinfo {author} {\bibfnamefont {K.~D.}\ \bibnamefont
  {{Irwin}}}, \bibinfo {author} {\bibfnamefont {C.~M.}\ \bibnamefont
  {{McKenney}}}, \bibinfo {author} {\bibfnamefont {D.}~\bibnamefont {{Li}}},
  \bibinfo {author} {\bibfnamefont {N.}~\bibnamefont {{Lourie}}}, \bibinfo
  {author} {\bibfnamefont {P.}~\bibnamefont {{Mauskopf}}}, \bibinfo {author}
  {\bibfnamefont {M.~R.}\ \bibnamefont {{Vissers}}}, \ and\ \bibinfo {author}
  {\bibfnamefont {Y.}~\bibnamefont {{Wang}}},\ }\bibfield  {title} {\enquote
  {\bibinfo {title} {{Optical Demonstration of THz, Dual-Polarization Sensitive
  Microwave Kinetic Inductance Detectors}},}\ }\href {\doibase
  10.1007/s10909-015-1434-3} {\bibfield  {journal} {\bibinfo  {journal}
  {Journal of Low Temperature Physics}\ }\textbf {\bibinfo {volume} {184}},\
  \bibinfo {pages} {173--179} (\bibinfo {year} {2016})},\ \Eprint
  {http://arxiv.org/abs/1603.02963} {arXiv:1603.02963 [astro-ph.IM]}
  \BibitemShut {NoStop}%
\bibitem [{\citenamefont {{Groppi}}\ \emph {et~al.}(2016)\citenamefont
  {{Groppi}}, \citenamefont {{Mauskopf}}, \citenamefont {{Ade}},\ and\
  \citenamefont {{Underhill}}}]{Groppi2016}%
  \BibitemOpen
  \bibfield  {author} {\bibinfo {author} {\bibfnamefont {C.~E.}\ \bibnamefont
  {{Groppi}}}, \bibinfo {author} {\bibfnamefont {P.~M.}\ \bibnamefont
  {{Mauskopf}}}, \bibinfo {author} {\bibfnamefont {P.~A.~R.}\ \bibnamefont
  {{Ade}}}, \ and\ \bibinfo {author} {\bibfnamefont {M.}~\bibnamefont
  {{Underhill}}},\ }\bibfield  {title} {\enquote {\bibinfo {title} {{Design and
  measurement of a direct-drillable smooth walled feedhorn at 1.2 THz for the
  next generation BLASTPol experiment}},}\ }in\ \href {\doibase
  10.1117/12.2231346} {\emph {\bibinfo {booktitle} {Millimeter, Submillimeter,
  and Far-Infrared Detectors and Instrumentation for Astronomy VIII}}},\
  \bibinfo {series} {Society of Photo-Optical Instrumentation Engineers (SPIE)
  Conference Series}, Vol.\ \bibinfo {volume} {9914},\ \bibinfo {editor}
  {edited by\ \bibinfo {editor} {\bibfnamefont {W.~S.}\ \bibnamefont
  {{Holland}}}\ and\ \bibinfo {editor} {\bibfnamefont {J.}~\bibnamefont
  {{Zmuidzinas}}}}\ (\bibinfo {year} {2016})\ p.\ \bibinfo {pages}
  {99142P}\BibitemShut {NoStop}%
\bibitem [{\citenamefont {{Meier Vald{\'e}s}}, \citenamefont {{Morris}},\ and\
  \citenamefont {{Demory}}(2021)}]{Meier2021}%
  \BibitemOpen
  \bibfield  {author} {\bibinfo {author} {\bibfnamefont {E.~A.}\ \bibnamefont
  {{Meier Vald{\'e}s}}}, \bibinfo {author} {\bibfnamefont {B.~M.}\ \bibnamefont
  {{Morris}}}, \ and\ \bibinfo {author} {\bibfnamefont {B.~O.}\ \bibnamefont
  {{Demory}}},\ }\bibfield  {title} {\enquote {\bibinfo {title} {{Monitoring
  precipitable water vapour in near real-time to correct near-infrared
  observations using satellite remote sensing}},}\ }\href {\doibase
  10.1051/0004-6361/202039629} {\bibfield  {journal} {\bibinfo  {journal}
  {\aap}\ }\textbf {\bibinfo {volume} {649}},\ \bibinfo {eid} {A132} (\bibinfo
  {year} {2021})},\ \Eprint {http://arxiv.org/abs/2103.05326} {arXiv:2103.05326
  [astro-ph.IM]} \BibitemShut {NoStop}%
\bibitem [{\citenamefont {{Golwala}}\ \emph {et~al.}(2012)\citenamefont
  {{Golwala}}, \citenamefont {{Bockstiegel}}, \citenamefont {{Brugger}},
  \citenamefont {{Czakon}}, \citenamefont {{Day}}, \citenamefont {{Downes}},
  \citenamefont {{Duan}}, \citenamefont {{Gao}}, \citenamefont {{Gill}},
  \citenamefont {{Glenn}}, \citenamefont {{Hollister}}, \citenamefont
  {{LeDuc}}, \citenamefont {{Maloney}}, \citenamefont {{Mazin}}, \citenamefont
  {{McHugh}}, \citenamefont {{Miller}}, \citenamefont {{Noroozian}},
  \citenamefont {{Nguyen}}, \citenamefont {{Sayers}}, \citenamefont
  {{Schlaerth}}, \citenamefont {{Siegel}}, \citenamefont {{Vayonakis}},
  \citenamefont {{Wilson}},\ and\ \citenamefont {{Zmuidzinas}}}]{Golwala2012}%
  \BibitemOpen
  \bibfield  {author} {\bibinfo {author} {\bibfnamefont {S.~R.}\ \bibnamefont
  {{Golwala}}}, \bibinfo {author} {\bibfnamefont {C.}~\bibnamefont
  {{Bockstiegel}}}, \bibinfo {author} {\bibfnamefont {S.}~\bibnamefont
  {{Brugger}}}, \bibinfo {author} {\bibfnamefont {N.~G.}\ \bibnamefont
  {{Czakon}}}, \bibinfo {author} {\bibfnamefont {P.~K.}\ \bibnamefont {{Day}}},
  \bibinfo {author} {\bibfnamefont {T.~P.}\ \bibnamefont {{Downes}}}, \bibinfo
  {author} {\bibfnamefont {R.}~\bibnamefont {{Duan}}}, \bibinfo {author}
  {\bibfnamefont {J.}~\bibnamefont {{Gao}}}, \bibinfo {author} {\bibfnamefont
  {A.~K.}\ \bibnamefont {{Gill}}}, \bibinfo {author} {\bibfnamefont
  {J.}~\bibnamefont {{Glenn}}}, \bibinfo {author} {\bibfnamefont {M.~I.}\
  \bibnamefont {{Hollister}}}, \bibinfo {author} {\bibfnamefont {H.~G.}\
  \bibnamefont {{LeDuc}}}, \bibinfo {author} {\bibfnamefont {P.~R.}\
  \bibnamefont {{Maloney}}}, \bibinfo {author} {\bibfnamefont {B.~A.}\
  \bibnamefont {{Mazin}}}, \bibinfo {author} {\bibfnamefont {S.~G.}\
  \bibnamefont {{McHugh}}}, \bibinfo {author} {\bibfnamefont {D.}~\bibnamefont
  {{Miller}}}, \bibinfo {author} {\bibfnamefont {O.}~\bibnamefont
  {{Noroozian}}}, \bibinfo {author} {\bibfnamefont {H.~T.}\ \bibnamefont
  {{Nguyen}}}, \bibinfo {author} {\bibfnamefont {J.}~\bibnamefont {{Sayers}}},
  \bibinfo {author} {\bibfnamefont {J.~A.}\ \bibnamefont {{Schlaerth}}},
  \bibinfo {author} {\bibfnamefont {S.}~\bibnamefont {{Siegel}}}, \bibinfo
  {author} {\bibfnamefont {A.~K.}\ \bibnamefont {{Vayonakis}}}, \bibinfo
  {author} {\bibfnamefont {P.~R.}\ \bibnamefont {{Wilson}}}, \ and\ \bibinfo
  {author} {\bibfnamefont {J.}~\bibnamefont {{Zmuidzinas}}},\ }\bibfield
  {title} {\enquote {\bibinfo {title} {{Status of MUSIC, the MUltiwavelength
  Sub/millimeter Inductance Camera}},}\ }in\ \href {\doibase 10.1117/12.926055}
  {\emph {\bibinfo {booktitle} {Millimeter, Submillimeter, and Far-Infrared
  Detectors and Instrumentation for Astronomy VI}}},\ \bibinfo {series}
  {Society of Photo-Optical Instrumentation Engineers (SPIE) Conference
  Series}, Vol.\ \bibinfo {volume} {8452},\ \bibinfo {editor} {edited by\
  \bibinfo {editor} {\bibfnamefont {W.~S.}\ \bibnamefont {{Holland}}}\ and\
  \bibinfo {editor} {\bibfnamefont {J.}~\bibnamefont {{Zmuidzinas}}}}\
  (\bibinfo {year} {2012})\ p.\ \bibinfo {pages} {845205},\ \Eprint
  {http://arxiv.org/abs/1211.0595} {arXiv:1211.0595 [astro-ph.IM]} \BibitemShut
  {NoStop}%
\bibitem [{\citenamefont {{Choi}}\ \emph {et~al.}(2018)\citenamefont {{Choi}},
  \citenamefont {{Austermann}}, \citenamefont {{Beall}}, \citenamefont
  {{Crowley}}, \citenamefont {{Datta}}, \citenamefont {{Duff}}, \citenamefont
  {{Gallardo}}, \citenamefont {{Ho}}, \citenamefont {{Hubmayr}}, \citenamefont
  {{Koopman}}, \citenamefont {{Li}}, \citenamefont {{Nati}}, \citenamefont
  {{Niemack}}, \citenamefont {{Page}}, \citenamefont {{Salatino}},
  \citenamefont {{Simon}}, \citenamefont {{Staggs}}, \citenamefont {{Stevens}},
  \citenamefont {{Ullom}},\ and\ \citenamefont {{Wollack}}}]{Choi2018}%
  \BibitemOpen
  \bibfield  {author} {\bibinfo {author} {\bibfnamefont {S.~K.}\ \bibnamefont
  {{Choi}}}, \bibinfo {author} {\bibfnamefont {J.}~\bibnamefont
  {{Austermann}}}, \bibinfo {author} {\bibfnamefont {J.~A.}\ \bibnamefont
  {{Beall}}}, \bibinfo {author} {\bibfnamefont {K.~T.}\ \bibnamefont
  {{Crowley}}}, \bibinfo {author} {\bibfnamefont {R.}~\bibnamefont {{Datta}}},
  \bibinfo {author} {\bibfnamefont {S.~M.}\ \bibnamefont {{Duff}}}, \bibinfo
  {author} {\bibfnamefont {P.~A.}\ \bibnamefont {{Gallardo}}}, \bibinfo
  {author} {\bibfnamefont {S.~P.}\ \bibnamefont {{Ho}}}, \bibinfo {author}
  {\bibfnamefont {J.}~\bibnamefont {{Hubmayr}}}, \bibinfo {author}
  {\bibfnamefont {B.~J.}\ \bibnamefont {{Koopman}}}, \bibinfo {author}
  {\bibfnamefont {Y.}~\bibnamefont {{Li}}}, \bibinfo {author} {\bibfnamefont
  {F.}~\bibnamefont {{Nati}}}, \bibinfo {author} {\bibfnamefont {M.~D.}\
  \bibnamefont {{Niemack}}}, \bibinfo {author} {\bibfnamefont {L.~A.}\
  \bibnamefont {{Page}}}, \bibinfo {author} {\bibfnamefont {M.}~\bibnamefont
  {{Salatino}}}, \bibinfo {author} {\bibfnamefont {S.~M.}\ \bibnamefont
  {{Simon}}}, \bibinfo {author} {\bibfnamefont {S.~T.}\ \bibnamefont
  {{Staggs}}}, \bibinfo {author} {\bibfnamefont {J.}~\bibnamefont {{Stevens}}},
  \bibinfo {author} {\bibfnamefont {J.}~\bibnamefont {{Ullom}}}, \ and\
  \bibinfo {author} {\bibfnamefont {E.~J.}\ \bibnamefont {{Wollack}}},\
  }\bibfield  {title} {\enquote {\bibinfo {title} {{Characterization of the
  Mid-Frequency Arrays for Advanced ACTPol}},}\ }\href {\doibase
  10.1007/s10909-018-1982-4} {\bibfield  {journal} {\bibinfo  {journal}
  {Journal of Low Temperature Physics}\ }\textbf {\bibinfo {volume} {193}},\
  \bibinfo {pages} {267--275} (\bibinfo {year} {2018})},\ \Eprint
  {http://arxiv.org/abs/1711.04841} {arXiv:1711.04841 [astro-ph.IM]}
  \BibitemShut {NoStop}%
\bibitem [{\citenamefont {{Sobrin}}\ \emph {et~al.}(2022)\citenamefont
  {{Sobrin}}, \citenamefont {{Anderson}}, \citenamefont {{Bender}},
  \citenamefont {{Benson}}, \citenamefont {{Dutcher}}, \citenamefont
  {{Foster}}, \citenamefont {{Goeckner-Wald}}, \citenamefont {{Montgomery}},
  \citenamefont {{Nadolski}}, \citenamefont {{Rahlin}}, \citenamefont {{Ade}},
  \citenamefont {{Ahmed}}, \citenamefont {{Anderes}}, \citenamefont
  {{Archipley}}, \citenamefont {{Austermann}}, \citenamefont {{Avva}},
  \citenamefont {{Aylor}}, \citenamefont {{Balkenhol}}, \citenamefont
  {{Barry}}, \citenamefont {{Thakur}}, \citenamefont {{Benabed}}, \citenamefont
  {{Bianchini}}, \citenamefont {{Bleem}}, \citenamefont {{Bouchet}},
  \citenamefont {{Bryant}}, \citenamefont {{Byrum}}, \citenamefont
  {{Carlstrom}}, \citenamefont {{Carter}}, \citenamefont {{Cecil}},
  \citenamefont {{Chang}}, \citenamefont {{Chaubal}}, \citenamefont {{Chen}},
  \citenamefont {{Cho}}, \citenamefont {{Chou}}, \citenamefont {{Cliche}},
  \citenamefont {{Crawford}}, \citenamefont {{Cukierman}}, \citenamefont
  {{Daley}}, \citenamefont {{de Haan}}, \citenamefont {{Denison}},
  \citenamefont {{Dibert}}, \citenamefont {{Ding}}, \citenamefont {{Dobbs}},
  \citenamefont {{Everett}}, \citenamefont {{Feng}}, \citenamefont
  {{Ferguson}}, \citenamefont {{Fu}}, \citenamefont {{Galli}}, \citenamefont
  {{Gambrel}}, \citenamefont {{Gardner}}, \citenamefont {{Gualtieri}},
  \citenamefont {{Guns}}, \citenamefont {{Gupta}}, \citenamefont {{Guyser}},
  \citenamefont {{Halverson}}, \citenamefont {{Harke-Hosemann}}, \citenamefont
  {{Harrington}}, \citenamefont {{Henning}}, \citenamefont {{Hilton}},
  \citenamefont {{Hivon}}, \citenamefont {{Holder}}, \citenamefont
  {{Holzapfel}}, \citenamefont {{Hood}}, \citenamefont {{Howe}}, \citenamefont
  {{Huang}}, \citenamefont {{Irwin}}, \citenamefont {{Jeong}}, \citenamefont
  {{Jonas}}, \citenamefont {{Jones}}, \citenamefont {{Khaire}}, \citenamefont
  {{Knox}}, \citenamefont {{Kofman}}, \citenamefont {{Korman}}, \citenamefont
  {{Kubik}}, \citenamefont {{Kuhlmann}}, \citenamefont {{Kuo}}, \citenamefont
  {{Lee}}, \citenamefont {{Leitch}}, \citenamefont {{Lowitz}}, \citenamefont
  {{Lu}}, \citenamefont {{Meyer}}, \citenamefont {{Michalik}}, \citenamefont
  {{Millea}}, \citenamefont {{Natoli}}, \citenamefont {{Nguyen}}, \citenamefont
  {{Noble}}, \citenamefont {{Novosad}}, \citenamefont {{Omori}}, \citenamefont
  {{Padin}}, \citenamefont {{Pan}}, \citenamefont {{Paschos}}, \citenamefont
  {{Pearson}}, \citenamefont {{Posada}}, \citenamefont {{Prabhu}},
  \citenamefont {{Quan}}, \citenamefont {{Reichardt}}, \citenamefont
  {{Riebel}}, \citenamefont {{Riedel}}, \citenamefont {{Rouble}}, \citenamefont
  {{Ruhl}}, \citenamefont {{Saliwanchik}}, \citenamefont {{Sayre}},
  \citenamefont {{Schiappucci}}, \citenamefont {{Shirokoff}}, \citenamefont
  {{Smecher}}, \citenamefont {{Stark}}, \citenamefont {{Stephen}},
  \citenamefont {{Story}}, \citenamefont {{Suzuki}}, \citenamefont {{Tandoi}},
  \citenamefont {{Thompson}}, \citenamefont {{Thorne}}, \citenamefont
  {{Tucker}}, \citenamefont {{Umilta}}, \citenamefont {{Vale}}, \citenamefont
  {{Vanderlinde}}, \citenamefont {{Vieira}}, \citenamefont {{Wang}},
  \citenamefont {{Whitehorn}}, \citenamefont {{Wu}}, \citenamefont
  {{Yefremenko}}, \citenamefont {{Yoon}},\ and\ \citenamefont
  {{Young}}}]{Sobrin2022}%
  \BibitemOpen
  \bibfield  {author} {\bibinfo {author} {\bibfnamefont {J.~A.}\ \bibnamefont
  {{Sobrin}}}, \bibinfo {author} {\bibfnamefont {A.~J.}\ \bibnamefont
  {{Anderson}}}, \bibinfo {author} {\bibfnamefont {A.~N.}\ \bibnamefont
  {{Bender}}}, \bibinfo {author} {\bibfnamefont {B.~A.}\ \bibnamefont
  {{Benson}}}, \bibinfo {author} {\bibfnamefont {D.}~\bibnamefont {{Dutcher}}},
  \bibinfo {author} {\bibfnamefont {A.}~\bibnamefont {{Foster}}}, \bibinfo
  {author} {\bibfnamefont {N.}~\bibnamefont {{Goeckner-Wald}}}, \bibinfo
  {author} {\bibfnamefont {J.}~\bibnamefont {{Montgomery}}}, \bibinfo {author}
  {\bibfnamefont {A.}~\bibnamefont {{Nadolski}}}, \bibinfo {author}
  {\bibfnamefont {A.}~\bibnamefont {{Rahlin}}}, \bibinfo {author}
  {\bibfnamefont {P.~A.~R.}\ \bibnamefont {{Ade}}}, \bibinfo {author}
  {\bibfnamefont {Z.}~\bibnamefont {{Ahmed}}}, \bibinfo {author} {\bibfnamefont
  {E.}~\bibnamefont {{Anderes}}}, \bibinfo {author} {\bibfnamefont
  {M.}~\bibnamefont {{Archipley}}}, \bibinfo {author} {\bibfnamefont {J.~E.}\
  \bibnamefont {{Austermann}}}, \bibinfo {author} {\bibfnamefont {J.~S.}\
  \bibnamefont {{Avva}}}, \bibinfo {author} {\bibfnamefont {K.}~\bibnamefont
  {{Aylor}}}, \bibinfo {author} {\bibfnamefont {L.}~\bibnamefont
  {{Balkenhol}}}, \bibinfo {author} {\bibfnamefont {P.~S.}\ \bibnamefont
  {{Barry}}}, \bibinfo {author} {\bibfnamefont {R.~B.}\ \bibnamefont
  {{Thakur}}}, \bibinfo {author} {\bibfnamefont {K.}~\bibnamefont {{Benabed}}},
  \bibinfo {author} {\bibfnamefont {F.}~\bibnamefont {{Bianchini}}}, \bibinfo
  {author} {\bibfnamefont {L.~E.}\ \bibnamefont {{Bleem}}}, \bibinfo {author}
  {\bibfnamefont {F.~R.}\ \bibnamefont {{Bouchet}}}, \bibinfo {author}
  {\bibfnamefont {L.}~\bibnamefont {{Bryant}}}, \bibinfo {author}
  {\bibfnamefont {K.}~\bibnamefont {{Byrum}}}, \bibinfo {author} {\bibfnamefont
  {J.~E.}\ \bibnamefont {{Carlstrom}}}, \bibinfo {author} {\bibfnamefont
  {F.~W.}\ \bibnamefont {{Carter}}}, \bibinfo {author} {\bibfnamefont {T.~W.}\
  \bibnamefont {{Cecil}}}, \bibinfo {author} {\bibfnamefont {C.~L.}\
  \bibnamefont {{Chang}}}, \bibinfo {author} {\bibfnamefont {P.}~\bibnamefont
  {{Chaubal}}}, \bibinfo {author} {\bibfnamefont {G.}~\bibnamefont {{Chen}}},
  \bibinfo {author} {\bibfnamefont {H.~M.}\ \bibnamefont {{Cho}}}, \bibinfo
  {author} {\bibfnamefont {T.~L.}\ \bibnamefont {{Chou}}}, \bibinfo {author}
  {\bibfnamefont {J.~F.}\ \bibnamefont {{Cliche}}}, \bibinfo {author}
  {\bibfnamefont {T.~M.}\ \bibnamefont {{Crawford}}}, \bibinfo {author}
  {\bibfnamefont {A.}~\bibnamefont {{Cukierman}}}, \bibinfo {author}
  {\bibfnamefont {C.}~\bibnamefont {{Daley}}}, \bibinfo {author} {\bibfnamefont
  {T.}~\bibnamefont {{de Haan}}}, \bibinfo {author} {\bibfnamefont {E.~V.}\
  \bibnamefont {{Denison}}}, \bibinfo {author} {\bibfnamefont {K.}~\bibnamefont
  {{Dibert}}}, \bibinfo {author} {\bibfnamefont {J.}~\bibnamefont {{Ding}}},
  \bibinfo {author} {\bibfnamefont {M.~A.}\ \bibnamefont {{Dobbs}}}, \bibinfo
  {author} {\bibfnamefont {W.}~\bibnamefont {{Everett}}}, \bibinfo {author}
  {\bibfnamefont {C.}~\bibnamefont {{Feng}}}, \bibinfo {author} {\bibfnamefont
  {K.~R.}\ \bibnamefont {{Ferguson}}}, \bibinfo {author} {\bibfnamefont
  {J.}~\bibnamefont {{Fu}}}, \bibinfo {author} {\bibfnamefont {S.}~\bibnamefont
  {{Galli}}}, \bibinfo {author} {\bibfnamefont {A.~E.}\ \bibnamefont
  {{Gambrel}}}, \bibinfo {author} {\bibfnamefont {R.~W.}\ \bibnamefont
  {{Gardner}}}, \bibinfo {author} {\bibfnamefont {R.}~\bibnamefont
  {{Gualtieri}}}, \bibinfo {author} {\bibfnamefont {S.}~\bibnamefont {{Guns}}},
  \bibinfo {author} {\bibfnamefont {N.}~\bibnamefont {{Gupta}}}, \bibinfo
  {author} {\bibfnamefont {R.}~\bibnamefont {{Guyser}}}, \bibinfo {author}
  {\bibfnamefont {N.~W.}\ \bibnamefont {{Halverson}}}, \bibinfo {author}
  {\bibfnamefont {A.~H.}\ \bibnamefont {{Harke-Hosemann}}}, \bibinfo {author}
  {\bibfnamefont {N.~L.}\ \bibnamefont {{Harrington}}}, \bibinfo {author}
  {\bibfnamefont {J.~W.}\ \bibnamefont {{Henning}}}, \bibinfo {author}
  {\bibfnamefont {G.~C.}\ \bibnamefont {{Hilton}}}, \bibinfo {author}
  {\bibfnamefont {E.}~\bibnamefont {{Hivon}}}, \bibinfo {author} {\bibfnamefont
  {G.~P.}\ \bibnamefont {{Holder}}}, \bibinfo {author} {\bibfnamefont {W.~L.}\
  \bibnamefont {{Holzapfel}}}, \bibinfo {author} {\bibfnamefont {J.~C.}\
  \bibnamefont {{Hood}}}, \bibinfo {author} {\bibfnamefont {D.}~\bibnamefont
  {{Howe}}}, \bibinfo {author} {\bibfnamefont {N.}~\bibnamefont {{Huang}}},
  \bibinfo {author} {\bibfnamefont {K.~D.}\ \bibnamefont {{Irwin}}}, \bibinfo
  {author} {\bibfnamefont {O.~B.}\ \bibnamefont {{Jeong}}}, \bibinfo {author}
  {\bibfnamefont {M.}~\bibnamefont {{Jonas}}}, \bibinfo {author} {\bibfnamefont
  {A.}~\bibnamefont {{Jones}}}, \bibinfo {author} {\bibfnamefont {T.~S.}\
  \bibnamefont {{Khaire}}}, \bibinfo {author} {\bibfnamefont {L.}~\bibnamefont
  {{Knox}}}, \bibinfo {author} {\bibfnamefont {A.~M.}\ \bibnamefont
  {{Kofman}}}, \bibinfo {author} {\bibfnamefont {M.}~\bibnamefont {{Korman}}},
  \bibinfo {author} {\bibfnamefont {D.~L.}\ \bibnamefont {{Kubik}}}, \bibinfo
  {author} {\bibfnamefont {S.}~\bibnamefont {{Kuhlmann}}}, \bibinfo {author}
  {\bibfnamefont {C.~L.}\ \bibnamefont {{Kuo}}}, \bibinfo {author}
  {\bibfnamefont {A.~T.}\ \bibnamefont {{Lee}}}, \bibinfo {author}
  {\bibfnamefont {E.~M.}\ \bibnamefont {{Leitch}}}, \bibinfo {author}
  {\bibfnamefont {A.~E.}\ \bibnamefont {{Lowitz}}}, \bibinfo {author}
  {\bibfnamefont {C.}~\bibnamefont {{Lu}}}, \bibinfo {author} {\bibfnamefont
  {S.~S.}\ \bibnamefont {{Meyer}}}, \bibinfo {author} {\bibfnamefont
  {D.}~\bibnamefont {{Michalik}}}, \bibinfo {author} {\bibfnamefont
  {M.}~\bibnamefont {{Millea}}}, \bibinfo {author} {\bibfnamefont
  {T.}~\bibnamefont {{Natoli}}}, \bibinfo {author} {\bibfnamefont
  {H.}~\bibnamefont {{Nguyen}}}, \bibinfo {author} {\bibfnamefont {G.~I.}\
  \bibnamefont {{Noble}}}, \bibinfo {author} {\bibfnamefont {V.}~\bibnamefont
  {{Novosad}}}, \bibinfo {author} {\bibfnamefont {Y.}~\bibnamefont {{Omori}}},
  \bibinfo {author} {\bibfnamefont {S.}~\bibnamefont {{Padin}}}, \bibinfo
  {author} {\bibfnamefont {Z.}~\bibnamefont {{Pan}}}, \bibinfo {author}
  {\bibfnamefont {P.}~\bibnamefont {{Paschos}}}, \bibinfo {author}
  {\bibfnamefont {J.}~\bibnamefont {{Pearson}}}, \bibinfo {author}
  {\bibfnamefont {C.~M.}\ \bibnamefont {{Posada}}}, \bibinfo {author}
  {\bibfnamefont {K.}~\bibnamefont {{Prabhu}}}, \bibinfo {author}
  {\bibfnamefont {W.}~\bibnamefont {{Quan}}}, \bibinfo {author} {\bibfnamefont
  {C.~L.}\ \bibnamefont {{Reichardt}}}, \bibinfo {author} {\bibfnamefont
  {D.}~\bibnamefont {{Riebel}}}, \bibinfo {author} {\bibfnamefont
  {B.}~\bibnamefont {{Riedel}}}, \bibinfo {author} {\bibfnamefont
  {M.}~\bibnamefont {{Rouble}}}, \bibinfo {author} {\bibfnamefont {J.~E.}\
  \bibnamefont {{Ruhl}}}, \bibinfo {author} {\bibfnamefont {B.}~\bibnamefont
  {{Saliwanchik}}}, \bibinfo {author} {\bibfnamefont {J.~T.}\ \bibnamefont
  {{Sayre}}}, \bibinfo {author} {\bibfnamefont {E.}~\bibnamefont
  {{Schiappucci}}}, \bibinfo {author} {\bibfnamefont {E.}~\bibnamefont
  {{Shirokoff}}}, \bibinfo {author} {\bibfnamefont {G.}~\bibnamefont
  {{Smecher}}}, \bibinfo {author} {\bibfnamefont {A.~A.}\ \bibnamefont
  {{Stark}}}, \bibinfo {author} {\bibfnamefont {J.}~\bibnamefont {{Stephen}}},
  \bibinfo {author} {\bibfnamefont {K.~T.}\ \bibnamefont {{Story}}}, \bibinfo
  {author} {\bibfnamefont {A.}~\bibnamefont {{Suzuki}}}, \bibinfo {author}
  {\bibfnamefont {C.}~\bibnamefont {{Tandoi}}}, \bibinfo {author}
  {\bibfnamefont {K.~L.}\ \bibnamefont {{Thompson}}}, \bibinfo {author}
  {\bibfnamefont {B.}~\bibnamefont {{Thorne}}}, \bibinfo {author}
  {\bibfnamefont {C.}~\bibnamefont {{Tucker}}}, \bibinfo {author}
  {\bibfnamefont {C.}~\bibnamefont {{Umilta}}}, \bibinfo {author}
  {\bibfnamefont {L.~R.}\ \bibnamefont {{Vale}}}, \bibinfo {author}
  {\bibfnamefont {K.}~\bibnamefont {{Vanderlinde}}}, \bibinfo {author}
  {\bibfnamefont {J.~D.}\ \bibnamefont {{Vieira}}}, \bibinfo {author}
  {\bibfnamefont {G.}~\bibnamefont {{Wang}}}, \bibinfo {author} {\bibfnamefont
  {N.}~\bibnamefont {{Whitehorn}}}, \bibinfo {author} {\bibfnamefont
  {W.~L.~K.}\ \bibnamefont {{Wu}}}, \bibinfo {author} {\bibfnamefont
  {V.}~\bibnamefont {{Yefremenko}}}, \bibinfo {author} {\bibfnamefont {K.~W.}\
  \bibnamefont {{Yoon}}}, \ and\ \bibinfo {author} {\bibfnamefont {M.~R.}\
  \bibnamefont {{Young}}},\ }\bibfield  {title} {\enquote {\bibinfo {title}
  {{The Design and Integrated Performance of SPT-3G}},}\ }\href {\doibase
  10.3847/1538-4365/ac374f} {\bibfield  {journal} {\bibinfo  {journal} {\apjs}\
  }\textbf {\bibinfo {volume} {258}},\ \bibinfo {eid} {42} (\bibinfo {year}
  {2022})},\ \Eprint {http://arxiv.org/abs/2106.11202} {arXiv:2106.11202
  [astro-ph.IM]} \BibitemShut {NoStop}%
\bibitem [{\citenamefont {{Grace}}\ \emph {et~al.}(2014)\citenamefont
  {{Grace}}, \citenamefont {{Beall}}, \citenamefont {{Bond}}, \citenamefont
  {{Cho}}, \citenamefont {{Datta}}, \citenamefont {{Devlin}}, \citenamefont
  {{D{\"u}nner}}, \citenamefont {{Fox}}, \citenamefont {{Gallardo}},
  \citenamefont {{Hasselfield}}, \citenamefont {{Henderson}}, \citenamefont
  {{Hilton}}, \citenamefont {{Hincks}}, \citenamefont {{Hlozek}}, \citenamefont
  {{Hubmayr}}, \citenamefont {{Irwin}}, \citenamefont {{Klein}}, \citenamefont
  {{Koopman}}, \citenamefont {{Li}}, \citenamefont {{Lungu}}, \citenamefont
  {{Newburgh}}, \citenamefont {{Nibarger}}, \citenamefont {{Niemack}},
  \citenamefont {{Maurin}}, \citenamefont {{McMahon}}, \citenamefont {{Naess}},
  \citenamefont {{Page}}, \citenamefont {{Pappas}}, \citenamefont {{Schmitt}},
  \citenamefont {{Sievers}}, \citenamefont {{Staggs}}, \citenamefont
  {{Thornton}}, \citenamefont {{Van Lanen}},\ and\ \citenamefont
  {{Wollack}}}]{Grace2014}%
  \BibitemOpen
  \bibfield  {author} {\bibinfo {author} {\bibfnamefont {E.}~\bibnamefont
  {{Grace}}}, \bibinfo {author} {\bibfnamefont {J.}~\bibnamefont {{Beall}}},
  \bibinfo {author} {\bibfnamefont {J.~R.}\ \bibnamefont {{Bond}}}, \bibinfo
  {author} {\bibfnamefont {H.~M.}\ \bibnamefont {{Cho}}}, \bibinfo {author}
  {\bibfnamefont {R.}~\bibnamefont {{Datta}}}, \bibinfo {author} {\bibfnamefont
  {M.~J.}\ \bibnamefont {{Devlin}}}, \bibinfo {author} {\bibfnamefont
  {R.}~\bibnamefont {{D{\"u}nner}}}, \bibinfo {author} {\bibfnamefont {A.~E.}\
  \bibnamefont {{Fox}}}, \bibinfo {author} {\bibfnamefont {P.}~\bibnamefont
  {{Gallardo}}}, \bibinfo {author} {\bibfnamefont {M.}~\bibnamefont
  {{Hasselfield}}}, \bibinfo {author} {\bibfnamefont {S.}~\bibnamefont
  {{Henderson}}}, \bibinfo {author} {\bibfnamefont {G.~C.}\ \bibnamefont
  {{Hilton}}}, \bibinfo {author} {\bibfnamefont {A.~D.}\ \bibnamefont
  {{Hincks}}}, \bibinfo {author} {\bibfnamefont {R.}~\bibnamefont {{Hlozek}}},
  \bibinfo {author} {\bibfnamefont {J.}~\bibnamefont {{Hubmayr}}}, \bibinfo
  {author} {\bibfnamefont {K.}~\bibnamefont {{Irwin}}}, \bibinfo {author}
  {\bibfnamefont {J.}~\bibnamefont {{Klein}}}, \bibinfo {author} {\bibfnamefont
  {B.}~\bibnamefont {{Koopman}}}, \bibinfo {author} {\bibfnamefont
  {D.}~\bibnamefont {{Li}}}, \bibinfo {author} {\bibfnamefont {M.}~\bibnamefont
  {{Lungu}}}, \bibinfo {author} {\bibfnamefont {L.}~\bibnamefont {{Newburgh}}},
  \bibinfo {author} {\bibfnamefont {J.~P.}\ \bibnamefont {{Nibarger}}},
  \bibinfo {author} {\bibfnamefont {M.~D.}\ \bibnamefont {{Niemack}}}, \bibinfo
  {author} {\bibfnamefont {L.}~\bibnamefont {{Maurin}}}, \bibinfo {author}
  {\bibfnamefont {J.}~\bibnamefont {{McMahon}}}, \bibinfo {author}
  {\bibfnamefont {S.}~\bibnamefont {{Naess}}}, \bibinfo {author} {\bibfnamefont
  {L.~A.}\ \bibnamefont {{Page}}}, \bibinfo {author} {\bibfnamefont
  {C.}~\bibnamefont {{Pappas}}}, \bibinfo {author} {\bibfnamefont {B.~L.}\
  \bibnamefont {{Schmitt}}}, \bibinfo {author} {\bibfnamefont {J.}~\bibnamefont
  {{Sievers}}}, \bibinfo {author} {\bibfnamefont {S.~T.}\ \bibnamefont
  {{Staggs}}}, \bibinfo {author} {\bibfnamefont {R.}~\bibnamefont
  {{Thornton}}}, \bibinfo {author} {\bibfnamefont {J.}~\bibnamefont {{Van
  Lanen}}}, \ and\ \bibinfo {author} {\bibfnamefont {E.~J.}\ \bibnamefont
  {{Wollack}}},\ }\bibfield  {title} {\enquote {\bibinfo {title} {{ACTPol:
  on-sky performance and characterization}},}\ }in\ \href {\doibase
  10.1117/12.2057243} {\emph {\bibinfo {booktitle} {Millimeter, Submillimeter,
  and Far-Infrared Detectors and Instrumentation for Astronomy VII}}},\
  \bibinfo {series} {Society of Photo-Optical Instrumentation Engineers (SPIE)
  Conference Series}, Vol.\ \bibinfo {volume} {9153},\ \bibinfo {editor}
  {edited by\ \bibinfo {editor} {\bibfnamefont {W.~S.}\ \bibnamefont
  {{Holland}}}\ and\ \bibinfo {editor} {\bibfnamefont {J.}~\bibnamefont
  {{Zmuidzinas}}}}\ (\bibinfo {year} {2014})\ p.\ \bibinfo {pages}
  {915310}\BibitemShut {NoStop}%
\bibitem [{\citenamefont {{Sobrin}}\ \emph {et~al.}(2018)\citenamefont
  {{Sobrin}}, \citenamefont {{Ade}}, \citenamefont {{Ahmed}}, \citenamefont
  {{Anderson}}, \citenamefont {{Avva}}, \citenamefont {{Basu Thakur}},
  \citenamefont {{Bender}}, \citenamefont {{Benson}}, \citenamefont
  {{Carlstrom}}, \citenamefont {{Carter}}, \citenamefont {{Cecil}},
  \citenamefont {{Chang}}, \citenamefont {{Cliche}}, \citenamefont
  {{Cukierman}}, \citenamefont {{de Haan}}, \citenamefont {{Ding}},
  \citenamefont {{Dobbs}}, \citenamefont {{Dutcher}}, \citenamefont
  {{Everett}}, \citenamefont {{Foster}}, \citenamefont {{Gallichio}},
  \citenamefont {{Gilbert}}, \citenamefont {{Groh}}, \citenamefont {{Guns}},
  \citenamefont {{Halverson}}, \citenamefont {{Harke-Hosemann}}, \citenamefont
  {{Harrington}}, \citenamefont {{Henning}}, \citenamefont {{Holzapfel}},
  \citenamefont {{Huang}}, \citenamefont {{Irwin}}, \citenamefont {{Jeong}},
  \citenamefont {{Jonas}}, \citenamefont {{Khaire}}, \citenamefont {{Kofman}},
  \citenamefont {{Korman}}, \citenamefont {{Kubik}}, \citenamefont
  {{Kuhlmann}}, \citenamefont {{Kuo}}, \citenamefont {{Lee}}, \citenamefont
  {{Lowitz}}, \citenamefont {{Meyer}}, \citenamefont {{Michalik}},
  \citenamefont {{Montgomery}}, \citenamefont {{Nadolski}}, \citenamefont
  {{Natoli}}, \citenamefont {{Nguyen}}, \citenamefont {{Noble}}, \citenamefont
  {{Novosad}}, \citenamefont {{Padin}}, \citenamefont {{Pan}}, \citenamefont
  {{Pearson}}, \citenamefont {{Posada}}, \citenamefont {{Quan}}, \citenamefont
  {{Rahlin}}, \citenamefont {{Ruhl}}, \citenamefont {{Sayre}}, \citenamefont
  {{Shirokoff}}, \citenamefont {{Smecher}}, \citenamefont {{Stark}},
  \citenamefont {{Story}}, \citenamefont {{Suzuki}}, \citenamefont
  {{Thompson}}, \citenamefont {{Tucker}}, \citenamefont {{Vanderlinde}},
  \citenamefont {{Vieira}}, \citenamefont {{Wang}}, \citenamefont
  {{Whitehorn}}, \citenamefont {{Yefremenko}}, \citenamefont {{Yoon}},\ and\
  \citenamefont {{Young}}}]{Sobrin2018}%
  \BibitemOpen
  \bibfield  {author} {\bibinfo {author} {\bibfnamefont {J.~A.}\ \bibnamefont
  {{Sobrin}}}, \bibinfo {author} {\bibfnamefont {P.~A.~R.}\ \bibnamefont
  {{Ade}}}, \bibinfo {author} {\bibfnamefont {Z.}~\bibnamefont {{Ahmed}}},
  \bibinfo {author} {\bibfnamefont {A.~J.}\ \bibnamefont {{Anderson}}},
  \bibinfo {author} {\bibfnamefont {J.~S.}\ \bibnamefont {{Avva}}}, \bibinfo
  {author} {\bibfnamefont {R.}~\bibnamefont {{Basu Thakur}}}, \bibinfo {author}
  {\bibfnamefont {A.~N.}\ \bibnamefont {{Bender}}}, \bibinfo {author}
  {\bibfnamefont {B.~A.}\ \bibnamefont {{Benson}}}, \bibinfo {author}
  {\bibfnamefont {J.~E.}\ \bibnamefont {{Carlstrom}}}, \bibinfo {author}
  {\bibfnamefont {F.~W.}\ \bibnamefont {{Carter}}}, \bibinfo {author}
  {\bibfnamefont {T.~W.}\ \bibnamefont {{Cecil}}}, \bibinfo {author}
  {\bibfnamefont {C.~L.}\ \bibnamefont {{Chang}}}, \bibinfo {author}
  {\bibfnamefont {J.~F.}\ \bibnamefont {{Cliche}}}, \bibinfo {author}
  {\bibfnamefont {A.}~\bibnamefont {{Cukierman}}}, \bibinfo {author}
  {\bibfnamefont {T.}~\bibnamefont {{de Haan}}}, \bibinfo {author}
  {\bibfnamefont {J.}~\bibnamefont {{Ding}}}, \bibinfo {author} {\bibfnamefont
  {M.~A.}\ \bibnamefont {{Dobbs}}}, \bibinfo {author} {\bibfnamefont
  {D.}~\bibnamefont {{Dutcher}}}, \bibinfo {author} {\bibfnamefont
  {W.}~\bibnamefont {{Everett}}}, \bibinfo {author} {\bibfnamefont
  {A.}~\bibnamefont {{Foster}}}, \bibinfo {author} {\bibfnamefont
  {J.}~\bibnamefont {{Gallichio}}}, \bibinfo {author} {\bibfnamefont
  {A.}~\bibnamefont {{Gilbert}}}, \bibinfo {author} {\bibfnamefont {J.~C.}\
  \bibnamefont {{Groh}}}, \bibinfo {author} {\bibfnamefont {S.~T.}\
  \bibnamefont {{Guns}}}, \bibinfo {author} {\bibfnamefont {N.~W.}\
  \bibnamefont {{Halverson}}}, \bibinfo {author} {\bibfnamefont {A.~H.}\
  \bibnamefont {{Harke-Hosemann}}}, \bibinfo {author} {\bibfnamefont {N.~L.}\
  \bibnamefont {{Harrington}}}, \bibinfo {author} {\bibfnamefont {J.~W.}\
  \bibnamefont {{Henning}}}, \bibinfo {author} {\bibfnamefont {W.~L.}\
  \bibnamefont {{Holzapfel}}}, \bibinfo {author} {\bibfnamefont
  {N.}~\bibnamefont {{Huang}}}, \bibinfo {author} {\bibfnamefont {K.~D.}\
  \bibnamefont {{Irwin}}}, \bibinfo {author} {\bibfnamefont {O.~B.}\
  \bibnamefont {{Jeong}}}, \bibinfo {author} {\bibfnamefont {M.}~\bibnamefont
  {{Jonas}}}, \bibinfo {author} {\bibfnamefont {T.~S.}\ \bibnamefont
  {{Khaire}}}, \bibinfo {author} {\bibfnamefont {A.~M.}\ \bibnamefont
  {{Kofman}}}, \bibinfo {author} {\bibfnamefont {M.}~\bibnamefont {{Korman}}},
  \bibinfo {author} {\bibfnamefont {D.~L.}\ \bibnamefont {{Kubik}}}, \bibinfo
  {author} {\bibfnamefont {S.}~\bibnamefont {{Kuhlmann}}}, \bibinfo {author}
  {\bibfnamefont {C.~L.}\ \bibnamefont {{Kuo}}}, \bibinfo {author}
  {\bibfnamefont {A.~T.}\ \bibnamefont {{Lee}}}, \bibinfo {author}
  {\bibfnamefont {A.~E.}\ \bibnamefont {{Lowitz}}}, \bibinfo {author}
  {\bibfnamefont {S.~S.}\ \bibnamefont {{Meyer}}}, \bibinfo {author}
  {\bibfnamefont {D.}~\bibnamefont {{Michalik}}}, \bibinfo {author}
  {\bibfnamefont {J.}~\bibnamefont {{Montgomery}}}, \bibinfo {author}
  {\bibfnamefont {A.}~\bibnamefont {{Nadolski}}}, \bibinfo {author}
  {\bibfnamefont {T.}~\bibnamefont {{Natoli}}}, \bibinfo {author}
  {\bibfnamefont {H.}~\bibnamefont {{Nguyen}}}, \bibinfo {author}
  {\bibfnamefont {G.~I.}\ \bibnamefont {{Noble}}}, \bibinfo {author}
  {\bibfnamefont {V.}~\bibnamefont {{Novosad}}}, \bibinfo {author}
  {\bibfnamefont {S.}~\bibnamefont {{Padin}}}, \bibinfo {author} {\bibfnamefont
  {Z.}~\bibnamefont {{Pan}}}, \bibinfo {author} {\bibfnamefont
  {J.}~\bibnamefont {{Pearson}}}, \bibinfo {author} {\bibfnamefont {C.~M.}\
  \bibnamefont {{Posada}}}, \bibinfo {author} {\bibfnamefont {W.}~\bibnamefont
  {{Quan}}}, \bibinfo {author} {\bibfnamefont {A.}~\bibnamefont {{Rahlin}}},
  \bibinfo {author} {\bibfnamefont {J.~E.}\ \bibnamefont {{Ruhl}}}, \bibinfo
  {author} {\bibfnamefont {J.~T.}\ \bibnamefont {{Sayre}}}, \bibinfo {author}
  {\bibfnamefont {E.}~\bibnamefont {{Shirokoff}}}, \bibinfo {author}
  {\bibfnamefont {G.}~\bibnamefont {{Smecher}}}, \bibinfo {author}
  {\bibfnamefont {A.~A.}\ \bibnamefont {{Stark}}}, \bibinfo {author}
  {\bibfnamefont {K.~T.}\ \bibnamefont {{Story}}}, \bibinfo {author}
  {\bibfnamefont {A.}~\bibnamefont {{Suzuki}}}, \bibinfo {author}
  {\bibfnamefont {K.~L.}\ \bibnamefont {{Thompson}}}, \bibinfo {author}
  {\bibfnamefont {C.}~\bibnamefont {{Tucker}}}, \bibinfo {author}
  {\bibfnamefont {K.}~\bibnamefont {{Vanderlinde}}}, \bibinfo {author}
  {\bibfnamefont {J.~D.}\ \bibnamefont {{Vieira}}}, \bibinfo {author}
  {\bibfnamefont {G.}~\bibnamefont {{Wang}}}, \bibinfo {author} {\bibfnamefont
  {N.}~\bibnamefont {{Whitehorn}}}, \bibinfo {author} {\bibfnamefont
  {V.}~\bibnamefont {{Yefremenko}}}, \bibinfo {author} {\bibfnamefont {K.~W.}\
  \bibnamefont {{Yoon}}}, \ and\ \bibinfo {author} {\bibfnamefont
  {M.}~\bibnamefont {{Young}}},\ }\bibfield  {title} {\enquote {\bibinfo
  {title} {{Design and characterization of the SPT-3G receiver}},}\ }in\ \href
  {\doibase 10.1117/12.2314366} {\emph {\bibinfo {booktitle} {Millimeter,
  Submillimeter, and Far-Infrared Detectors and Instrumentation for Astronomy
  IX}}},\ \bibinfo {series} {Society of Photo-Optical Instrumentation Engineers
  (SPIE) Conference Series}, Vol.\ \bibinfo {volume} {10708},\ \bibinfo
  {editor} {edited by\ \bibinfo {editor} {\bibfnamefont {J.}~\bibnamefont
  {{Zmuidzinas}}}\ and\ \bibinfo {editor} {\bibfnamefont {J.-R.}\ \bibnamefont
  {{Gao}}}}\ (\bibinfo {year} {2018})\ p.\ \bibinfo {pages} {107081H},\ \Eprint
  {http://arxiv.org/abs/1809.00032} {arXiv:1809.00032 [astro-ph.IM]}
  \BibitemShut {NoStop}%
\bibitem [{\citenamefont {{Ade}}\ \emph {et~al.}(2006)\citenamefont {{Ade}},
  \citenamefont {{Pisano}}, \citenamefont {{Tucker}},\ and\ \citenamefont
  {{Weaver}}}]{Ade2006}%
  \BibitemOpen
  \bibfield  {author} {\bibinfo {author} {\bibfnamefont {P.~A.~R.}\
  \bibnamefont {{Ade}}}, \bibinfo {author} {\bibfnamefont {G.}~\bibnamefont
  {{Pisano}}}, \bibinfo {author} {\bibfnamefont {C.}~\bibnamefont {{Tucker}}},
  \ and\ \bibinfo {author} {\bibfnamefont {S.}~\bibnamefont {{Weaver}}},\
  }\bibfield  {title} {\enquote {\bibinfo {title} {{A review of metal mesh
  filters}},}\ }in\ \href {\doibase 10.1117/12.673162} {\emph {\bibinfo
  {booktitle} {Millimeter and Submillimeter Detectors and Instrumentation for
  Astronomy III}}},\ \bibinfo {series} {Society of Photo-Optical
  Instrumentation Engineers (SPIE) Conference Series}, Vol.\ \bibinfo {volume}
  {6275},\ \bibinfo {editor} {edited by\ \bibinfo {editor} {\bibfnamefont
  {J.}~\bibnamefont {{Zmuidzinas}}}, \bibinfo {editor} {\bibfnamefont {W.~S.}\
  \bibnamefont {{Holland}}}, \bibinfo {editor} {\bibfnamefont {S.}~\bibnamefont
  {{Withington}}}, \ and\ \bibinfo {editor} {\bibfnamefont {W.~D.}\
  \bibnamefont {{Duncan}}}}\ (\bibinfo {year} {2006})\ p.\ \bibinfo {pages}
  {62750U}\BibitemShut {NoStop}%
\bibitem [{\citenamefont {Zmuidzinas}(2012)}]{zmuidzinas_2012}%
  \BibitemOpen
  \bibfield  {author} {\bibinfo {author} {\bibfnamefont {J.}~\bibnamefont
  {Zmuidzinas}},\ }\bibfield  {title} {\enquote {\bibinfo {title}
  {Superconducting microresonators: Physics and applications},}\ }\href
  {\doibase 10.1146/annurev-conmatphys-020911-125022} {\bibfield  {journal}
  {\bibinfo  {journal} {Annual Review of Condensed Matter Physics}\ }\textbf
  {\bibinfo {volume} {3}},\ \bibinfo {pages} {169–214} (\bibinfo {year}
  {2012})},\ \bibinfo {note} {funding by NASA}\BibitemShut {NoStop}%
\bibitem [{\citenamefont {{Gao}}\ \emph {et~al.}(2008)\citenamefont {{Gao}},
  \citenamefont {{Daal}}, \citenamefont {{Martinis}}, \citenamefont
  {{Vayonakis}}, \citenamefont {{Zmuidzinas}}, \citenamefont {{Sadoulet}},
  \citenamefont {{Mazin}}, \citenamefont {{Day}},\ and\ \citenamefont
  {{Leduc}}}]{Gao2008}%
  \BibitemOpen
  \bibfield  {author} {\bibinfo {author} {\bibfnamefont {J.}~\bibnamefont
  {{Gao}}}, \bibinfo {author} {\bibfnamefont {M.}~\bibnamefont {{Daal}}},
  \bibinfo {author} {\bibfnamefont {J.~M.}\ \bibnamefont {{Martinis}}},
  \bibinfo {author} {\bibfnamefont {A.}~\bibnamefont {{Vayonakis}}}, \bibinfo
  {author} {\bibfnamefont {J.}~\bibnamefont {{Zmuidzinas}}}, \bibinfo {author}
  {\bibfnamefont {B.}~\bibnamefont {{Sadoulet}}}, \bibinfo {author}
  {\bibfnamefont {B.~A.}\ \bibnamefont {{Mazin}}}, \bibinfo {author}
  {\bibfnamefont {P.~K.}\ \bibnamefont {{Day}}}, \ and\ \bibinfo {author}
  {\bibfnamefont {H.~G.}\ \bibnamefont {{Leduc}}},\ }\bibfield  {title}
  {\enquote {\bibinfo {title} {{A semiempirical model for two-level system
  noise in superconducting microresonators}},}\ }\href {\doibase
  10.1063/1.2937855} {\bibfield  {journal} {\bibinfo  {journal} {Applied
  Physics Letters}\ }\textbf {\bibinfo {volume} {92}},\ \bibinfo {eid} {212504}
  (\bibinfo {year} {2008})},\ \Eprint {http://arxiv.org/abs/0804.0467}
  {arXiv:0804.0467 [cond-mat.supr-con]} \BibitemShut {NoStop}%
\bibitem [{\citenamefont {{Sayers}}\ \emph {et~al.}(2010)\citenamefont
  {{Sayers}}, \citenamefont {{Golwala}}, \citenamefont {{Ade}}, \citenamefont
  {{Aguirre}}, \citenamefont {{Bock}}, \citenamefont {{Edgington}},
  \citenamefont {{Glenn}}, \citenamefont {{Goldin}}, \citenamefont {{Haig}},
  \citenamefont {{Lange}}, \citenamefont {{Laurent}}, \citenamefont
  {{Mauskopf}}, \citenamefont {{Nguyen}}, \citenamefont {{Rossinot}},\ and\
  \citenamefont {{Schlaerth}}}]{Sayers2010}%
  \BibitemOpen
  \bibfield  {author} {\bibinfo {author} {\bibfnamefont {J.}~\bibnamefont
  {{Sayers}}}, \bibinfo {author} {\bibfnamefont {S.~R.}\ \bibnamefont
  {{Golwala}}}, \bibinfo {author} {\bibfnamefont {P.~A.~R.}\ \bibnamefont
  {{Ade}}}, \bibinfo {author} {\bibfnamefont {J.~E.}\ \bibnamefont
  {{Aguirre}}}, \bibinfo {author} {\bibfnamefont {J.~J.}\ \bibnamefont
  {{Bock}}}, \bibinfo {author} {\bibfnamefont {S.~F.}\ \bibnamefont
  {{Edgington}}}, \bibinfo {author} {\bibfnamefont {J.}~\bibnamefont
  {{Glenn}}}, \bibinfo {author} {\bibfnamefont {A.}~\bibnamefont {{Goldin}}},
  \bibinfo {author} {\bibfnamefont {D.}~\bibnamefont {{Haig}}}, \bibinfo
  {author} {\bibfnamefont {A.~E.}\ \bibnamefont {{Lange}}}, \bibinfo {author}
  {\bibfnamefont {G.~T.}\ \bibnamefont {{Laurent}}}, \bibinfo {author}
  {\bibfnamefont {P.~D.}\ \bibnamefont {{Mauskopf}}}, \bibinfo {author}
  {\bibfnamefont {H.~T.}\ \bibnamefont {{Nguyen}}}, \bibinfo {author}
  {\bibfnamefont {P.}~\bibnamefont {{Rossinot}}}, \ and\ \bibinfo {author}
  {\bibfnamefont {J.}~\bibnamefont {{Schlaerth}}},\ }\bibfield  {title}
  {\enquote {\bibinfo {title} {{Studies of Millimeter-wave Atmospheric Noise
  above Mauna Kea}},}\ }\href {\doibase 10.1088/0004-637X/708/2/1674}
  {\bibfield  {journal} {\bibinfo  {journal} {\apj}\ }\textbf {\bibinfo
  {volume} {708}},\ \bibinfo {pages} {1674--1691} (\bibinfo {year} {2010})},\
  \Eprint {http://arxiv.org/abs/0904.3943} {arXiv:0904.3943 [astro-ph.IM]}
  \BibitemShut {NoStop}%
\bibitem [{\citenamefont {{Errard}}\ \emph {et~al.}(2015)\citenamefont
  {{Errard}}, \citenamefont {{Ade}}, \citenamefont {{Akiba}}, \citenamefont
  {{Arnold}}, \citenamefont {{Atlas}}, \citenamefont {{Baccigalupi}},
  \citenamefont {{Barron}}, \citenamefont {{Boettger}}, \citenamefont
  {{Borrill}}, \citenamefont {{Chapman}}, \citenamefont {{Chinone}},
  \citenamefont {{Cukierman}}, \citenamefont {{Delabrouille}}, \citenamefont
  {{Dobbs}}, \citenamefont {{Ducout}}, \citenamefont {{Elleflot}},
  \citenamefont {{Fabbian}}, \citenamefont {{Feng}}, \citenamefont {{Feeney}},
  \citenamefont {{Gilbert}}, \citenamefont {{Goeckner-Wald}}, \citenamefont
  {{Halverson}}, \citenamefont {{Hasegawa}}, \citenamefont {{Hattori}},
  \citenamefont {{Hazumi}}, \citenamefont {{Hill}}, \citenamefont
  {{Holzapfel}}, \citenamefont {{Hori}}, \citenamefont {{Inoue}}, \citenamefont
  {{Jaehnig}}, \citenamefont {{Jaffe}}, \citenamefont {{Jeong}}, \citenamefont
  {{Katayama}}, \citenamefont {{Kaufman}}, \citenamefont {{Keating}},
  \citenamefont {{Kermish}}, \citenamefont {{Keskitalo}}, \citenamefont
  {{Kisner}}, \citenamefont {{Le Jeune}}, \citenamefont {{Lee}}, \citenamefont
  {{Leitch}}, \citenamefont {{Leon}}, \citenamefont {{Linder}}, \citenamefont
  {{Matsuda}}, \citenamefont {{Matsumura}}, \citenamefont {{Miller}},
  \citenamefont {{Myers}}, \citenamefont {{Navaroli}}, \citenamefont
  {{Nishino}}, \citenamefont {{Okamura}}, \citenamefont {{Paar}}, \citenamefont
  {{Peloton}}, \citenamefont {{Poletti}}, \citenamefont {{Puglisi}},
  \citenamefont {{Rebeiz}}, \citenamefont {{Reichardt}}, \citenamefont
  {{Richards}}, \citenamefont {{Ross}}, \citenamefont {{Rotermund}},
  \citenamefont {{Schenck}}, \citenamefont {{Sherwin}}, \citenamefont
  {{Siritanasak}}, \citenamefont {{Smecher}}, \citenamefont {{Stebor}},
  \citenamefont {{Steinbach}}, \citenamefont {{Stompor}}, \citenamefont
  {{Suzuki}}, \citenamefont {{Tajima}}, \citenamefont {{Takakura}},
  \citenamefont {{Tikhomirov}}, \citenamefont {{Tomaru}}, \citenamefont
  {{Whitehorn}}, \citenamefont {{Wilson}}, \citenamefont {{Yadav}},\ and\
  \citenamefont {{Zahn}}}]{Errard2015}%
  \BibitemOpen
  \bibfield  {author} {\bibinfo {author} {\bibfnamefont {J.}~\bibnamefont
  {{Errard}}}, \bibinfo {author} {\bibfnamefont {P.~A.~R.}\ \bibnamefont
  {{Ade}}}, \bibinfo {author} {\bibfnamefont {Y.}~\bibnamefont {{Akiba}}},
  \bibinfo {author} {\bibfnamefont {K.}~\bibnamefont {{Arnold}}}, \bibinfo
  {author} {\bibfnamefont {M.}~\bibnamefont {{Atlas}}}, \bibinfo {author}
  {\bibfnamefont {C.}~\bibnamefont {{Baccigalupi}}}, \bibinfo {author}
  {\bibfnamefont {D.}~\bibnamefont {{Barron}}}, \bibinfo {author}
  {\bibfnamefont {D.}~\bibnamefont {{Boettger}}}, \bibinfo {author}
  {\bibfnamefont {J.}~\bibnamefont {{Borrill}}}, \bibinfo {author}
  {\bibfnamefont {S.}~\bibnamefont {{Chapman}}}, \bibinfo {author}
  {\bibfnamefont {Y.}~\bibnamefont {{Chinone}}}, \bibinfo {author}
  {\bibfnamefont {A.}~\bibnamefont {{Cukierman}}}, \bibinfo {author}
  {\bibfnamefont {J.}~\bibnamefont {{Delabrouille}}}, \bibinfo {author}
  {\bibfnamefont {M.}~\bibnamefont {{Dobbs}}}, \bibinfo {author} {\bibfnamefont
  {A.}~\bibnamefont {{Ducout}}}, \bibinfo {author} {\bibfnamefont
  {T.}~\bibnamefont {{Elleflot}}}, \bibinfo {author} {\bibfnamefont
  {G.}~\bibnamefont {{Fabbian}}}, \bibinfo {author} {\bibfnamefont
  {C.}~\bibnamefont {{Feng}}}, \bibinfo {author} {\bibfnamefont
  {S.}~\bibnamefont {{Feeney}}}, \bibinfo {author} {\bibfnamefont
  {A.}~\bibnamefont {{Gilbert}}}, \bibinfo {author} {\bibfnamefont
  {N.}~\bibnamefont {{Goeckner-Wald}}}, \bibinfo {author} {\bibfnamefont
  {N.~W.}\ \bibnamefont {{Halverson}}}, \bibinfo {author} {\bibfnamefont
  {M.}~\bibnamefont {{Hasegawa}}}, \bibinfo {author} {\bibfnamefont
  {K.}~\bibnamefont {{Hattori}}}, \bibinfo {author} {\bibfnamefont
  {M.}~\bibnamefont {{Hazumi}}}, \bibinfo {author} {\bibfnamefont
  {C.}~\bibnamefont {{Hill}}}, \bibinfo {author} {\bibfnamefont {W.~L.}\
  \bibnamefont {{Holzapfel}}}, \bibinfo {author} {\bibfnamefont
  {Y.}~\bibnamefont {{Hori}}}, \bibinfo {author} {\bibfnamefont
  {Y.}~\bibnamefont {{Inoue}}}, \bibinfo {author} {\bibfnamefont {G.~C.}\
  \bibnamefont {{Jaehnig}}}, \bibinfo {author} {\bibfnamefont {A.~H.}\
  \bibnamefont {{Jaffe}}}, \bibinfo {author} {\bibfnamefont {O.}~\bibnamefont
  {{Jeong}}}, \bibinfo {author} {\bibfnamefont {N.}~\bibnamefont {{Katayama}}},
  \bibinfo {author} {\bibfnamefont {J.}~\bibnamefont {{Kaufman}}}, \bibinfo
  {author} {\bibfnamefont {B.}~\bibnamefont {{Keating}}}, \bibinfo {author}
  {\bibfnamefont {Z.}~\bibnamefont {{Kermish}}}, \bibinfo {author}
  {\bibfnamefont {R.}~\bibnamefont {{Keskitalo}}}, \bibinfo {author}
  {\bibfnamefont {T.}~\bibnamefont {{Kisner}}}, \bibinfo {author}
  {\bibfnamefont {M.}~\bibnamefont {{Le Jeune}}}, \bibinfo {author}
  {\bibfnamefont {A.~T.}\ \bibnamefont {{Lee}}}, \bibinfo {author}
  {\bibfnamefont {E.~M.}\ \bibnamefont {{Leitch}}}, \bibinfo {author}
  {\bibfnamefont {D.}~\bibnamefont {{Leon}}}, \bibinfo {author} {\bibfnamefont
  {E.}~\bibnamefont {{Linder}}}, \bibinfo {author} {\bibfnamefont
  {F.}~\bibnamefont {{Matsuda}}}, \bibinfo {author} {\bibfnamefont
  {T.}~\bibnamefont {{Matsumura}}}, \bibinfo {author} {\bibfnamefont {N.~J.}\
  \bibnamefont {{Miller}}}, \bibinfo {author} {\bibfnamefont {M.~J.}\
  \bibnamefont {{Myers}}}, \bibinfo {author} {\bibfnamefont {M.}~\bibnamefont
  {{Navaroli}}}, \bibinfo {author} {\bibfnamefont {H.}~\bibnamefont
  {{Nishino}}}, \bibinfo {author} {\bibfnamefont {T.}~\bibnamefont
  {{Okamura}}}, \bibinfo {author} {\bibfnamefont {H.}~\bibnamefont {{Paar}}},
  \bibinfo {author} {\bibfnamefont {J.}~\bibnamefont {{Peloton}}}, \bibinfo
  {author} {\bibfnamefont {D.}~\bibnamefont {{Poletti}}}, \bibinfo {author}
  {\bibfnamefont {G.}~\bibnamefont {{Puglisi}}}, \bibinfo {author}
  {\bibfnamefont {G.}~\bibnamefont {{Rebeiz}}}, \bibinfo {author}
  {\bibfnamefont {C.~L.}\ \bibnamefont {{Reichardt}}}, \bibinfo {author}
  {\bibfnamefont {P.~L.}\ \bibnamefont {{Richards}}}, \bibinfo {author}
  {\bibfnamefont {C.}~\bibnamefont {{Ross}}}, \bibinfo {author} {\bibfnamefont
  {K.~M.}\ \bibnamefont {{Rotermund}}}, \bibinfo {author} {\bibfnamefont
  {D.~E.}\ \bibnamefont {{Schenck}}}, \bibinfo {author} {\bibfnamefont {B.~D.}\
  \bibnamefont {{Sherwin}}}, \bibinfo {author} {\bibfnamefont {P.}~\bibnamefont
  {{Siritanasak}}}, \bibinfo {author} {\bibfnamefont {G.}~\bibnamefont
  {{Smecher}}}, \bibinfo {author} {\bibfnamefont {N.}~\bibnamefont {{Stebor}}},
  \bibinfo {author} {\bibfnamefont {B.}~\bibnamefont {{Steinbach}}}, \bibinfo
  {author} {\bibfnamefont {R.}~\bibnamefont {{Stompor}}}, \bibinfo {author}
  {\bibfnamefont {A.}~\bibnamefont {{Suzuki}}}, \bibinfo {author}
  {\bibfnamefont {O.}~\bibnamefont {{Tajima}}}, \bibinfo {author}
  {\bibfnamefont {S.}~\bibnamefont {{Takakura}}}, \bibinfo {author}
  {\bibfnamefont {A.}~\bibnamefont {{Tikhomirov}}}, \bibinfo {author}
  {\bibfnamefont {T.}~\bibnamefont {{Tomaru}}}, \bibinfo {author}
  {\bibfnamefont {N.}~\bibnamefont {{Whitehorn}}}, \bibinfo {author}
  {\bibfnamefont {B.}~\bibnamefont {{Wilson}}}, \bibinfo {author}
  {\bibfnamefont {A.}~\bibnamefont {{Yadav}}}, \ and\ \bibinfo {author}
  {\bibfnamefont {O.}~\bibnamefont {{Zahn}}},\ }\bibfield  {title} {\enquote
  {\bibinfo {title} {{Modeling Atmospheric Emission for CMB Ground-based
  Observations}},}\ }\href {\doibase 10.1088/0004-637X/809/1/63} {\bibfield
  {journal} {\bibinfo  {journal} {\apj}\ }\textbf {\bibinfo {volume} {809}},\
  \bibinfo {eid} {63} (\bibinfo {year} {2015})},\ \Eprint
  {http://arxiv.org/abs/1501.07911} {arXiv:1501.07911 [astro-ph.IM]}
  \BibitemShut {NoStop}%
\end{thebibliography}%

\end{document}